\documentclass[a4paper,11pt]{article}
\pdfoutput=1

\usepackage{jheppub}

\usepackage[T1]{fontenc}

\usepackage{amsmath}
\usepackage{amssymb}
\usepackage{amsthm}
\usepackage{mathrsfs}
\usepackage{physics}
\usepackage{bbm}
\usepackage{enumitem}
\usepackage{cancel}

\newtheorem{theorem}{Theorem}
\newtheorem{lemma}{Lemma}
\newtheorem*{lemma*}{Lemma}

\newtheorem{corollary}{Corollary}

\newtheorem{proposition}{Proposition}
\newtheorem*{proposition*}{Proposition}
\theoremstyle{definition}
\newtheorem{definition}{Definition}

\theoremstyle{remark}
\newtheorem{remark}{Remark}
\usepackage{tikz}

\title{\boldmath Proof of the QNEC, without assumptions and under a weight} 

\author{Thomas Faulkner}

\affiliation{Department of Physics, University of Illinois Urbana-Champaign}

\emailAdd{tomf@illinois.edu}
\abstract{
We complete a proof of the quantum null energy condition (QNEC) in the mathematical framework of Half-Sided Translations and Modular Inclusions (HST/HSMI) 
removing an assumption that applied to previous proofs.
Here we establish a theorem, which when translated to a physical setting implies that
 the relative entropy comparing the state of fields outside a stationary black hole 
with a Hartle-Hawking (HH) state is always a convex function of the horizon cut.  
Previous proofs only applied when a
purification
with finite null energy was available.
We also obtain a QNEC generalization in which the HH Hilbert space is replaced with one based on a weight which is plausibly of relevance to evaporating black holes in flat space.
}

\begin{document}

\maketitle
\flushbottom

\section{Introduction}
\label{sec:intro}

The quantum null energy condition (QNEC)~\cite{Bousso:2015wca,Koeller:2015qmn,Balakrishnan:2017bjg}, a quantum corrected bound on the null energy density, can be precisely studied using an operator algebra structure
called a Half-Sided Translation (HST) which arises from a corresponding structure called a Half-Sided Modular Inclusion (HSMI) \cite{wiesbrock1993half,borchers1992cpt,araki2005extension}.
 In this structure there is a one parameter family of nested von Neumann algebras acting on the same Hilbert space $\mathscr{H}$, defined via an order reversing map
from $\mathbb{R} \ni V \rightarrow \mathcal{A}_V \subset \mathcal{B}(\mathscr{H})$.
Furthermore, these algebras are translates of each other $V \rightarrow V+a$ under the adjoint action of a unitary $U(a) = e^{-i a P}$
with positive generator $P$.  They also admit a common cyclic and separating vector $\eta \in \mathscr{H}$ that is annihilated by $P$, allowing for
a relation to the modular data of Tomita-Takesaki theory.
A physically motivated quantity to study is the relative entropy~\cite{araki1976relative,araki1977relative} with respect to $\eta$ restricted
to $\mathcal{A}_V$:
\begin{equation}
\label{eq:intro-D}
D(\rho|_V) \equiv  D(\rho|_{\mathcal{A}_V} | \omega_{\eta}|_{\mathcal{A}_V})
\end{equation}
and the QNEC is the statement that $D(\rho|_V)$ is a convex function of $V$ for all states $\rho$ from $\mathcal{B}(\mathscr{H})_\star$. 
In this paper our first task will be to prove convexity without further assumptions. 
 We use essentially the same proof strategy as in \cite{Ceyhan:2018zfg}, taking the opportunity to streamline, simplify and improve some of the intermediate steps, along with the introduction of an additional approximation scheme (Theorem~\ref{thm:fullapprox}) that removes the assumed existence of a purification with finite $P$ expectation. A more recent proof \cite{Hollands:2025glm} made more restrictive assumptions on the state
 in question allowing for a slick derivation of the QNEC. However, as is the case for both \cite{Ceyhan:2018zfg} and \cite{Hollands:2025glm}, a full proof cannot be claimed until these assumptions are removed.\footnote{Both proofs work for a dense set of states; however, since relative entropy is an unbounded quantity, this does not imply the inequality for all states. Otherwise of course this would have then been the claim in those papers.}
For other developments, explicit calculations and consequences of the QNEC see an incomplete set of papers \cite{Leichenauer:2018obf,qnecsat,Hollands:2019coherent,Hollands:2025modular,Longo:2018obd,Longo:2019coherent,Ciolli:2019wave,Ciolli:2021curved,Morinelli:2021nullplane,
Mezei:2019sla,Kibe:2021qjy,Kibe:2024icu,Casini:2023kyj,
Lashkari:2018nsl,Moosa:2020jwt,Roy:2022RenyiQNEC,Kibe:2026wsg,Kibe:2026offdiagonal,Kibe:2026idb,faulkner-renyi}.

Convexity goes beyond the well-known monotonicity of relative entropy~\cite{uhlmann1977relative,ohya-petz}, and thus a natural question arises: under what
conditions should convexity along a one parameter family of algebras be expected? Perhaps the HSMI setting is the only possibility. In order to probe this question further we should try to generalize these results as much as possible. The other main goal of this paper
is to prove a QNEC result using a generalization of the HST/HSMI structure where the faithful state $\omega_\eta(\cdot)$ 
on the family $\mathcal{A}_V$ is replaced by a normal semi-finite faithful (nsf) weight $\varphi$, where its GNS representation is appropriately (for a weight) cyclic and separating on the family $\mathcal{A}_V$. 

This generalization
of the HSMI structure has been known since the work of Araki-Zsido~\cite{araki2005extension}. In this paper we will introduce a HST structure for weights which turns out to be slightly more general than the HSMI structure, in contrast to the state case. The HST is the appropriate one for studying the QNEC. 

Physical motivation for the HST/HSMI structure for weights has only started to become apparent.
In particular weights have naturally arisen in the context of semiclassical gravity in black hole and cosmological backgrounds~\cite{Witten:2021unn,Chandrasekaran:2022cip,Chandrasekaran:2022eqq,Faulkner:2024gst,Speranza:2025joj,Cao:2025els}, where they play an essential physical role.

For example a weight naturally arises in the Hilbert space of an evaporating black hole in flat space
when restricting to the algebra of operators outside the black hole. In this setup one neglects back-reaction treating only QFT on the fixed fully extended Schwarzschild background.
The quantum fields act on a Hilbert space disjoint from the one containing the Hartle-Hawking state since all normalizable states are out of equilibrium. The existence of a weight whose modular flow is the isometry of this spacetime was argued for in \cite{Chen:2024rpx,Chen:2026zxc}. A physical argument can also be made that this weight is semifinite on the subalgebras $\mathcal{A}_V$. Together with geometric modular flow and additivity, this would allow the Araki-Zsido HSMI structure theorems to apply.
We leave this more physical argument to the conclusion Section~\ref{sec:conclusions} since it lies outside of the main tone of this paper. 

We now present our main results. In order to study properties of the relative entropy in the HSMI setting it is useful to introduce: 
\begin{equation}
\label{eq:intro-theta}
\Theta(\rho |_V) = \inf_{ \phi \in \mathscr{H}: \omega_\phi|_V = \rho|_V} \big\| \sqrt{P} \left| \phi \right> \big\|^2
\end{equation}
where we assign the value $\big\| \sqrt{P} \left| \phi \right> \big\| = \infty$ when $\phi \notin \mathcal{D}(\sqrt{P})$. The notation suggests this quantity is related to the quantum version of the ``null expansion'' and in fact
it has analogous properties. It is possible to relate $\Theta$ to the derivative of $D$ and this implies the QNEC.

We distill this into the first main theorem. Define the asymptotic relative entropy $D_\infty(\rho) = \lim_{V \rightarrow \infty} D(\rho|_V)$, where the limit always
exists by monotonicity of relative entropy.\footnote{If this is infinite then the QNEC
is trivial in the sense that $D(\rho|_V) = \infty$ is convex as a function valued in the extended real line.} Otherwise:
\begin{theorem}
\label{thm:mainthm}
Consider a normalized state $\rho \in (\mathcal{B}(\mathscr{H}))_{\star}$ with $D_\infty(\rho) < \infty$. Then for all $V$:
\begin{enumerate}
\item $V \rightarrow D(\rho|_{V}) \in \mathbb{R}_{\geq 0} \cup \{ \infty\}$
is a monotonically non-increasing, convex, right-continuous function. 
On open sets where $D(\rho|_V) < \infty$ this function is continuously differentiable.
\item $V \rightarrow \Theta(\rho|_{V}) \in \mathbb{R}_{\geq 0} \cup \{ \infty\}$ is a monotonically non-increasing, right-continuous function of $V$. On open sets where $\Theta(\rho|_V) < \infty$ this function is continuous.
\item They satisfy the integral relation with $V_1 < V_2$:
\begin{equation}
\label{eq:mainintegral}
D(\rho|_{V_1})= D(\rho|_{V_2})
+ 2\pi \int_{V_1}^{V_2} d V\Theta(\rho|_{V})
\end{equation}
where both sides can be infinite. 
\end{enumerate} 
\end{theorem}

This theorem was established in \cite{Ceyhan:2018zfg} with the additional assumption that $\Theta(\rho|_{V_0}) < \infty$
for some $V_0$ after which we established the QNEC for all $V > V_0$. 
Here we remove this assumption, and as a corollary show that
if $D(\rho|_{V_0}) < \infty$ then in fact $\Theta(\rho|_V) < \infty$ for all $V > V_0$, see Lemma~\ref{lem:infinity}. It follows that any possible issue with the assumption in \cite{Ceyhan:2018zfg}
happens at most at one cut with finite relative entropy for any given state $\rho$.

The weight case, Theorem~\ref{thm:mainthmweight}, is a similar result to Theorem~\ref{thm:mainthm} when $\Theta(\rho|_{V_0})$ is assumed finite. Otherwise, for general states, Theorem~\ref{thm:weight-convexity} shows that we retain convexity on intervals where $D(\rho|_{V})$ is finite, but we lose control of $\Theta$.

The outline of this paper is as follows. We start with some background explaining HSMI and some preliminary results
on $\Theta$ in Section~\ref{sec:prelim}. We then revisit the proof of the QNEC and strengthen it in Section~\ref{sec:streamline}.
Section~\ref{sec:genweight} is dedicated to the weight generalization where we also derive several new results
on the relative entropy with respect to a weight that might be of independent interest. We note that this part of the paper is considerably more technically
involved than for the vanilla QNEC, simply due to the difficulties of working with a relative entropy which is now unbounded
above and below.  We conclude in Section~\ref{sec:conclusions}. Several proofs and details are left to the Appendices. 

\emph{Comment on notation:} Throughout the paper we use a streamlined notation where often $\mathcal{A}_V$ is simply denoted as $V$ and $\mathcal{A}_V'$
as $V'$ (without loss of generality we choose Haag dual families so this notation makes sense). That is, 
 the vector and the induced states are interchangeable
via the usual relations. For example $\rho|_V \equiv \rho|_{\mathcal{A}_V}$ where $\rho$ is assumed to be a normal
state for a larger subregion and $\psi|_V \equiv \omega_{\psi}|_V$ for the positive normal linear functional $\omega_\psi(\cdot) = \left< \psi \right| \cdot \left| \psi \right>$.

\section{Some preliminaries}
\label{sec:prelim}

We start with a summary of the HSMI structure in Section~\ref{sec:hsmi}, then we remind the reader of some essential
results in modular theory in Section~\ref{sec:states} including the fact that for different states both modular flow and modular conjugation are equivalent up to inner automorphisms. Finally in Section~\ref{sec:theta} we prove various properties satisfied by $\Theta(\rho)$.

\subsection{Half sided translations and modular inclusions - basics}
\label{sec:hsmi}

We start by explaining the structure of a HST, which is equivalent to the HSMI structure, but in a form
directly applicable to the QNEC discussion. We quote a basic theorem that establishes equivalent conditions leading to HSMIs:
\begin{theorem}
\label{thm:hsmi}
Consider a von Neumann algebra $\mathcal{A} \subset \mathcal{B}(\mathscr{H})$ in standard
form $(\mathcal{A}, \mathscr{H}, J, \mathscr{P}^\natural)$ with cyclic and separating $\eta \in \mathscr{P}^\natural$, normalized as $\| \eta \| = 1$, and where $\mathscr{H}$ furnishes a strongly continuous unitary representation of $\mathbb{R}$, that is $U(a)$ for $a \in \mathbb{R}$, satisfying:
\begin{equation}
\label{eq:Uainclusion}
U(a)^\dagger \mathcal{A} U(a) \subset \mathcal{A} \qquad a > 0
\end{equation}
Consider the following conditions:
\begin{itemize}
    \item[(a)] $U(a) \eta = \eta$
\item[(b)] The generator of $U(a) = e^{- i a P}$ is positive $P \geq 0$ 
\item[(c)] $\Delta^{-is} U(a) \Delta^{is} = U( a e^{2\pi s}), \,J U(a) J = U(-a)$
\end{itemize}
where $J,\Delta$ are the  Tomita-Takesaki modular data for $\eta$ and $\mathcal{A}$. 
Then $(a)+(b) \iff (c)$. 
\end{theorem}
\begin{proof}
See \cite{wiesbrock1993half,borchers1992cpt,Florig:1998ne}. We will also provide a proof of the generalization of this theorem to the weight case, Theorem~\ref{thm:hsmi-weight}, proved in Appendix~\ref{app:hsmi-proof}, of which the above is a special case.
\end{proof}

\begin{definition}
    \label{def:hst}
The space of $P$'s satisfying the assumptions of the above theorem will be defined as $\mathcal{TR}(\mathcal{A},\sigma)$
where $\sigma \in  \mathcal{A}_{\star}^{+}$ is faithful, with representative $\eta \in \mathscr{P}^\natural$ in the standard form. 
\end{definition}

This structure is intrinsic to the subalgebra along with a joint modular structure which is the usual context for the HSMI structure:
\begin{definition}
We say that $\mathcal{B} \subset_\sigma \mathcal{A}$ forms a HSMI for faithful $\sigma \in \mathcal{A}_{\star}^{+}$
if modular flow satisfies $\sigma_{\eta;\mathcal{A}}^s(\mathcal{B}) \subset \mathcal{B}$ for all $s \geq 0$ and $\eta \in \mathscr{P}^\natural_\mathcal{A}$ (with $\| \eta \| = 1$),
representing $\sigma = \omega_\eta|_{\mathcal{A}}$, is cyclic for $\mathcal{B}$.
The space of such von Neumann subalgebras will be denoted as $\mathcal{HS}(\mathcal{A},\sigma)$
and in short, we say that  $\mathcal{B} \in \mathcal{HS}(\mathcal{A},\sigma)$ forms a HSMI. 
\end{definition}

\begin{remark}
It is a well-known theorem due to Wiesbrock \cite{wiesbrock1993half}, see~\cite{borchers1997lattice} that the space of translations $\mathcal{TR}(\mathcal{A},\sigma)$ is bijectively equivalent
to $\mathcal{HS}(\mathcal{A},\sigma)$ via $\mathcal{B} =U(-1) \mathcal{A}U(1)$ and the inverse $2\pi P = \overline{\ln \Delta_{\eta;\mathcal{B}} - \ln \Delta_{\eta;\mathcal{A}}}$.
As such it is common to refer to these structures collectively as HSMIs. However, as we will see in Section~\ref{sec:weighthsmi}, this is no longer true for half sided translations and modular inclusions with respect to a weight.
\end{remark}

We define the one parameter family of von Neumann algebras $\mathcal{A}_V = U(V)^\dagger \mathcal{A} U(V)$
for all $V \in \mathbb{R}$ and these have a joint cyclic and separating vector $\eta \in \mathscr{H}$. Translations then act on
the algebras as $U(a)^\dagger \mathcal{A}_V U(a) = \mathcal{A}_{V+a} \subset \mathcal{A}_V$ for $a > 0$.
The algebra in (c) of Theorem~\ref{thm:hsmi} is satisfied for any cut:
\begin{equation}
\label{eq:hsmi-alg1}
\Delta_{V}^{-is} U(a) \Delta_{V}^{is}
=   U( e^{2\pi s}a)\,,  \qquad J_V U(a) J_V = U(-a)
\end{equation}
where $J_V, \Delta_{V}$ are the modular data for $\mathcal{A}_V$
and $\eta$. Translating the modular operators via their definition on the algebra gives:
\begin{equation}
 U(a)^\dagger J_V  U(a) = J_{V+a} \,, \qquad U(a)^\dagger \Delta_V^{1/2}  U(a) = \Delta_{V+a}^{1/2}
\end{equation}
since $\eta$ is invariant under translations. Combined with \eqref{eq:hsmi-alg1} this leads to the identities:
\begin{equation}
\label{eq:hsmi-alg2}
J_{V_1} J_{V_2} = U(2(V_2-V_1))\,, \qquad \Delta_{V_1}^{-is} \Delta_{V_2}^{is} = U((1-e^{2\pi s}) (V_2-V_1) )
\end{equation}
for all $V_{1,2} \in \mathbb{R}$. We define modular flow with respect to $\omega_\eta|_V$ and modular conjugation as:\footnote{For those interested in keeping track of signs
we use physicist conventions for the half sided translations and modular flow.}
\begin{equation}
\sigma_{V}^{s}(\cdot) = \Delta_{V}^{-is}( \cdot )\Delta_{V}^{is} : \mathcal{A}_V \rightarrow \mathcal{A}_V
\,, \qquad j_V( \cdot) = J_V( \cdot )J_V : \mathcal{A}_V \rightarrow \mathcal{A}_V'
\end{equation}

Defining $\mathcal{A}_{V'} = \mathcal{A}_V'$ then these also form a half-sided modular inclusion in the opposite direction
with the same positive generator $P$ satisfying $U(a)^\dagger \mathcal{A}_{V'} U(a) =\mathcal{A}_{(V+a)'} \subset \mathcal{A}_{V'}$ for $a < 0$. 
And it is useful to recall the following identities:
\begin{equation}
J_{V'} = J_V \qquad \Delta_{V'}^{1/2} = \Delta_{V}^{-1/2}
\end{equation}
where the left hand side is the modular data for $\mathcal{A}_V'$ and $\eta$. 

We will give a general treatment of the QNEC based only on the existence of a half-sided modular inclusion. 
Further refinements of this structure are possible and well studied, but none of these are necessary for our results. 

\subsection{States, cocycles and the natural cone}

\label{sec:states}

In this section we consider states and cocycles for the algebra $\mathcal{A}$ (and its commutant). All the results also apply to the algebras $\mathcal{A}_V$ with appropriate changes to the notation.

Recall that the natural self dual cone can be constructed given some cyclic and separating $\eta$:
\begin{equation}
\mathscr{P}^\natural = \overline{ \Delta_{\eta}^{1/4} \mathcal{A}^+ \left| \eta \right> }  
\end{equation}
where $\mathcal{A}^+$ denotes the non-negative operators in $\mathcal{A}$, and so the standard form $(\mathcal{A}, \mathscr{H}, J,\mathscr{P}^\natural )$ can be constructed
from $\eta$. 
Note that the natural cone for the commutant algebra is the same cone:  $\mathscr{P}'^\natural = J  \mathscr{P}^\natural =  \mathscr{P}^\natural$.
It is well known that for any normal state $\rho$ on $\mathcal{A}$ there is a unique vector representative
denoted $\xi_{\rho} \in \mathscr{P}^\natural$.

In this section we will instead work with a fixed $\psi \in \mathscr{H}$ and for much of the notation $\psi$ will be implicit. 
There are partial isometries $\theta' \in \mathcal{A}', \theta \in \mathcal{A}$ that take $\psi$ to the canonical cone $\xi_{\omega_\psi|_{\mathcal{A}}} = \theta' \psi$
and the corresponding canonical cone vector for the commutant state $\xi_{\omega_\psi|_{\mathcal{A}'}} = \theta \psi$
which lives in the same cone but is a different vector. 
The $\theta, \theta'$ have final and initial support determined by the support projections for the corresponding states:
\begin{equation}
    \label{eq:thetasupport}
(\theta')^\dagger \theta' = \pi' \qquad  \theta' (\theta')^\dagger =  j( \pi) \equiv J (\pi) J
\end{equation} 
and similarly for $\theta$. That is the relevant support projections are defined as:
\begin{equation}
\pi = \overline{\mathcal{A}' \psi} \in \mathcal{A} \qquad \pi' = \overline{\mathcal{A} \psi} \in \mathcal{A}'
\end{equation}

\begin{lemma}
\label{lem:jtheta}
The two partial isometries are related:
\begin{equation}
j( \theta^\dagger) = \theta'
\end{equation}
\end{lemma}
\begin{proof}
Consider $j (\theta^\dagger) \psi =
j( \theta^\dagger) \theta^\dagger \xi_{\omega_\psi|_{\mathcal{A}'}} \in \mathscr{P}^\natural$
which can be checked by taking an inner product with other states in the natural cone and checking positivity. 
Hence $j( \theta^\dagger) \psi  = \theta' \psi$ and so
by the consistency of the support projections it follows by cyclicity of $\mathcal{A}$ for $\psi$ on the appropriate support that $j( \theta^\dagger) = \theta'$. 
\end{proof}

We will need the modular operators $J_{\psi}, \Delta_{\psi}$ and the relative modular operators $\Delta_{\eta,\psi}$, $\Delta_{\psi,\eta}$ for $\mathcal{A}$ and $\psi$ relative to $\eta$.
In particular we need a version of relative Tomita-Takesaki theory that allows $\psi$ to have non unit support projections, see for example \cite{Ceyhan:2018zfg}. Modular flow for the $\psi$ state is defined as:
\begin{equation}
 \sigma_{\psi}^s(a) \equiv \Delta_{\psi,\eta}^{-is} a \Delta_{\psi,\eta}^{is}   \qquad  \sigma_{\psi}'^s(a') \equiv \Delta_{\eta,\psi}^{is} a' \Delta_{\eta,\psi}^{-is}  
\end{equation}
and there is also a relative version of modular flow:
\begin{equation}
\sigma_{\eta,\psi}^{s}(a) \equiv  \Delta^{-is}  a \Delta_{\psi,\eta}^{is} \qquad \sigma_{\eta,\psi}'^{s}(a') \equiv  \Delta^{is}  a' \Delta_{\eta,\psi}^{-is}
\end{equation}
where $a \in \mathcal{A}$ and $a' \in \mathcal{A}'$. 
While we do not include a full derivation, the 
properties that we need are summarized in the following lemma:
\begin{lemma}
\label{lem:flow}
The modular flow $\sigma_{\psi}^s$ and relative modular flow 
$\sigma_{\eta,\psi}^{s}$ are maps from the algebra $\mathcal{A}$ to itself. $\sigma_{\psi}^s$  restricts to a one parameter automorphism group on $\pi \mathcal{A} \pi$, which is the modular group for this reduced algebra. They additionally satisfy:
\begin{equation}
\label{eq:refflow}
\sigma_{\psi}^s(a) \pi' = \Delta_{\psi}^{-is} a \Delta_{\psi}^{is}
  \qquad \sigma_{\eta,\psi}^s(a) \pi' = \Delta_{\eta,\psi}^{-is} a \Delta_{\psi}^{is}
\end{equation}
Similar statements apply for the commutant flows with identities:
\begin{equation}
\sigma_{\psi}'^s(a') \pi = \Delta_{\psi}^{is} a' \Delta_{\psi}^{-is}
  \qquad \sigma_{\eta,\psi}'^s(a') \pi = \Delta_{\psi,\eta}^{is} a' \Delta_{\psi}^{-is}
\end{equation}
And also:
\begin{equation}
\sigma'^{s}(a') \pi = \Delta_{\psi,\eta}^{is} a' \Delta_{\psi,\eta}^{-is}     \qquad   \sigma^{s}(a)  \pi' = \Delta_{\eta,\psi}^{-is} a \Delta_{\eta,\psi}^{is} 
\end{equation}
\end{lemma}
\begin{remark}
A useful mnemonic for these identities is that they arise from canceling the flow on the commutant side of the operator $a$ or $a'$ so we can always
replace the state there with any other state - at least, that is, once accounting for the support projections
that arise from this cancelation. 
\end{remark}

The cocycle partial isometries are defined as:
\begin{align}
u_s \equiv  (D \omega_\eta |_{\mathcal{A}} :D \omega_\psi|_{\mathcal{A}})_s & = \sigma_{\eta,\psi}^s(1) = \Delta^{-is} \Delta_{\psi,\eta}^{is}  \\
 u'_s \equiv  (D \omega_\eta|_{\mathcal{A}'} :D \omega_\psi|_{\mathcal{A}'})_s
&=\sigma_{\eta,\psi}'^{s}(1)  =  \Delta^{is} \Delta_{\eta,\psi}^{-is}
\end{align}
and it is easy to use Lemma~\ref{lem:flow} to show that they are contained respectively in $\mathcal{A}, \mathcal{A}'$.
These are partial isometries with initial and final supports:
\begin{equation}
(u_s)^\dagger u_s = \pi \qquad u_s (u_s)^\dagger = \sigma^s(\pi)
\end{equation}
and similarly for $\mathcal{A}'$. 

As is well known, the cocycles satisfy the following 1-cocycle relation $u_{t+s} = \sigma^t( u_s) u_t$. 
Here the main result we will utilize is that the product of two modular operators can be written in terms of a product of cocycles.
There is a similar relation for the modular conjugation operators.
\begin{lemma}
\label{lem:twoproduct}
The product of two modular flows is equal to the following product of cocycles local to the algebras:
\begin{equation}
\Delta^{-is}  \Delta_{\psi}^{is} = u'_{-s} u_s 
\end{equation}
Similarly the product of two modular conjugation operators also has such a decomposition:
\begin{equation}
J J_{\psi} =  \theta \theta' 
\end{equation}
\end{lemma}
\begin{proof}
Set $a=1$ in the second equation of \eqref{eq:refflow} in Lemma~\ref{lem:flow}. We see that $\pi'  u_s = \Delta_{\eta,\psi}^{-is} \Delta_{\psi}^{is}$.
Then we have:
\begin{equation}
u'_{-s} u_s  =  u'_{-s} \pi' u_s  =  \Delta^{-is} \pi' \Delta_{\psi}^{is}  =  \Delta^{-is} \Delta_{\psi}^{is} 
\end{equation}

Consider $\widetilde{J}_\psi = (\theta')^\dagger J \theta'$
since the right hand side satisfies the following characterization of a modular conjugation operator for
$\psi$ and the algebra $\mathcal{A}_\pi = \pi \mathcal{A} \pi \pi'$ (for which $\psi$ is cyclic and separating) 
\begin{equation}
\widetilde{J}_\psi \psi = \psi \qquad \widetilde{J}_\psi \mathcal{A}_\pi \widetilde{J}_\psi= \mathcal{A}_\pi' \qquad \left< \psi \right| \widetilde{J}_\psi a \widetilde{J}_\psi a \left| \psi  \right> \geq 0 \quad \forall  a\in \mathcal{A}_\pi 
\end{equation}
(see \cite{araki1974some}), it must be that  $\widetilde{J}_\psi = J_{\psi}$.
Combining this equation with Lemma~\ref{lem:jtheta} gives the result. 
\end{proof}

For a unitary $W$ we use the standard notation ${\rm Ad}_W(x) \equiv W x W^\dagger$.
\begin{lemma}
\label{lem:normcont}
For a normal state $\rho \in \mathcal{A}_\star^+$, then: 
\begin{equation}
\lim_{b \rightarrow 0^+} \| \rho - \rho \circ {\rm Ad}_{U(b)^\dagger}|_{\mathcal{A}}\| = 0
\end{equation}
\end{lemma}
\begin{proof}
We work with $\xi_{\rho} \in \mathscr{P}^\natural$ and note that $U(b) \xi_{\rho}$ induces the state
$\rho \circ {\rm Ad}_{U(b)^\dagger}$ on $\mathcal{A}$ for $b>0$, so we have the bound:
\begin{equation}
 \| \rho - \rho \circ {\rm Ad}_{U(b)^\dagger}|_{\mathcal{A}}\| \leq 2 \left(\rho(1)^2 - |\left< \xi_\rho \right| U(b) \left| \xi_\rho \right>|^2 \right)^{1/2}
 \end{equation} 
which follows from monotonicity of the linear functional norm to $\mathcal{B}(\mathscr{H})$ and then converting
this into the fidelity of two pure states. Strong continuity of $U(b)$ as $b \rightarrow 0$ gives the result. 
\end{proof}

\subsection{Properties of the quantum expansion}

\label{sec:theta}

Here we study some properties of $\Theta$ introduced in \eqref{eq:intro-theta} that we recall here for the algebra $\mathcal{A}$: 
\begin{equation}
\Theta(\rho) = \inf_{ \phi \in \mathscr{H} :  \omega_\phi|_{\mathcal{A}} = \rho }\| \sqrt{P} \phi \|^2 
\end{equation}
Note that while the variation is taken in a specific representation, the actual definition is intrinsic to the state
$\rho \in \mathcal{A}_\star^+$ at least after the von Neumann subalgebra $\mathcal{A}_1 \subset_\sigma \mathcal{A}$, furnishing a HSMI, is specified.
This is because the representation is the standard form, which is unique up to unitary equivalence, and we can define the operator $P \in \mathcal{TR}(\mathcal{A},\sigma)$ on the right hand side using the subalgebra
structure $\mathcal{A}_1 \in \mathcal{HS}(\mathcal{A},\sigma)$ using the jointly cyclic and separating $\eta$ which is the unique vector on the standard form
 representing $\sigma$. 
A more complete notation would make at least one of these structures clear:
\begin{equation}
\Theta(\cdot|_{\mathcal{A}|\mathcal{A}_1})_\sigma 
: \mathcal{A}_\star^+ \rightarrow [0,\infty]
\end{equation}
for the corresponding $P \in \mathcal{TR}(\mathcal{A},\sigma) \leftrightarrow \mathcal{A}_1 \in \mathcal{HS}(\mathcal{A},\sigma)$.
Focusing on the latter then for example we have unitary equivariance and scaling:
\begin{equation}
\Theta(\rho|_{\mathcal{A}|\mathcal{A}_a })_\sigma =
a \Theta(\rho_u|_{\mathcal{A}|u (\mathcal{A}_1) u^\dagger })_{\sigma_u} 
\end{equation}
for any unitary $u \in \mathcal{U}(\mathcal{A})$ and $a \geq 0$, where $\rho_u = \rho(u^\dagger \cdot u)$ and $\sigma_u = \sigma(u^\dagger \cdot u)$.

We can apply the same notation to the $V$ family determined by a fixed $P$ so:
\begin{equation}
 \Theta(\cdot |_{\mathcal{A}_V|\mathcal{A}_{V+1}})_{\sigma_V}: \rho \in (\mathcal{A}_V)_\star^+ \rightarrow [0,\infty]
 \end{equation}
and in this case the associated inclusion is $\mathcal{A}_{V+1} \subset \mathcal{A}_V$
and state $\sigma_V = \omega_\eta|_{\mathcal{A}_V}$. Since this specific one parameter family of HSMI will be fixed in this paper we can 
return to our original notation used in the introduction where for any globally defined normal state $\rho \in \mathcal{B}(\mathscr{H})^+_\star$ we have:
\begin{equation}
\Theta(\rho|_V) \equiv \Theta(\rho|_{\mathcal{A}_V}) \equiv \Theta(\rho|_{\mathcal{A}_V|\mathcal{A}_{V+1}})_{\sigma_V}
\end{equation}

We start with two computational tools for finding the infimum:
\begin{lemma}
Suppose that $\Theta(\rho)$ is finite, then the following formulas apply:
\begin{itemize}
\item[(a)] $\Theta(\rho) = \lim_{s \rightarrow -\infty} \big\| \sqrt{P} (D \omega_\eta|_{\mathcal{A}'}:D\omega_\psi|_{\mathcal{A}'})_s \left| \psi \right> \|^2$
for any $\psi$ such that $\omega_\psi|_\mathcal{A} = \rho$ and $\psi \in \mathcal{D}(\sqrt{P})$. 
\item[(b)] $\Theta(\rho) = \frac{1}{2} \| \sqrt{P}  \left| \xi_{\rho} \right>\|^2$ where $\xi_{\rho}$ is the standard
purification in $\mathscr{P}^\natural$.  
\end{itemize}
\label{lem:calc}
\end{lemma}
\begin{remark}
This represents an improvement on Lemma~3 in \cite{Ceyhan:2018zfg} where $ \left| \xi_{\rho} \right> \in \mathcal{D}(\sqrt{P})$
was proven but the exact expectation value $ \| \sqrt{P}  \left| \xi_{\rho} \right>\|$ was only estimated.
\end{remark}
For the proof of this Lemma we need some preliminaries. Define $P_\psi \equiv \left< \psi \right| P \left| \psi \right>$.

\begin{lemma} \label{lem:sum} Consider two partial isometries $w \in \mathcal{A}$ and $w' \in \mathcal{A}'$ such that $w^\dagger w \psi = \psi$ and 
 $(w')^\dagger w' \psi = \psi$. Assuming that $\psi,w\psi,w'\psi,ww'\psi \in \mathcal{D}(\sqrt{P})$, then: 
\begin{equation}
P_{w w' \psi } + P_\psi = P_{w \psi} + P_{w' \psi}
\end{equation}
\end{lemma}
\begin{proof} See Appendix~\ref{app:prop1}. \end{proof}

The tricky part is to prove finiteness. We work with a $\psi$ satisfying the assumptions in Lemma~\ref{lem:calc}(a). We use
the modular theory quantities defined in Section~\ref{sec:states} then:

\begin{proposition} 
\label{prop:finite}
Suppose that $\psi \in \mathcal{D}(\sqrt{P})$, then for all $s \in \mathbb{R}$,  $u_s' \psi \in \mathcal{D}(\sqrt{P})$
and also $\theta' \psi, \theta \psi \in \mathcal{D}(\sqrt{P})$. 
\end{proposition}
\begin{proof} This was established in \cite{Ceyhan:2018zfg}.
We will also give a proof of this for the weight case and the state case is a special case. See Propositions~\ref{prop:finite-cocycle} and~\ref{prop:finite-cone}.
\end{proof}

\begin{proof}[Proof of Lemma~\ref{lem:calc}.]

We start by considering the following cocycle relation:
\begin{equation}
 \theta \theta' = J J_\psi
\end{equation}
When this is applied to both Lemma~\ref{lem:sum} and Proposition~\ref{prop:finite} we find:
\begin{align}
P_{\theta \psi} + P_{\theta' \psi} & = 2 P_\psi
\label{eq:recalderiv}
\end{align}
Which gives the bound:
\begin{equation}
P_\psi \geq \frac{1}{2} P_{\theta' \psi} = \frac{1}{2} P_{\xi_{\rho}}
\end{equation}
so that $\Theta(\rho) \geq \frac{1}{2} P_{\xi_{\rho}}$.

Also it is not hard to see that the two states in the natural cone $\mathscr{P}^\natural$ associated to $u_s' \psi$ are $ \theta' \psi$ and $\Delta^{is} \theta \psi$ respectively for
$\mathcal{A}$ and $\mathcal{A}'$ where we used Lemma~\ref{lem:flow} along
with the easily established fact that $\Delta^{is} \mathscr{P}^\natural = \mathscr{P}^\natural$. Applied to Lemma~\ref{lem:sum} this gives:
\begin{equation}
e^{2\pi s} P_{\theta \psi} + P_{\theta' \psi} = 2 P_{u_s' \psi}
\end{equation}
by the HSMI relation \eqref{eq:hsmi-alg1}.
This implies that:
\begin{equation}
\Theta(\rho) \leq \inf_s P_{u_s' \psi}  = \lim_{s \rightarrow - \infty}  P_{u_s' \psi} = \frac{1}{2} P_{\theta' \psi}
=  \frac{1}{2} P_{\xi_{\rho}}
\end{equation}
and hence equality.

\end{proof}
Some basic results follow:
\begin{lemma}
We have $\Theta(Z \rho) = Z \Theta(\rho)$ for any $Z > 0$. Also, suppose that $\Theta(\rho) = 0$, then $\rho = \omega_\phi|_{\mathcal{A}}$ for some $\phi \in {\rm ker} P$. 
\end{lemma}
\begin{proof}
The first statement is obvious. The second follows from Lemma~\ref{lem:calc}(b)
with $\phi = \xi_\rho$. 
\end{proof}
We can also prove lower semi-continuity of $\Theta(\cdot)$ in the state:
\begin{lemma}
\label{lem:cont}
For any sequence of states $\rho_i \rightarrow \rho \in \mathcal{A}_\star^+$ (in linear functional norm), then:
\begin{equation}
\liminf_i \Theta(\rho_i) \geq \Theta(\rho)
\end{equation}
that is $\Theta : \mathcal{A}_\star^+\rightarrow \mathbb{R}_{\geq 0} \cup \{ \infty\} $ is lower semicontinuous (using the linear functional norm topology on $ \mathcal{A}_\star$).
It follows that $\Theta(\rho|_V)$ as a function of $V$ is continuous from the right.
\end{lemma}
\begin{proof} We may assume the left hand side is finite and can thus remove any $\rho_i$ from the sequence for which $\Theta(\rho_i) = \infty$. 
From \cite{haagerup1975standard,araki1974some} we know that $\| \xi_\rho - \xi_{\rho_i} \| \rightarrow 0$. Then we have the spectral representation of $P$:
\begin{equation}
\| \sqrt{P} \xi_{\rho_i} \|^2 = \int_0^\infty \left< \xi_{\rho_i} \right| dE_p(P) \left| \xi_{\rho_i} \right> p
\geq \int_0^\Lambda \left< \xi_{\rho_i} \right| dE_p(P) \left| \xi_{\rho_i} \right> p
\end{equation}
So that dominated convergence gives:
\begin{equation}
\liminf_i \| \sqrt{P} \xi_{\rho_i} \|^2 
\geq \lim_i  \int_0^\Lambda \left< \xi_{\rho_i} \right| dE_p(P) \left| \xi_{\rho_i} \right> p
=\int_0^\Lambda \left< \xi_\rho \right| dE_p(P) \left| \xi_\rho \right> p
\end{equation}
and taking $\Lambda \rightarrow \infty$ with monotone convergence gives the result (the right hand side might diverge in which case we conclude the left hand
side is infinite).

We also clearly have monotonicity in $V$. Setting $\rho_i = \rho \circ {\rm Ad}_{U(a_i)^\dagger}$ in the semicontinuity result, which converges in norm by Lemma~\ref{lem:normcont}, for $a_i \rightarrow 0$
with $a_i > 0$ and combining with monotonicity gives:
\begin{equation}
\Theta(\rho|_V)\geq \limsup_i \Theta(\rho|_{V+a_i}) \geq \liminf_i \Theta(\rho|_{V+a_i}) \geq \Theta(\rho|_V)
\end{equation}
where we used the strong continuity of $U(a_i)$.
Hence the limit exists when approaching from the right where it is continuous. 
\end{proof}

We now give a variational formula for $\Theta$ over convex sums:
\begin{lemma}
    \label{lem:preconvex}
Given a finite set of mutually orthogonal isometries $v_i' \in \mathcal{A}'$ such that $(v_i')^\dagger v_j' = \delta_{ij}$
and such that $\sum_i v_i' (v_i')^\dagger = 1$, then:
\begin{equation}
    \label{eq:preconvex}
\Theta(\rho) = \inf_{\phi \in \mathscr{H}: \omega_\phi|_{\mathcal{A}} = \rho} \sum_i \left< \phi \right|  v_i' P (v_i')^\dagger \left| \phi \right>
\end{equation}
\end{lemma}
\begin{proof}
Let $k$ equal the number of terms in the sum. 
Construct the unitary $U = \sum_i \left| i \right> (v_i')^\dagger$ which acts $U : \mathscr{H} \rightarrow \mathscr{H} \otimes \mathbb{C}^k$. 
It is easy to check that $U$ is a unitary and an intertwiner $(a \otimes 1_k) U = U a$ for all $a \in \mathcal{A}$.
We can write the right hand side as:
\begin{equation}
\inf_{\phi \in \mathscr{H}: \omega_\phi|_{\mathcal{A}} = \rho} \sum_i \left< \phi \right|  v_i' P (v_i')^\dagger \left| \phi \right> = \inf_{u' \in \mathcal{V}_{\xi_\rho}(\mathcal{A}')} \left< \Psi \right| U (u')^\dagger U^\dagger( P \otimes 1_k) U u' U^\dagger \left| \Psi\right>
\end{equation}
where $\Psi = U \xi_\rho$ and, for a von Neumann algebra $\mathcal{M}$ and a vector $\xi$, $\mathcal{V}_\xi(\mathcal{M})$ denotes the set of partial isometries $u \in \mathcal{M}$ with $u^\dagger u \xi = \xi$, so that the vectors $u' \xi_\rho$ with $u' \in \mathcal{V}_{\xi_\rho}(\mathcal{A}')$ are exactly the $\phi$ with $\omega_\phi|_{\mathcal{A}} = \rho$. Now $U u' U^\dagger$ is a general partial isometry in $\mathcal{V}_{\Psi}((\mathcal{A} \otimes 1)')$. If we enlarge by an extra $k$-bit 
then we can write this as a quantum expansion on an enlarged system with cyclic and separating $H = \eta \otimes \sum_i k^{-1/2} \left| i \right> \otimes \left| i \right>$
generating the HSMI for $\mathcal{A}_1 \otimes 1_k \otimes \mathcal{B}(\mathbb{C}^k) \subset \mathcal{A} \otimes 1_k \otimes \mathcal{B}(\mathbb{C}^k)$. The translation generator is $P \otimes 1_k \otimes 1_k$ with vector state $\Psi \otimes \left| 1 \right>$ and so the right hand side equals
a quantum expansion on this enlarged system which we can compute using the standard
purification Lemma~\ref{lem:calc} (b) which is simply $\xi_\rho \otimes \left|1 \right> \otimes \left|1 \right> $ so
\begin{equation}
= \Theta(\omega_{\Psi \otimes \left| 1 \right>}|_{\mathcal{A} \otimes 1_k \otimes \mathcal{B}(\mathbb{C}^k)}) = \frac{1}{2}\left< \xi_\rho \right| P \left| \xi_\rho \right>
= \Theta(\rho|_{\mathcal{A}})
\end{equation}
as long as $\Theta(\omega_{\Psi \otimes \left| 1 \right>}|_{\mathcal{A} \otimes 1_k \otimes \mathcal{B}(\mathbb{C}^k)}) < \infty$ (the right hand side of \eqref{eq:preconvex}) in which case $\Theta(\rho) < \infty$ (the left hand side of \eqref{eq:preconvex}). Conversely if $\Theta(\rho) < \infty$ then it must also be that $\Theta(\omega_{\Psi \otimes \left| 1 \right>}|_{\mathcal{A} \otimes 1_k \otimes \mathcal{B}(\mathbb{C}^k)}) < \infty$ since the standard purification is one such purification. 
\end{proof}

Using a similar proof technique we can also show:
\begin{lemma}
    \label{lem:convex}
The quantum expansion $\Theta(\rho)$ is a convex function of $\rho \in \mathcal{A}_\star^+$.
\end{lemma}
\begin{proof}
    Consider a finite convex sum of states $\rho = \sum_i p_i \rho_i$.
Using the same enlarged HSMI structure as in the proof of Lemma~\ref{lem:preconvex}, we consider:
\begin{equation}
\Psi = \sum_i \sqrt{p_i} \phi_i \otimes \left| i\right> \otimes \left| 1 \right>
\end{equation}
for $\phi_i$ any purification of $\rho_i = \omega_{\phi_i}|_{\mathcal{A}}$.
Then $\Psi$ purifies $\rho\otimes \left< 1 \right| \cdot \left| 1 \right>$ on $
\mathcal{A} \otimes 1_k \otimes \mathcal{B}(\mathbb{C}^k)$. Again the standard purification
is $\xi_\rho \otimes \left| 1 \right> \otimes \left| 1 \right>$ so that:
\begin{equation}
\Theta(\rho) = \Theta(\omega_\Psi|_{\mathcal{A} \otimes 1_k \otimes \mathcal{B}(\mathbb{C}^k)}) 
\leq \left< \Psi \right| P \otimes 1_k \otimes 1_k \left| \Psi \right> = \sum_i p_i \left< \phi_i \right| P \left| \phi_i \right>
\end{equation}
Minimizing over $\phi_i$ gives the answer.
\end{proof}

Finally we note here that it is possible to restrict to unitaries in the variational formula:
\begin{lemma}
The following variational formula applies whenever the right hand side is finite:
\begin{equation}
    \label{eq:lessmore}
\Theta(\psi|_{\mathcal{A}}) = \inf_{u' \in \mathcal{U}(\mathcal{A}')} \| \sqrt{P} u' \left| \psi \right> \|^2
\end{equation}
\end{lemma}
\begin{proof}
Redefine $\psi \rightarrow w' \psi$ for some unitary $w' \in \mathcal{A}'$ such that
the new $\psi \in \mathcal{D}(\sqrt{P})$. 
We still have the inequality $\leq$ in \eqref{eq:lessmore} directly from the definition of $\Theta$, since $u' \psi$ purifies $\psi|_{\mathcal{A}}$ for every $u' \in \mathcal{U}(\mathcal{A}')$.
The opposite inequality $\geq$ in \eqref{eq:lessmore} is usually found with a cocycle $u_s'$. 
However since the cocycle may have non-trivial support projections we need to find a unitary $u'_s$ such that $u'_s \pi_\psi(\mathcal{A}')
= (D \omega_\eta|_{\mathcal{A}'} : D\omega_\psi|_{\mathcal{A}'})_s$ so we can still use this in the variational formula
as the lower bound for $s \rightarrow - \infty$. Construct the state $\rho' = (\omega_\psi(\cdot)
+ \omega_\eta( (1-\pi_\psi(\mathcal{A}')) \cdot (1-\pi_\psi(\mathcal{A}')))|_{\mathcal{A}'}$ which is faithful so that $u'_s = 
(D\omega_\eta|_{\mathcal{A}'} : D\rho')_s$ is unitary and satisfies the required relation to the optimal cocycle. We can then prove $\geq$ with Lemma~\ref{lem:calc} given that $\psi \in \mathcal{D}(\sqrt{P})$. 
\end{proof}

\section{Streamlining the proof of the QNEC}
\label{sec:streamline}

\subsection{Assuming \texorpdfstring{$\Theta < \infty$}{finite quantum expansion}}
\label{sec:finite}

The two basic elements in the proof of the QNEC pertain to the commutant algebra. In particular we introduce the following quantities
defined for states on the commutant von Neumann algebras:
\begin{align}
D(\psi|_{V'}) &\equiv D(\psi|\eta; \mathcal{A}_{V}') \\
\Theta(\psi|_{V'})& \equiv \Theta(\omega_\psi|_{V'}) \equiv
\inf_{  \phi \in \mathscr{H}: \omega_\phi|_{V'} = \omega_\psi|_{V'} } \big\| \sqrt{ P} \left| \phi \right> \big\|^2
\end{align}
where we use a (hopefully obvious) notation explained in the introduction. 

These are subject to the duality results:
\begin{lemma}
\label{lem:duality}
Given some $\psi \in \mathscr{H}$ for which $\psi \in \mathcal{D}(\sqrt{P})$, then
both $\Theta(\psi|_V)$ and $\Theta(\psi|_{V'})$ are finite with:
\begin{equation}
\label{eq:sumT}
\| \sqrt{P} \left| \psi \right> \|^2 = \Theta(\psi|_V) + \Theta(\psi|_{V'})
\end{equation}
\end{lemma}
\begin{proof}
We recall that $2P_\psi = P_{\theta \psi} + P_{\theta' \psi}$, see derivation of \eqref{eq:recalderiv}. Also we know that 
$P_{\theta' \psi}  =2 \Theta(\psi|_V)$ in Lemma~\ref{lem:calc} and the obvious commutant version
of this result $P_{\theta \psi}  =2 \Theta(\psi|_{V'})$.
\end{proof}

And also:
\begin{proposition}
\label{prop:sum}
Consider two cuts $V_1 < V_2$. Given some $\psi \in \mathscr{H}$,
then:
\begin{equation}
\label{eq:sumS}
2\pi  (V_2-V_1) \| \sqrt{P} \left| \psi \right> \|^2 + D(\psi|_{V_2}) + D(\psi|_{V_1'})  = D(\psi|_{V_1}) +  D(\psi|_{V_2'}) 
\end{equation}
holds where both sides could be $\infty$. 
\end{proposition}
\begin{remark}
This represents an improvement on Lemma~1 in \cite{Ceyhan:2018zfg} where this was proven only when all quantities except $D(\psi|_{V_2'})$
were assumed finite.
\end{remark}
\begin{proof}
Assuming all the relative entropies and null energy are finite easily gives the sum rule via derivatives of the cocycle relation for relative entropy,
see the weight case,  Proposition~\ref{prop:sum-weight}, for a demonstration of this.
We only have to deal with the $\infty$ cases.

Without loss of generality we can take $\| \psi \| = 1$. Consider
\begin{align}
h_V(z) &\equiv \left< \psi \right| (1- \Delta_{\eta,\psi;V}^z)\left| \psi \right>\,, \quad h_{V'}(z) \equiv \left< \psi \right| (1-\Delta_{\psi,\eta;V}^{-z})\left| \psi \right>  \\
\| \psi \|^2 - f(z) &\equiv \left< \psi \right| \Delta_{\eta,\psi;V_1}^z \Delta_{\psi,\eta;V_2}^{-z} \left| \psi \right>
=  \left< \psi \right| \Delta_{\psi,\eta;V_1}^{-z} \exp( -ia P (1-e^{i 2\pi z })) \Delta_{\eta,\psi;V_2}^z \left| \psi \right>
\end{align}
where $a = (V_2-V_1) > 0$. 
The functions $h_V(z),h_{V'}(z), f(z)$ are analytic in the strip $0 < {\rm Re} z <1/2$ and continuous in the closure
from standard results in modular theory.
For the equality in the second line, we clearly have equality along $z=is$ via Lemma~\ref{lem:flow} and \eqref{eq:hsmi-alg2},
 and both have analytic continuations into the strip $0 < {\rm Re} z < 1/2$, continuous in the closure,
where they must agree by the edge of the wedge theorem.

Assume that the left hand side of \eqref{eq:sumS} is infinite and that the right hand side
is finite, aiming for a contradiction. Monotonicity implies that $ D(\psi|_{V_2}),D(\psi|_{V_1'}) <\infty$ so that these assumptions entail $\| \sqrt{P} \psi \| =\infty$. 

 Parameterizing $i(1 - e^{i 2\pi z }) = x$ and picking real $x >0$, we see that:
\begin{equation}
\lim_{x \rightarrow 0^+} |h_P(z(x))|/x = a \| \sqrt{P} \left| \psi \right> \|^2  = \infty \qquad h_P(z) \equiv \left< \psi \right|  (1-  e^{-ia P (1-e^{i 2\pi z})} ) \left| \psi \right>
\end{equation}
due to monotone convergence of $\left<\psi \right| ( 1- e^{- a x P})/x \left| \psi \right>$ which is faithful to the infinite value of the null energy. 

We then consider the inequality:
\begin{equation}
\label{eq:testin1}
|  |h_P(z)|- |f(z) - h_{V_1'}(z) - h_{V_2}(z)|  |^2 \leq 16 q(z) |h_P(z)|  +8  {\rm Re} h_{V_1'}(z) {\rm Re}h_{V_2}(z) 
\end{equation}
with $ q(z) =  {\rm Re}(h_{V_1'}(z)+h_{V_2}(z))$. This is derived using Cauchy-Schwarz 
along with 
\begin{align}
 \left< \psi \right| (1-\Delta_{\eta,\psi;V}^z) (1-\Delta_{\eta,\psi;V}^{z^\star}) \left| \psi \right>
& \leq 2{\rm Re} h_V(z) \\
 \left< \psi \right| (1-e^{ ia P (1-e^{-i 2\pi z^\star })}) (1-e^{-ia P (1-e^{i 2\pi z })} ) \left| \psi \right>
& \leq 2{\rm Re} h_P(z)  \leq 2 |h_P(z) | \\
\| e^{ -ia P (1-e^{i 2\pi z })}\|  \leq 1
\end{align}
which hold along $z = z(x)$.
Dividing \eqref{eq:testin1} by $x^2$, all quantities except $h_P(z)/x$ have finite limits along $z(x)$ and $ x\rightarrow 0^+$ which then leads to a contradiction
between the divergence of $|h_P(z)|/x$ and the inequality. 

Now assume that the right hand side of \eqref{eq:sumS} is $\infty$ but the left hand side is finite. 
We have a similar inequality, now for real $z > 0$:
\begin{equation}
\label{eq:testin2}
| h_{V_1}(z) + h_{V_2'}(z) - f(z) |^2  \leq  (2 h_{V_1}(z) - h_{V_1}(2z)) (2 h_{V_2'}(z) - h_{V_2'}(2z))
\end{equation}
Note that $h_{V}(z)$ is real and non-negative for real $z$ and $2 h_{V}(z) - h_{V}(2z) \geq 0$.
Finiteness of the left hand side of the equality implies that $f(z)/z$ converges as $z \rightarrow 0^+$. 
On the right hand side of \eqref{eq:testin2} we drop the negative term multiplying $h_{V_2'}(2z)$ yielding a new inequality. We then do the same with the term multiplying $h_{V_1}(2z)$. Averaging the two resulting inequalities gives:
\begin{equation}
0\leq | h_{V_1}(z) - h_{V_2'}(z)|^2  + |f(z)|^2  \leq  h_{V_1}(z) (2{\rm Re} f(z) - h_{V_2'}(2z))
+ h_{V_2'}(z) (2{\rm Re} f(z) -h_{V_1}(2z))
\end{equation}
implying that at least one of the two terms $2 {\rm Re} f(z) - h_{V_2'}(2z)$ or $2{\rm Re} f(z) -h_{V_1}(2z)$
is non-negative.
Since $f(z)/z$ converges and $h_{V}(2z)/z$ monotonically increases as $z \rightarrow 0^+$ (again see \cite{ohya-petz})
it must be the case that one of the relative entropies $D(\psi|_{V_1})$ or $D(\psi|_{V_2'})$ is finite.
Assume, without loss of generality, that the latter is finite, then $\lim h_{V_1}(z)/z = D(\psi|_{V_1}) = \infty$
by assumption. But the first inequality \eqref{eq:testin2} can now be simplified to
\begin{equation}
\label{eq:testin3}
| h_{V_1}(z) + h_{V_2'}(z) - f(z) |^2/z^2 \leq 4 (h_{V_1}(z)/z) (h_{V_2'}(z)/z)
\end{equation}
dropping terms multiplying $ h_{V_1}(2z)$ and then those multiplying $h_{V_2'}(2z)$ (the first can be dropped because $2h_{V_2'}(z) - h_{V_2'}(2z) \geq 0$ and the latter can be dropped
because $h_{V_1}(z) \geq 0$).
But \eqref{eq:testin3} leads to a contradiction since $f(z)/z$ and $h_{V_2'}(z)/z$ have finite limits while $h_{V_1}(z)/z$ diverges.
\end{proof}

\begin{corollary}
The sum rules combine as:
\begin{equation}
 D(\psi|_{V_1}) +D(\psi|_{V_2'})   = D(\psi|_{V_2}) + D(\psi|_{V_1'}) +2\pi(V_2-V_1) 
 \max \begin{cases}  \Theta(\psi|_{V_1})+\Theta(\psi|_{V_1'}) \\ \Theta(\psi|_{V_2})+\Theta(\psi|_{V_2'})  \end{cases}
\end{equation}
This relation is unambiguous in the presence of $\infty$ values - both sides could be $=\infty$.
\end{corollary}
\begin{proof}
If the left hand side is finite then Proposition~\ref{prop:sum} implies that $P_\psi < \infty$ and the right hand relative entropies are finite.
Also $P_\psi < \infty$ implies the expansions are finite from Lemma~\ref{lem:duality}. If the right hand side is finite then we can apply
Theorem~\ref{thm:mainthm} (this corollary will not be used anywhere in this paper, so the proof of this theorem does not circularly rely on this) 
to conclude the integral relation $D(\psi|_{V_1})
= D(\psi|_{V_2})+ 2\pi \int_{V_1}^{V_2} d V \Theta(\psi|_{V}) \leq D(\psi|_{V_2})+ 2\pi (V_2-V_1) \Theta(\psi|_{V_1}) $
and similarly for $D(\psi|_{V_2'}) <\infty$. 
\end{proof}

\begin{lemma}
\label{lem:deriv}
Suppose that $\psi \in \mathcal{D}(\sqrt{P})$ and that $D(\psi|_{V_1}) < \infty$, then the following right derivative exists:
\begin{equation}
\label{eq:deriv}
- \partial_V^+ D(\psi|_{V})  =2 \pi \Theta(\psi|_{V})
\end{equation}
for all $V \geq V_1$. 
\end{lemma}
\begin{proof}
Since the derivative is taken from the right,
without loss of generality we can replace $\psi \rightarrow \psi = \xi_{\omega_\psi|_{V_1}}$. 
The symmetry of states in the natural cone then implies that $ D(\psi|_{V}) \leq  D(\psi|_{V_1}) = D(\psi|_{V_1'})< \infty$,
and Proposition~\ref{prop:finite} means that we maintain $\psi \in \mathcal{D}(\sqrt{P})$. 
Also Proposition~\ref{prop:sum} implies that $D(\psi|_{V'}) < \infty$.

Now consider the implications of the sum rules \eqref{eq:sumS} and \eqref{eq:sumT}
for the state $\psi_s = \Delta_{\eta,\psi;V_1}^{is} \psi$
which has an induced state satisfying $\omega_{\psi_s}|_{V_1} = \omega_{\psi} \circ \sigma_{V_1}^s$
and also $\omega_{\psi_s}|_{V_1'} = \omega_{\psi}|_{V_1'}$. Now since modular flow holds fixed relative entropy (restricting to the same algebra), we have:
\begin{align}
 \| \sqrt{P} \left| \psi_s \right> \|^2  &= e^{2\pi s} \Theta(\psi|_{V_1}) + \Theta(\psi|_{V_1'})  \\ D(\psi_s|_{V_1})  &= D(\psi|_{V_1})  
  \,,\,\, D(\psi_s|_{V_1'})  = D(\psi|_{V_1'})  
   \,,\,\, D(\psi_s|_{V}) =  D(\psi|_{V_2})  
   \label{eq:focuslast}
\end{align}
where $V_2= V_1 + (V-V_1)e^{2\pi s} > V_1$ so that all of the relative entropies in the second line are finite. 
To derive the last equality in \eqref{eq:focuslast} we note, via \eqref{eq:hsmi-alg2}, that
\begin{equation}
\sigma_{V_1}^s = {\rm Ad}_{ U( (V-V_1)(1-e^{2\pi s}))} \circ \sigma_{V}^s\,,
\end{equation}
use the independence of the $V$-relative entropy under $ \sigma_{V}^s$-modular flow and translate the algebra 
from $V$ to $V_2 = V- (V-V_1)(1-e^{2\pi s})$.
From \eqref{eq:sumS} we can infer via monotonicity $D(\psi_s|_{V'}) - D(\psi_s|_{V_1'}) \geq 0$ that:
\begin{equation}
\label{eq:sbound1}
 2\pi(V_2-V_1)  \left( \Theta(\psi|_{V_1}) + e^{-2\pi s} \Theta(\psi|_{V'_1}) \right) 
\geq  \big( D(\psi|_{V_1}) - D(\psi|_{V_2}) \big)
\end{equation}
We can pick $V$ to vary with $s$ via $V(s) = V_1 + (V_2-V_1) e^{-2\pi s}$,
for $V_2 > V_1$, allowing us to hold $V_2$ fixed while maintaining $V(s)>V_1$ that was used to derive these
inequalities. A similar implication arises from the cocycle state on the complement acting at $V_2'$.
Equivalently this can be inferred by the replacement $V_2 \leftrightarrow V_1'$ and $V_1 \leftrightarrow V_2'$ (holding fixed $V_2-V_1$). 
\begin{equation}
 2\pi(V_2-V_1)  \left( \Theta(\psi|_{V'_2}) + e^{-2\pi s} \Theta(\psi|_{V_2}) \right) 
\geq  \big( D(\psi|_{V_2'}) - D(\psi|_{V_1'}) \big)
\end{equation}
One further application
of the sum rule \eqref{eq:sumS} for $V_1,V_2$ and
\eqref{eq:sumT} for $V \rightarrow V_2$ gives:
\begin{equation}
\label{eq:sbound2}
 2\pi(V_2-V_1) ( 1- e^{-2\pi s} ) \Theta(\psi|_{V_2}) 
\leq \big( D(\psi|_{V_1}) - D(\psi|_{V_2}) \big)
\end{equation}
At large $s$ we have from \eqref{eq:sbound1} and \eqref{eq:sbound2}:
\begin{equation}
    \label{eq:to-deriv}
2\pi  \Theta(\psi|_{V_2}) \leq \frac{ D(\psi|_{V_1}) - D(\psi|_{V_2}) }{V_2-V_1}
\leq 2\pi  \Theta(\psi|_{V_1})
\end{equation}
We can take the limit $V_2 \rightarrow V_1^+$ and use right continuity of $\Theta$
established in Lemma~\ref{lem:cont} to finish the proof. 
\end{proof}
\begin{theorem}
 \label{thm:initialproof}
The statements of Theorem~\ref{thm:mainthm} hold assuming $\Theta(\rho|_{V_0}), D(\rho|_{V_0}) < \infty$,
restricted to future cuts $V \geq V_0$.
\end{theorem}
\begin{proof}
Under these assumptions $D,\Theta$ are finite for all $V \geq V_0$. 
Note that monotonicity of $\Theta$ is clear by the variational definition \eqref{eq:intro-theta}. Monotonicity of $D$ is well known \cite{uhlmann1977relative,ohya-petz}.
We pick a $\psi = \xi_{\rho|_{V_0}}$ which satisfies the assumptions in Lemma~\ref{lem:deriv} with $V_1= V \geq V_0$. 
The integral relation \eqref{eq:mainintegral} for $V_1,V_2 \geq V_0$ follows from 
integrating \eqref{eq:deriv}.
Convexity of $D$
follows from the integral relation and monotonicity of $\Theta$. Continuity of $D$ for $V \geq V_0$ follows from the sum rule Proposition~\ref{prop:sum}
for $ \xi_{\rho|_{V_0}}$ which implies Lipschitz continuity of $D$. Right continuity of $\Theta$ was proven in Lemma~\ref{lem:cont} while left continuity follows from an application of the duality relation Lemma~\ref{lem:duality}
with $\psi = \xi_{\rho|_{V_0}}$ so that $\Theta(\rho|_{V}) = \| \sqrt{P} \xi_{\rho|_{V_0}}\|^2 - 
 \Theta(\xi_{\rho|_{V_0}}|_{V'})$ so the right hand side is left continuous by Lemma~\ref{lem:cont}.  
 Since the right derivative \eqref{eq:deriv} is now continuous and $D(\rho|_V)$ is continuous, this implies that $D(\rho|_V)$ is continuously differentiable.
\end{proof}  

\subsection{Smoothing and removing assumptions}
\label{sec:smooth}

We will need to be able to approximate any state with one having finite relative entropy and expansion and with continuous limits:
\begin{theorem}
\label{thm:fullapprox}
Given some $\rho \in \mathcal{A}_\star^+$ there exists a sequence of states $\rho_n \in \mathcal{A}_\star^+$
such that $\rho_n \rightarrow \rho$ in the linear functional norm and such that: $\Theta(\rho_n), D(\rho_n) < \infty$ with:
\begin{equation}
\label{eq:toprovelim}
 \lim_n D(\rho_n) = D(\rho) \qquad \lim_n \Theta(\rho_n) = \Theta(\rho)
\end{equation}
where it is possible the right hand side of either of these equations is $=\infty$. 
\end{theorem}
\begin{proof}
See Appendix~\ref{app:approximation}. 
\end{proof}

Now we have:
\begin{lemma}
\label{lem:infinity}
Given $\rho \in (\mathcal{B}(\mathscr{H}))^+_\star$, then:
\begin{enumerate}
\item Suppose that $D(\rho|_{V_0}) < \infty$, then
for all $V > V_0$ we have $\Theta(\rho|_V) < \infty$. 
\item Conversely, suppose that $\Theta(\rho|_{V_0}) < \infty$ and there is some $V_1 > V_0$ for which  $D(\rho|_{V_1}) < \infty$,
then  $D(\rho|_{V_0})  \leq D(\rho|_{V_1}) +   2\pi \Theta(\rho|_{V_0})(V_1-V_0) < \infty$.
\end{enumerate}
\end{lemma}
\begin{proof}
(1) Applying Theorem~\ref{thm:fullapprox} we find $\rho_n \in (\mathcal{A}_{V_0})_\star^+$ and $\rho_n \rightarrow \rho|_{V_0}$ and
the integral relation \eqref{eq:mainintegral} applies to $\rho_n$ since it satisfies the assumptions of Theorem~\ref{thm:initialproof}. That is:
\begin{equation}
\label{eq:tocontra}
D(\rho_n|_{V_0}) = D(\rho_n|_V) + 2\pi \int_{V_0}^V d \tilde{V} \Theta(\rho_n|_{\tilde{V}})
\end{equation}
Suppose, aiming for a contradiction, that $\Theta(\rho|_V) = \infty$ for some window $V_0 \leq V < V_\infty$
for some $V_\infty > V_0$. The limit $n \rightarrow \infty$ of the integral in \eqref{eq:tocontra} will diverge (by Lemma~\ref{lem:cont} and
Fatou's lemma), and hence the right hand
side diverges, while
the left hand side has a finite limit. 

(2) Again we use the same approximations from Theorem~\ref{thm:fullapprox}. Then note that:
\begin{equation}
 \liminf_n( D(\rho_n|_{V_0}) - D(\rho_n|_{V_1})  ) \leq 2\pi \liminf_n \Theta(\rho_n|_{V_0})(V_1-V_0)
 =  2\pi \Theta(\rho|_{V_0})(V_1-V_0)
\end{equation}
using monotonicity of $\Theta$. 
Semicontinuity of the relative entropy
difference, established in the weight case (for which the state case is a special case) in Lemma~\ref{lem:weight-semi},
implies that $D(\rho|_{V_0}) \leq D(\rho|_{V_1}) +   2\pi \Theta(\rho|_{V_0})(V_1-V_0) < \infty$ as required. 
\end{proof}

Consider the sets $V_D = \{ V: D(\rho|_V) < \infty\}$ and $V_\Theta = \{ V: \Theta(\rho|_V) < \infty\}$.
Assume that $V_D$ is not empty then using Lemma~\ref{lem:infinity} we get (1) $V_D \subset \overline{V_\Theta}$ and  (2) $V_\Theta \subset V_D$. Monotonicity
implies these sets are of the form $\{V_D,V_\Theta \}=\{ [V_\star,\infty),(V_\star,\infty )\}$  or $= \{ (V_\star,\infty),(V_\star,\infty )\}$
or $= \{[V_\star,\infty),[V_\star,\infty )\}$ for some $V_\star < \infty$, possibly $=-\infty$. 

We can now give a proof of Theorem~\ref{thm:mainthm} with the assumptions as stated:
\begin{proof}
(1) and (2) are mostly evident but we briefly summarize the argument here. 
Recall that 
for both $D,\Theta$ we allow for values in the extended real line and continuity refers to this. 
The continuity statement for $D$ 
inside $V > V_\star$ follows from convexity there (established already in Theorem~\ref{thm:initialproof}, see below) and monotonicity
and lower semicontinuity gives right continuity to $V=V_\star$. 
Right continuity for $\Theta$ follows from monotonicity
and lower semicontinuity.

Convexity of $D$ allows for infinite values of the relative entropy and, thanks
to monotonicity, is only
a non-trivial statement if all the relative entropies are finite, say in $[V_1, V_2]$, in which case we have already established:
\begin{equation}
D(\rho|_{V_p}) \leq p D(\rho|_{V}) + (1-p) D(\rho|_{V_2}) \qquad V_p = p V + (1-p) V_2 
\end{equation}
for $V_1 < V < V_p < V_2$ 
in Theorem~\ref{thm:initialproof} since Lemma~\ref{lem:infinity} (1)
establishes $D(\rho|_{V_1}) < \infty \implies \Theta(\rho|_V) < \infty$ giving the required assumptions for Theorem~\ref{thm:initialproof}. Taking the limit $V \rightarrow V_1$ using right
continuity gives the general convexity statement.

For (3) if the left hand side of \eqref{eq:mainintegral} is finite for some $V_1$ then, via Lemma~\ref{lem:infinity} and
 Theorem~\ref{thm:initialproof}, we
 have established the integral relation \eqref{eq:mainintegral} between $\tilde{V}_1$ and $V_2$
 where $V_1 < \tilde{V_1} < V_2$. We can take the limit $\tilde{V}_1 \rightarrow V_1$ from the right by right continuity and
 on the right hand side the integral in \eqref{eq:mainintegral} varies continuously with $\tilde{V}_1$ thanks to monotone convergence (establishing integrability). 

If the right hand side of (3) is finite then we know
$\Theta(\rho|_{V}) < \infty$ for any $V > V_1$ implying (via Lemma~\ref{lem:infinity}) that $D(\rho|_{\tilde{V}_1})$ is finite
for all $V_1 < \tilde{V}_1 < V_2$ and equal to the integral relation \eqref{eq:mainintegral} between $\tilde{V}_1$ and $V_2$.
We again take the limit from the right using right continuity to give \eqref{eq:mainintegral} between $V_1$ and $V_2$.

The infinite cases are the contrapositive of the above so they have also been established.\end{proof}

\section{Generalizing to a weight}
\label{sec:genweight}

In this section we turn to a generalization where the half sided modular inclusion is based on a translationally invariant weight and not a state.
In Section~\ref{sec:weightentropy} we summarize some properties of weights and in particular relative entropies with respect to such a weight.
We find various computational techniques for such a relative entropy and prove a monotonicity result that is more general than has previously appeared. 
In Section~\ref{sec:weighthsmi} we discuss the generalization of the HSMI to a weight originally worked out by Araki-Zsido. We prove the half sided translation theorem which seems to be new.
Then in Sections~\ref{sec:sumrule} and~\ref{sec:finite-weight} we turn to proving two main ingredients (Propositions~\ref{prop:sum-weight}, \ref{prop:finite-cocycle} and~\ref{prop:finite-cone}) that allow us to generalize the QNEC to a weight.

\subsection{Weights, relative entropy and monotonicity}
\label{sec:weightentropy}

A weight is an unbounded version of a state, see \cite{takesaki2003theory} for more details than we shall provide. We only consider faithful normal semi-finite (nsf) weights in this paper, so we often refer to these simply as weights.
For all positive elements of $\mathcal{A}$ we have the map:
\begin{equation}
\label{eq:aroundweight}
    \sigma : \mathcal{A}^+ \rightarrow \mathbb{R}_{\geq 0} \cup \{ \infty \}
\end{equation}
which we demand is linear on positive elements. Define $\mathfrak{N}_\sigma = \{ n : \sigma(n^\dagger n ) < \infty \}$ which forms a left ideal of $\mathcal{A}$.
The linear span $\mathfrak{M}_\sigma = {\rm span} \left( \mathfrak{N}_\sigma^\dagger \mathfrak{N}_\sigma \right) $ is equal to the linear span
of finite positive elements for $\sigma$. 
Normality requires the map to be lower-semicontinuous in the ultraweak topology on $\mathcal{A}^+$. This means that $\liminf \sigma(a^+_i) \geq \sigma(a^+)$ whenever
$a^+_i \rightarrow a^+$ ultraweakly. Semi-finiteness is simply the fact that $\mathfrak{M}_\sigma$ is ultraweakly dense in $\mathcal{A}$ and faithfulness is the obvious generalization from the state case, $\sigma(a^+) = 0 \implies a^+ =0$.
Note that if the identity makes an appearance in $\mathfrak{M}_\sigma$ then the weight is actually a state (that is, a bounded weight). 

We assume the representation of $\mathcal{A}$ on $\mathscr{H}$ is in standard form with $\sigma$ giving rise to a semi-cyclic representation:
\begin{equation}
n \in \mathfrak{N}_\sigma \rightarrow \eta_\sigma(n) \in \mathscr{H}
\end{equation} 
with the defining representation of $a \in \mathcal{A}$ acting as $a \eta_\sigma(n) = \eta_\sigma( a n)$.  The left ideal property of $\mathfrak{N}_\sigma$ is important here.  The linear span of this map is dense $\overline{\eta_\sigma(\mathfrak{N}_\sigma)} = \mathscr{H}$,
that is the representation is ``semi-cyclic''. The modular data $J, \Delta_\sigma$ are defined in the obvious way. They satisfy:
\begin{equation}
J \Delta_\sigma^{1/2} \eta_\sigma(m) = \eta_\sigma(m^\dagger) \qquad \forall m \in \mathfrak{N}_\sigma \cap \mathfrak{N}_\sigma^\dagger
\end{equation}
We can define the commutant weight $\sigma' = \sigma(J \cdot J)$ on $\mathcal{A}'$ and with domain $\mathfrak{N}_{\sigma'}$. And the commutant representation $\eta_{\sigma'} = J\eta_\sigma(J\cdot J)$. One identity that we will need every so often is compatibility of the commutant and the algebra:
$n \eta_{\sigma'}(n') = n' \eta_\sigma(n)$ for $n \in \mathfrak{N}_\sigma$ and $n' \in \mathfrak{N}_{\sigma'}$  
which can be established with the more sophisticated machinery of left/right Hilbert algebras. See for example \cite{takesaki2003theory}.

The positive natural self-dual cone $\mathscr{P}^\natural$ can be constructed as usual:
\begin{equation}
\mathscr{P}^\natural = \overline{\Delta_\sigma^{1/4} \eta_\sigma(n^+)} \qquad n^+ \in \mathfrak{N}_\sigma^+
\end{equation}
 and for any normal state $\rho \in \mathcal{A}_\star^+$ there
is a vector $\xi_\rho \in \mathscr{P}^\natural$ that represents this state, see \cite{haagerup1975standard}.

Given some vector $\psi$, as in the state case of Section~\ref{sec:states} we can construct support projections $\pi, \pi'$
and partial isometries $\theta'$ ($\theta$) taking $\psi$ to $\mathscr{P}^\natural$ preserving the state on $\mathcal{A}$ (respectively $\mathcal{A}'$).
The relative modular operator for a weight and $\psi$ satisfies:
\begin{equation}
\label{eq:relmod-weight}
J \Delta_{\eta_\sigma, \psi}^{1/2} \left( n \left| \psi \right> + \left| \xi' \right>\right) = \eta_\sigma(\pi n^\dagger)  \qquad \forall n^\dagger \in \mathfrak{N}_\sigma,\,
\xi' \in (1- \pi') \mathscr{H} 
\end{equation}
where for simplicity we have taken $\psi \in\mathscr{P}^\natural$ and 
where as usual we allow for non-trivial support projections which satisfy $\pi' = J \pi J$ in the natural cone case.
Cocycles can be defined as in the state case, and satisfy the same properties as Lemma~\ref{lem:flow}.
These also satisfy the product relation locally relating products of modular data for different weights and states,
analogous to Lemma~\ref{lem:twoproduct}.

We will of course be interested in computing relative entropies. A big difference is that $\psi$ may not be in the domain of $\Delta_{\eta_{\sigma},\psi}^{1/2}$ so 
the usual formula will not result in a positive quantity. Given $\rho \in \mathcal{A}_\star^+$ we construct $\psi = \xi_\rho$ and then define:
\begin{equation}
    \label{eq:specpm}
D^\pm(\rho |\sigma) =  \int_0^{\mp \infty} \lambda \left< \psi \right| dE_\lambda(\ln\Delta_{\eta_\sigma, \psi} ) \left| \psi \right> \in [0,\infty]
\end{equation}
then if one of these is finite then we say that the relative entropy is well defined and set:
\begin{equation}
    \label{eq:defplusminus}
D(\rho |\sigma ) = D^+( \rho|\sigma) - D^-(\rho|\sigma)
\end{equation}
where this takes values in $\mathbb{R} \cup \{-\infty,\infty\}$. 
We can also regularize these as follows:
\begin{equation}
\label{eq:dbp}
D_{\beta}^+(\rho |\sigma) \equiv 
\int_{1/\beta}^{\beta} d \beta' \left< \psi \right| \left( \frac{1}{\beta'+ \Delta_{\eta_\sigma,\psi} }- \frac{1}{\beta'+1}  \right) E_{[0,1]}(\Delta_{\eta_\sigma,\psi}) \left| \psi \right> 
\end{equation}
where $E_{[a,b]}(X)$ denotes the spectral projection on the interval $[a,b]$ and we note that this along with $X= \Delta_{\eta_\sigma,\psi}$ itself always commutes with $\pi'$ which determines the non-zero support of the $\Delta_{\eta_\sigma,\psi}$. This means we don't have to be too careful to include such projections here since they
are absorbed by the vectors $\psi$. 
Also:
\begin{equation}
\label{eq:dbm}
D_{\beta}^-(\rho|\sigma) \equiv
\int_{1/\beta}^{\beta} d \beta' \left< \psi \right| \left( -\frac{1}{\beta'+ \Delta_{\eta_\sigma,\psi}}+ \frac{1}{\beta'+1}  \right) E_{[1,\infty]}(\Delta_{\eta_\sigma,\psi}) \left| \psi \right> 
\end{equation}
which are always well defined and positive. Note that we don't overcount the point $1$ in the spectral representation since the function it multiplies vanishes there (so any delta function measure would integrate to zero). Define also $D_\beta(\rho |\sigma) = D^+_\beta(\rho|\sigma)  - D^-_\beta(\rho|\sigma)  $.

We have the following calculational tools that we summarize as follows:

\begin{lemma}
\label{lem:calculational}
For $\rho \in \mathcal{A}_\star^+$ the relative entropy $D$ and partial versions $D^\pm$ with respect to an unbounded nsf weight satisfy (below
$\psi \in \mathscr{H}$ is any vector for which $\omega_\psi|_{\mathcal{A}} = \rho$):
\begin{enumerate}
\item Suppose that $D(\rho|\sigma) $ is well defined and finite, $-\infty< D( \rho |\sigma) < \infty$, then
the cocycle formula applies, that is the following derivative exists:
\begin{equation}
D( \rho |\sigma) =-i \left. \frac{d}{d s} \left< \psi \right| ( D\sigma: D \rho)_s \left| \psi \right>\right|_{s=0} = -i \left. \frac{d}{d s} \left< \psi \right| \Delta_{\eta_\sigma,\psi}^{-is} \left| \psi \right>\right|_{s=0}
\end{equation}
\item Suppose that $\psi \in \mathcal{D}(\Delta_{\eta_\sigma,\psi}^{\theta})$ for some $\theta > 0$, then the relative entropy is well defined, $>-\infty$, and if we normalize ($\rho(1) = 1$), then
\begin{equation}
\label{eq:theta}
D( \rho |\sigma)  = - \lim_{\theta \rightarrow 0^+} \frac{1}{\theta} \ln \left< \psi \right| \Delta_{\eta_\sigma, \psi}^{\theta} \left| \psi \right> 
= \lim_{\theta \rightarrow 0^+} \frac{1}{\theta}  \left< \psi \right|(1- \Delta_{\eta_\sigma, \psi}^{\theta}) \left| \psi \right> 
\end{equation} 
where the limit on $\theta$ is monotonically non-decreasing and faithful to the possible $+\infty$ value of the relative entropy. 
\item We have monotone convergence:
\begin{equation}
\lim_{\beta \rightarrow \infty} D_{\beta}^{\pm}(\rho|\sigma) = D^{\pm}(\rho|\sigma )
\end{equation}
and these limits are faithful to the possible infinite value. 
\item If the relative entropy is well defined, then:
\begin{equation}
\lim_{\beta \rightarrow \infty} D_{\beta}(\rho|\sigma) = D(\rho|\sigma)
\end{equation}
\item $D_\beta^\pm(\rho|\sigma) \,\, (D^\pm(\rho|\sigma))$ are continuous (lower semicontinuous) in the norm topology on $\rho \in \mathcal{A}_\star^+$.
\end{enumerate}
\end{lemma}
\begin{proof}
(1) Following Ohya-Petz we use the estimate $| \lambda- i (\exp(-is \lambda)-1)/s | \leq 2 | \lambda |$ 
in the spectral representation of $ \ln\Delta_{\eta_\sigma,\psi}$ where the right hand side integrates to $2(D^+ +D^-)$ which is finite by assumption,
so the dominated convergence theorem allows us to take the pointwise limit on $s\rightarrow 0$ inside the integral which proves the derivative formula. (2)
The domain condition implies the integral defining $D^-(\rho|\sigma)$ converges so the relative entropy is well defined and $> -\infty$.
Monotonicity of \eqref{eq:theta} in $\theta$ follows from Jensen's inequality on the spectral measure of $\Delta_{\eta_\sigma,\psi}^{\theta_1}$ for the convex function $x \rightarrow -x^{\theta_2/\theta_1}$
for $\theta_2 < \theta_1$. Jensen's inequality applied to the function $x \rightarrow - \ln x$ gives an upper bound on the limit $\theta \rightarrow 0$ when the relative
entropy is finite. Along with the elementary bound this gives:
\begin{equation}
\label{eq:tochain5}
D(\rho|\sigma) \geq - \lim_{\theta \rightarrow 0^+} \frac{1}{\theta} \ln \left< \psi \right| \Delta_{\eta_\sigma, \psi}^{\theta} \left| \psi \right>  
\geq \lim_{\theta \rightarrow 0^+} \frac{1}{\theta}  \left< \psi \right|(1- \Delta_{\eta_\sigma, \psi}^{\theta}) \left| \psi \right> 
\end{equation}
Projecting the expectation on the right of \eqref{eq:tochain5} with $E_{[0,1]}(\Delta_{\eta_\sigma,\psi})$ the resulting integrand for the spectral measure of $\Delta_{\eta_\sigma,\psi}  $ is monotonically increasing and positive as $\theta \rightarrow 0$ limiting to $D^+(\rho|\sigma)$ by the monotone convergence theorem. Similarly projection on $E_{[1,\infty]}(\Delta_{\eta_\sigma,\psi})$ and multiplying by $-1$
gives a monotonically decreasing positive integrand, with finite integral at fixed $\theta$, allowing another application of monotone convergence
to $D^-(\rho|\sigma)$. This proves equality through the chain \eqref{eq:tochain5}. (3) Straightforward application of monotone convergence.
(4) Follows from (3) and the well defined condition which guarantees at least one
of the terms in (3) has a finite limit and hence there are no ambiguities in combining the limits. 
(5) Take $\rho_n \in \mathcal{A}_\star^+$ with $\rho_n \rightarrow \rho$. This result follows from the following continuity
result for modular operators:
\begin{equation}
so-\lim_n  f(\Delta_{\eta_\sigma,\xi_{\rho_n}} )\pi' =f( \Delta_{\eta_\sigma,\xi_\rho} ) \pi'
\end{equation}
for a continuous bounded function $f$ on $\mathbb{R}_{\geq 0}$ such that the limit $\lim_{x\rightarrow \infty} f(x)$ is well defined. 
This can be found in Ohya-Petz when $\sigma$ is a state, for the weight case we establish this in Appendix~\ref{app:rescont}.
Using the bounded continuous functions:
\begin{equation}
\label{eq:deffpm}
 f_\beta^\pm(x) =\int_{1/\beta}^{\beta} d \beta' \left( \frac{1}{x+\beta'} -\frac{1}{1+\beta'}\right)\theta(\mp(x-1))
\end{equation}
where $\theta$ is the unit step function, along with $(\pi'-1) \xi_{\rho_n} \rightarrow 0$
we can get the continuity statements for $D_\beta^\pm$. 
Lower semicontinuity follows from:
\begin{equation}
\liminf_n D^\pm(\rho_n |\sigma) \geq \lim_n D_\beta^\pm(\rho_n |\sigma) = D_\beta^\pm(\rho |\sigma)
\end{equation}
after which we send $\beta \rightarrow \infty$ using (3). 
\end{proof}

Our next task is to discuss monotonicity of relative entropy. We consider a von Neumann subalgebra $\mathcal{B}\subset \mathcal{A}$ and the following class of weights on it:
\begin{definition}
\label{def:subweight}
Let $\varsigma$ be an nsf weight on $\mathcal{B}$. 
We call $\varsigma$ an nsf subweight of $\sigma$ if additionally:
\begin{equation}
\mathfrak{N}_\varsigma \subset \mathfrak{N}_\sigma \cap \mathcal{B}
\,,\qquad \varsigma|_{\mathfrak{M}_\varsigma} = \sigma|_{\mathfrak{M}_\varsigma} 
\label{eq:agree}
\end{equation}
and we write $\varsigma \subset \sigma$.
We will call $\varsigma$ an nsf cyclic subweight of $\sigma$ if additionally 
$\eta_\sigma(\mathfrak{N}_\varsigma)$ is dense in $\mathscr{H}$. 
\end{definition}

An example of an nsf subweight is simply the restriction of $\sigma$ to $\mathcal{B}$
assuming that $\sigma|_{\mathcal{B}}$ remains semifinite (it must be faithful and normal).
In this paper we track a more general class of subweights where the existence of such an nsf (cyclic) subweight $\varsigma$ implies that $\sigma|_{\mathcal{B} }$ is semifinite (and $\eta_\sigma|_{\mathcal{B}\cap \mathfrak{N}_\sigma}$ is cyclic on $\mathscr{H}$) and in fact:
\begin{equation}
\sigma|_{\mathcal{B}} \leq \varsigma
\end{equation}
We will now prove monotonicity in this more general setting.

For $\varsigma$ an nsf cyclic subweight of $\sigma$ the GNS representation of $\varsigma$
can be taken on $\mathscr{H}$ such that the semicyclic representation $\eta_\varsigma : \mathfrak{N}_\varsigma \rightarrow \mathscr{H}$ satisfies $\eta_\varsigma = \eta_\sigma|_{\mathfrak{N}_\varsigma }$ due to $\eqref{eq:agree}$. In other words $\mathscr{H}$ is still in standard form for $\eta_\varsigma$.
Also the Tomita operators agree on a dense subspace:
\begin{equation}
S_\varsigma \eta_\varsigma(n) = S_\sigma \eta_{\varsigma}(n) 
\qquad n \in \mathfrak{N}_\varsigma \cap \mathfrak{N}_\varsigma^\dagger
\end{equation}
which is a core for the left hand side, so 
the graph closure of the Tomita operator gives the proper extension $\overline{S}_\varsigma \subset \overline{S}_\sigma$. The polar decomposition is $\overline{S}_\varsigma = J_\mathcal{B} \Delta_\varsigma^{1/2}$ with the usual properties. In particular there is a new positive self dual natural
cone that we will label $\mathscr{P}^\natural_\mathcal{B}$.
This cone and $J_\mathcal{B}$ are different to $\mathscr{P}^\natural$ and $J$. We will now introduce notation that labels only the algebra, leaving the weight/subweight as understood:
 \begin{equation}
 \Delta_{\sigma} \equiv \Delta_{\mathcal{A}}\,, \quad \Delta_\varsigma \equiv \Delta_{\mathcal{B}} \,,\quad
J \equiv J_\mathcal{A}\,, \quad \mathscr{P}^\natural \equiv \mathscr{P}_\mathcal{A}^\natural 
 \end{equation}
along with:
 \begin{equation}
 \mathfrak{N}_\sigma = \mathfrak{N}_\mathcal{A} \,,\quad  \mathfrak{N}_\varsigma = \mathfrak{N}_\mathcal{B}  \qquad \eta_\sigma = \eta_{\mathcal{A}} \qquad \eta_\varsigma = \eta_{\mathcal{B}}  
 \end{equation}
 Also consider the relative modular data 
 that we defined previously associated to $\mathcal{A}$, $\sigma$ and $\psi \in \mathscr{H}$. Since we will momentarily introduce relative modular data for the subalgebra we will also label these now as:
 \begin{equation}
  \Delta_{\eta_\sigma, \psi} \equiv  \Delta_{\eta_\mathcal{A}, \psi;\mathcal{A}}\,, \quad \pi \equiv\pi_\mathcal{A}\,,\quad  \pi'\equiv \pi'_\mathcal{A}\,, \quad \theta\equiv\theta_\mathcal{A}\,,\quad  \theta'\equiv\theta'_\mathcal{A}
 \end{equation}

We can also define relative entropies for the subalgebra.
Start with the state $\rho = \omega_{\psi}|_{\mathcal{A}}$ further restricted to $\rho|_{\mathcal{B}}$; we will consider the relative entropy of this state with respect to the subweight $\varsigma$.
 We use the relative modular operator for $\xi_{\rho|_{\mathcal{B}}} \in \mathscr{P}^\natural_{\mathcal{B}}$
\begin{equation}
J_{\mathcal{B}} \Delta_{\eta_\mathcal{B}, \xi_{\rho|_{\mathcal{B}}} ;\mathcal{B}}^{1/2} \left( n \left| \xi_{\rho|_{\mathcal{B}}}  \right> + \left| \xi' \right>\right) = \eta_\mathcal{B}(\pi_{\mathcal{B}} n^\dagger)  \qquad \forall n^\dagger \in \mathfrak{N}_{\mathcal{B}},\,
\xi' \in (1- J_{\mathcal{B}} \pi_{\mathcal{B}} J_{\mathcal{B}} ) \mathscr{H}
\end{equation}
where we again allow for non-trivial support projections $\pi_{\mathcal{B}}$ for the state $\rho$. In general a vector $\psi \in \mathscr{P}^\natural_\mathcal{A}$ will not be in the corresponding
natural cone for $\mathcal{B}$ where we can use a partial isometry to fix this $ \theta_{\mathcal{B}}' \psi  = \xi_{\rho|_{\mathcal{B}}} \in \mathscr{P}^\natural_\mathcal{B}$  where
$(\theta_{\mathcal{B}}')^\dagger \theta_{\mathcal{B}}' = \pi_{\mathcal{B}}'$ and 
$\theta_{\mathcal{B}}' (\theta_{\mathcal{B}}')^\dagger
= J_{\mathcal{B}} \pi_{\mathcal{B}}J_{\mathcal{B}}$ after which we can use the 
relation $\Delta_{\eta_{\mathcal{B}},\psi ;\mathcal{B}} = (\theta_{\mathcal{B}}')^\dagger \Delta_{\eta_{\mathcal{B}},  \xi_{\rho|_{\mathcal{B}}}  ;\mathcal{B}} \theta_{\mathcal{B}}'$ to define
the more general modular operators. 

We then have the following monotonicity result on relative entropy:
\begin{lemma}
\label{lem:mono-weight}
For a von Neumann subalgebra $\mathcal{B} \subset \mathcal{A}$ with an nsf weight $\sigma$ and an nsf cyclic subweight $\varsigma$ for $\sigma$,
then if the relative entropies $D(\rho|\sigma),D(\rho|_{\mathcal{B}}|\varsigma)$ are well defined for both $\mathcal{A}$ and $\mathcal{B}$, monotonicity applies
for all $\rho \in \mathcal{A}_\star^+$:
\begin{equation}
    \label{eq:provemono}
D(\rho|_{\mathcal{B}}|\varsigma) \leq D(\rho|\sigma)
\end{equation}
Furthermore, for a general state we have the bounds:
\begin{align}
    \label{eq:plusmono}
D^+(\rho|_{\mathcal{B}}|\varsigma)
\leq \rho(1) 2  \ln 2 + D^+(\rho| \sigma)\,\\
D^-(\rho|\sigma)
\leq \rho(1) 2  \ln 2 + D^-(\rho|_{\mathcal{B}}| \varsigma)\,
\label{eq:minusmono}
\end{align}
where either side could be infinite. 
Hence if $D(\rho|\sigma)$ is well defined and $<\infty$, 
then we can conclude that $ D(\rho|_{\mathcal{B}}|\varsigma)$ is well defined and $<\infty$.
Similarly if we know that $ D(\rho|_{\mathcal{B}}|\varsigma)$ is well defined and $> -\infty$, then
we can conclude that $D(\rho|\sigma)$ is well defined and $> -\infty$.
\end{lemma}
\begin{proof}
The proper extension applies for the closed unbounded antilinear operators for $\psi \in \mathscr{P}^\natural_{\mathcal{A}}$
\begin{equation}
\label{eq:twoops}
\overline{\pi_\mathcal{A} S_{\mathcal{B}}} 
\subset S_{\mathcal{A}} \pi_\mathcal{B}'
\qquad S_{\mathcal{B}} \equiv J_\mathcal{B}\theta_{\mathcal{B}}' \Delta_{\eta_\mathcal{B},\psi;\mathcal{B}}^{1/2}  \,,\quad
S_{\mathcal{A}} \equiv J_{\mathcal{A}} \Delta_{\eta_\mathcal{A},\psi;\mathcal{A}}^{1/2}
\end{equation}
where one can check that these operators agree on the dense domain $n \psi + \xi'$ where $\xi' \in (1-\pi_\mathcal{B}')\mathscr{H}$
and $n^\dagger \in \mathfrak{N}_\mathcal{B}$ (with image $\pi_\mathcal{A} \eta_{\mathcal{B}}( \mathfrak{N}_\mathcal{B})$) which is a core for the left hand side. 
This implies for all $\beta > 0$
\begin{align}
\frac{1}{\beta + \Delta_{\eta_{\mathcal{B}},\psi;\mathcal{B}}} \pi_{\mathcal{B}}'
& \leq \frac{1}{\beta + S_{\mathcal{B}}^\dagger \overline{\pi_\mathcal{A} S_{\mathcal{B}} } } \pi_{\mathcal{B}}'  \leq
\frac{1}{\beta+ \pi_{\mathcal{B}}'   \Delta_{\eta_{\mathcal{A}},\psi;\mathcal{A}} \pi_{\mathcal{B}}'  } \pi_{\mathcal{B}}' 
\leq \pi_{\mathcal{B}}'  \frac{1}{\beta +   \Delta_{\eta_{\mathcal{A}},\psi;\mathcal{A}}   }\pi_{\mathcal{B}}'
\label{eq:resbound}
\end{align}
where in the first inequality we used $\pi_{\mathcal{A}} \leq \pi_{\mathcal{B}}$ so that
$ S_{\mathcal{B}}^\dagger \overline{\pi_\mathcal{A} S_{\mathcal{B}}}
\leq  \Delta_{\eta_{\mathcal{B}},\psi;\mathcal{B}}$, in the second inequality we used the inequality on
 projections to the two closed subspaces of the graph of the two operators in \eqref{eq:twoops} and
in the third inequality we used the operator Jensen inequality.
Taking expectations in the state $\psi$ and integrating gives:
\begin{equation}
D_{\beta}^{+}(\rho|\sigma) - D_{\beta}^{-}(\rho|\sigma) 
\geq D_{\beta}^{+}(\rho|_{\mathcal{B}}|\varsigma) - D_{\beta}^{-}(\rho|_{\mathcal{B}}|\varsigma)
\end{equation}
If both relative entropies are well defined and finite we can take the limit $\beta \rightarrow \infty$ freely to prove monotonicity \eqref{eq:provemono}.

Consider now \eqref{eq:plusmono}.
In $D_\beta^+(\rho|_{\mathcal{B}}|\varsigma)$ we estimate the part of the $\beta'$
integral from $(1,\beta)$ as:
\begin{equation}
D_\beta^+(\rho|_{\mathcal{B}}|\varsigma)
\leq \rho(1) \ln 2 + \int_{1/\beta}^1 d \beta' \left< \psi \right| \left( \frac{1}{\beta'+ \Delta_{\eta_\mathcal{B},\psi;\mathcal{B}} }- \frac{1}{\beta'+1}  \right) E_{[0,1]}(\Delta_{\eta_\mathcal{B},\psi;\mathcal{B}}) \left| \psi \right>  
\end{equation}
we can then drop the second term in the bracket inside the integral
after which we can drop the projection $E_{[0,1]}(\Delta_{\eta_\mathcal{B},\psi;\mathcal{B}})$
and use the bound in \eqref{eq:resbound} to find:
\begin{align}
D_\beta^+(\rho|_{\mathcal{B}}|\varsigma)
&\leq \rho(1) \ln 2 + \int_{1/\beta}^1 d \beta' \left< \psi \right|  \frac{1}{\beta'+ \Delta_{\eta_\mathcal{A},\psi;\mathcal{A}} } \left| \psi \right>   \\
&= - \rho(1) \ln( \frac{1}{4}\left( \frac{1}{\beta}+1 \right)) + \int_{1/\beta}^1 d \beta' \left< \psi \right| \left( \frac{1}{\beta'+ \Delta_{\eta_\mathcal{A},\psi;\mathcal{A}} } - \frac{1}{\beta' + 1} \right) \left| \psi \right> \\
& \leq\rho(1) 2  \ln 2  + \int_{1/\beta}^1 d \beta' \left< \psi \right| \left( \frac{1}{\beta'+ \Delta_{\eta_\mathcal{A},\psi;\mathcal{A}} } - \frac{1}{\beta' + 1} \right) E_{[0,1]}(\Delta_{\eta_\mathcal{A},\psi;\mathcal{A}}) \left| \psi \right>   \\
&\leq \rho(1) 2  \ln 2 + D_\beta^+(\rho| \sigma) 
\end{align} 
where in the second inequality we dropped the projection $(1- E_{[0,1]}(\Delta_{\eta_\mathcal{A},\psi;\mathcal{A}}))$ since it contributes a negative amount to the integral. The last inequality
extends the $\beta'$ integral back to $(1/\beta,\beta)$. 
Taking the limit $\beta \rightarrow \infty$ gives the bound \eqref{eq:plusmono} which is faithful to possible values of $\infty$.

For \eqref{eq:minusmono} we follow essentially the same steps summarized below
without commentary:
\begin{align}
D_\beta^-(\rho |\sigma)
& \leq \rho(1) \ln 2 + \int_{1}^\beta \frac{d \beta'}{\beta'} \left< \psi \right| \left( \frac{\Delta_{\eta_\mathcal{A},\psi;\mathcal{A}}}{\beta'+ \Delta_{\eta_\mathcal{A},\psi;\mathcal{A}} }- \frac{1}{\beta'+1}  \right) E_{[1,\infty]}(\Delta_{\eta_\mathcal{A},\psi;\mathcal{A}}) \left| \psi \right>   \\
&\leq \rho(1) \ln 2 + \int_{1}^\beta \frac{d \beta'}{\beta'} \left< \psi \right| \frac{\Delta_{\eta_\mathcal{B},\psi;\mathcal{B}}}{\beta'+ \Delta_{\eta_\mathcal{B},\psi;\mathcal{B}}} \left| \psi \right>   \\
&= - \rho(1) \ln( \frac{1}{4}\left( \frac{1}{\beta}+1 \right)) + \int_{1}^\beta d \beta' \left< \psi \right| \left( -\frac{1}{\beta'+ \Delta_{\eta_\mathcal{B},\psi;\mathcal{B}} } + \frac{1}{\beta' + 1} \right) \left| \psi \right> \\
& \leq\rho(1) 2  \ln 2  + \int_{1}^\beta d \beta' \left< \psi \right| \left( -\frac{1}{\beta'+ \Delta_{\eta_\mathcal{B},\psi;\mathcal{B}} } + \frac{1}{\beta' + 1} \right) E_{[1,\infty]}(\Delta_{\eta_\mathcal{B},\psi;\mathcal{B}}) \left| \psi \right>   \\
&\leq \rho(1) 2  \ln 2 + D_\beta^-(\rho|_{\mathcal{B}}| \varsigma) 
\end{align} 
Taking the limit $\beta \rightarrow \infty$ gives the bound \eqref{eq:minusmono} which is faithful to possible values of $\infty$.

If $D(\rho|\sigma)$ is well defined 
and finite this implies that $D^+(\rho|\sigma) < \infty$ and so from \eqref{eq:plusmono} we conclude
that $D^+(\rho|_{\mathcal{B}}|\varsigma)$ is finite so that $D(\rho|_{\mathcal{B}}|\varsigma)$ is well defined (possibly $-\infty$). Similarly if $D(\rho|_{\mathcal{B}}|\varsigma)$ is well defined  and finite then we can use \eqref{eq:minusmono} to show that $D(\rho|\sigma)$ is well defined (possibly $+\infty$). 

If $D(\rho|\sigma)$ is well defined but $=-\infty$ then \eqref{eq:minusmono} establishes
that $D(\rho|_{\mathcal{B}}|\varsigma) = -\infty$ and if $D(\rho|_{\mathcal{B}}|\varsigma)$ is well defined but $=\infty$ then \eqref{eq:plusmono} establishes that $D(\rho|\sigma) = \infty$.
\end{proof}

Next we prove a semicontinuity result for the relative entropy difference:

\begin{lemma}
\label{lem:weight-semi}
Consider a net $\rho_\iota \in \mathcal{A}_\star^+$ converging to $\lim_\iota \rho_\iota = \rho \in  \mathcal{A}_\star^+$ in the linear functional norm sense.
For a von Neumann subalgebra $\mathcal{B} \subset \mathcal{A}$ with an nsf weight $\sigma$ and an nsf cyclic subweight $\varsigma$ for $\sigma$, suppose that the relative entropies $D(\rho_\iota|\sigma)$ and $D(\rho_\iota|_{\mathcal{B}}|\varsigma)$ are well defined and finite 
for sufficiently large $\iota$. If $D(\rho|\sigma),D(\rho|_{\mathcal{B}}|\varsigma)$ are also well defined with $D(\rho|\sigma) > -\infty$
and $D(\rho|_{\mathcal{B}}|\varsigma) <\infty$, then
\begin{equation}
\label{eq:toprovesemi}
\liminf_\iota \left( D(\rho_\iota |\sigma)  - D(\rho_\iota|_{\mathcal{B}} |\varsigma)  \right) 
\geq D(\rho|\sigma)  - D(\rho|_{\mathcal{B}}|\varsigma)  \geq 0
\end{equation}
\end{lemma}
\begin{proof}
We have the formula for the relative entropy difference in the case that both relative entropies are well defined and finite:
\begin{align}
D(\rho_\iota |\sigma)  - D(\rho_\iota|_{\mathcal{B}} |\varsigma) 
&=\lim_{\beta \rightarrow \infty} D_{\beta}(\rho_\iota|\sigma)  - D_{\beta}(\rho_\iota|_{\mathcal{B}}|\varsigma)  \\
&\geq \int_{1/\beta}^{\beta} d\beta' \left< \psi_\iota \right| \frac{1}{\beta'+ \Delta_{\eta_{\mathcal{A}},\psi_\iota;\mathcal{A}}} -\frac{1}{\beta'+ \Delta_{\eta_{\mathcal{B}},\psi_\iota;\mathcal{B}}}
\left| \psi_\iota \right>  \\
& = \left< \psi_\iota \right| f_\beta(\Delta_{\eta_{\mathcal{A}},\psi_\iota;\mathcal{A}}) -f_\beta(\Delta_{\eta_{\mathcal{B}},\psi_\iota;\mathcal{B}})
\left| \psi_\iota \right> 
\end{align}
where $\psi_\iota = \xi_{\rho_{\iota}} \in \mathscr{P}^\natural_{\mathcal{A}}$ and $f_\beta(x) = f_\beta^++ f_\beta^-$ defined in \eqref{eq:deffpm}. The bound is
due to positivity in \eqref{eq:resbound}.
But the modular operators vary continuously with the state for bounded continuous functions $f_\beta$ (when acting respectively on $\pi_\mathcal{A}' \mathscr{H}$ and $\pi_\mathcal{B}' \mathscr{H}$), see Appendix~\ref{app:rescont}.\footnote{This continuity result applies to convergence along a net $\iota$,
since we have stated it in a way that avoids the use of dominated convergence, which does not work for convergence along a net $\iota$. Instead
the essential step uses Stone-Weierstrass.} Hence:
\begin{align}
\liminf_\iota D(\rho_\iota |\sigma)  - D(\rho_\iota|_{\mathcal{B}} |\varsigma) 
& \geq \int_{1/\beta}^{\beta} d\beta' \left< \psi \right| \frac{1}{\beta'+ \Delta_{\eta_{\mathcal{A}},\psi;\mathcal{A}}} -\frac{1}{\beta'+ \Delta_{\eta_{\mathcal{B}},\psi;\mathcal{B}}} \left| \psi \right>\\
& = D_\beta(\rho |\sigma)  - D_\beta(\rho|_{\mathcal{B}} |\varsigma) 
\end{align}
where $\psi = \xi_{\rho}$. We can take $\beta \rightarrow \infty$ on the difference, using Lemma~\ref{lem:calculational}(4),
because the assumptions avoid any $\infty-\infty$ ambiguity.  We arrive at \eqref{eq:toprovesemi}.
\end{proof}

\subsection{Weights, HSTs and HSMIs}
\label{sec:weighthsmi}

The weight case for a half sided modular inclusion has largely been worked out in \cite{araki2005extension}. 
We focus on the half sided translation case where there is a more
limited version of Theorem~\ref{thm:hsmi}:
\begin{theorem}
\label{thm:hsmi-weight}
Consider a von Neumann algebra $\mathcal{A} \subset \mathcal{B}(\mathscr{H})$ in standard
form (with modular conjugation $J$) and based on a faithful normal semi-finite weight $\sigma$, and where $\mathscr{H}$ furnishes a strongly continuous unitary representation of $\mathbb{R}$, that is $U(a)$ for $a \in \mathbb{R}$, satisfying:
\begin{equation}
\label{eq:Uainclusion-weight}
U(a)^\dagger \mathcal{A} U(a) \subset \mathcal{A} \qquad \forall\,a > 0
\end{equation}
Consider the following extra conditions:
\begin{itemize}
    \item[(a)] The semicyclic representation is invariant: $
    U(-a) \mathfrak{N}_\sigma U(a) \subset \mathfrak{N}_\sigma$ with,
\begin{equation}
    \label{eq:weight-semi}
    U(-a) \eta_\sigma(n) = \eta_\sigma( U(-a) n U(a)) \qquad \forall n \in \mathfrak{N}_\sigma\,, \quad a>0 
\end{equation}
\item[(b)] The generator $P$ of $U(a) = e^{- i aP}$ is positive 

\item[(c)] The following algebra applies:
\begin{equation}
\Delta_{\sigma}^{-is} U(a) \Delta_{\sigma}^{is} = U( a e^{2\pi s})\,,\quad J U(a) J = U(-a)
\end{equation}
\end{itemize}
Then (a)+(b) $\iff$ (a)+(c). If $\sigma(1) < \infty$, then $(a)+(b) \iff (c)$. 
\end{theorem}
\begin{proof}
See Appendix~\ref{app:hsmi-proof}.
\end{proof}

We give the weight version of Definition~\ref{def:hst}:
\begin{definition}
    \label{def:tr-weight}
The space of $P$'s satisfying the assumptions (a,b,c) of the above theorem will be defined as $\mathcal{TR}(\mathcal{A},\sigma)$
where $\sigma$ is an nsf weight for $\mathcal{A}$ and $\mathscr{H}$ is the standard form Hilbert space for $\sigma$. 
\end{definition}

\begin{remark}
Note that the assumptions defining $\mathcal{TR}(\mathcal{A},\sigma)$ give only
 $\sigma(U(-a) m U(a))= \sigma(m) $ for $m \in \mathfrak{M}_\sigma$ which gives:
 \begin{equation}
    \label{eq:boundtranslate}
 \sigma|_{\mathcal{A}_a} \leq  \sigma|_{\mathcal{A}}(U(a) \cdot U(-a)) 
 \end{equation}
 with equality for $U(-a)\mathfrak{M}_\sigma U(a) \subset \mathcal{A}_a$. In general
 the left hand side can be defined on more elements of $\mathcal{A}_a$ and so could be a proper extension. In the state case this is not possible.
\end{remark}

The weight version of the HSMI now reads:
\begin{definition}
\label{def:hsmi-weight}
We say that $\mathcal{B} \subset_\sigma \mathcal{A}$ forms a HSMI for nsf weight $\sigma$ if modular flow satisfies $\sigma_{\sigma;\mathcal{A}}^s(\mathcal{B}) \subset \mathcal{B}$ for all $s \geq 0$, $\sigma$ is nsf restricted to $\mathcal{B}$ and the semicyclic representation of $\sigma$ is cyclic for $\mathcal{B} \cap \mathfrak{N}_\sigma$.
The space of such von Neumann subalgebras will be denoted as $\mathcal{HS}(\mathcal{A},\sigma)$
and in short, we say that  $\mathcal{B} \in \mathcal{HS}(\mathcal{A},\sigma)$ forms a HSMI. 
\end{definition}

Given a HSMI there is always a corresponding half sided translation \cite{araki2005extension}
with:
\begin{equation}
    \label{eq:promise}
\widetilde{U}(a) = e^{ -i a \widetilde{P}}\,, \qquad 2\pi \widetilde{P} = \overline{\ln \Delta_{\sigma|_{\mathcal{B}} } 
- \ln \Delta_{\sigma} }
\end{equation}
and this generates a nested set of algebras that interpolate $\mathcal{B} \subset \mathcal{A}$. 

\begin{theorem}
    \label{thm:trtohm}
The map $P \rightarrow e^{i P} \mathcal{A} e^{- i P}$ from $\mathcal{TR}(\mathcal{A},\sigma)$ to $\mathcal{HS}(\mathcal{A},\sigma)$ is surjective. Restricting
to half sided translations for which $\sigma|_{\mathcal{A}_a} = \sigma( U(a) \cdot U(-a))$
for all $a\geq 0$,  this map becomes a bijection where the inverse is given by $\mathcal{B} \rightarrow \widetilde{P}$ in \eqref{eq:promise}. 
\end{theorem}
\begin{proof}
See Appendix~\ref{app:hsmi-proof}. The proof relies on Lemma~\ref{lem:onlyc} given below. 
\end{proof}

Although the half sided translation definition for a weight cannot obviously forgo requiring the covariance of the semi-cyclic representation (a) in Theorem~\ref{thm:hsmi-weight}, we can proceed with only (c) of Theorem~\ref{thm:hsmi-weight} along with an extra condition on the definition domain of the weight. 
Then, as long as we are happy to replace the translation operator with one that preserves the same one parameter family of translated algebras, we still arrive at the structure of a half sided translation/modular inclusion:
\begin{lemma}
    \label{lem:onlyc}
Consider a von Neumann algebra $\mathcal{A} \subset \mathcal{B}(\mathscr{H})$
with nsf weight $\sigma$ and translations $U(a)$ satisfying the properties in the preamble and (c) of Theorem~\ref{thm:hsmi-weight} as well as:
\begin{equation}
    \label{eq:condN}
U(-a) \mathfrak{N}_\sigma U(a) \subset \mathfrak{N}_\sigma\,, \qquad a > 0
\end{equation}
then $\mathcal{A}_a \equiv U(-a) \mathcal{A} U(a)\subset_\sigma \mathcal{A}$ for $a > 0$ is a HSMI, i.e. $\mathcal{A}_a \in \mathcal{HS}(\mathcal{A},\sigma)$, with the same translated algebras $\widetilde{U}(-a)\mathcal{A}\widetilde{U}(a)  = 
\mathcal{A}_a$ for all $a \in \mathbb{R}$ where we define $\widetilde{U}(a)$ using \eqref{eq:promise} for $\mathcal{B} = \mathcal{A}_1$.
\end{lemma} 
\begin{proof}
See Appendix~\ref{app:hsmi-proof}.
\end{proof}

Given some half-sided translation structure for the weight $\sigma$, i.e. $P \in \mathcal{TR}(\mathcal{A},\sigma)$ given in Definition~\ref{def:tr-weight} we can use this 
to construct the translated family of algebras:
\begin{equation}
\mathcal{A}_V \equiv U(V)^\dagger \mathcal{A} U(V) \,, \quad V \in \mathbb{R}
\end{equation}
along with a family of translated nsf weights on $\mathcal{A}_V$
\begin{equation}
\varphi_V \equiv \sigma( U(V) \cdot U(-V)) \,, \quad V \in \mathbb{R}
\end{equation}
which are finite on $\mathfrak{M}_V \equiv U(-V) \mathfrak{M}_\sigma U(V)$. We see that $\mathscr{H}$ forms a semicyclic representation for $\mathcal{A}_V$ with 
\begin{equation}
\eta_V \equiv  U(V)^\dagger \eta_\sigma( U(V) \cdot U(V)^\dagger)\,, \quad V \in \mathbb{R}
\end{equation} 
defined on $\mathfrak{N}_V \equiv U(-V) \mathfrak{N}_\sigma U(V)$
and dense in $\mathscr{H}$. Now the HST structure implies that:
\begin{equation}
    \label{eq:weused}
\mathcal{A}_{V_1} \subset \mathcal{A}_{V_2}\,, \qquad \varphi_{V_1} \subset \varphi_{V_2}
\end{equation}
for $V_1 > V_2$. The subweight $\varphi_{V_1} \subset \varphi_{V_2}$ is an nsf cyclic subweight, see Definition~\ref{def:subweight}. 

In fact we can define a different weight on the family $\mathcal{A}_V$
via restriction and this is also an nsf cyclic subweight.
For $V > 0$ we can define $\varphi|_V = \sigma|_{\mathcal{A}_V}$.
For the other $V$ we can construct a new translation operator $\widetilde{P}$
using the inverse of the map in Theorem~\ref{thm:trtohm} and then define
the common weight $\varphi$ on $\bigcup_V \mathcal{A}_V^+$ as follows:
\begin{equation}
\varphi : m \in \bigcup_V \mathcal{A}_V^+ \rightarrow \begin{cases} \sigma(\widetilde{U}(V)  m \widetilde{U}(-V) ) & \exists V :  m \in \widetilde{U}(-V) \mathfrak{M}^+_\sigma \widetilde{U}(V)
\\ \infty & {\rm otherwise} \end{cases}
\end{equation}
Note that if there are multiple $V$ that apply in the top case then the answer is consistent 
due to $\widetilde{P}$ translation invariance of the weight shown in \cite{araki2005extension}.
Furthermore the restriction of $\varphi$ to $\mathcal{A}_V^+$ for all $V \in \mathbb{R}$ satisfies \cite{araki2005extension}:
\begin{equation}
\varphi|_{V} = \sigma( \widetilde{U}(V) \cdot \widetilde{U}(V)^\dagger)
\end{equation}
and this extends to an nsf weight which is finite on the linear span of the definition domain $\mathfrak{M}|_V = \widetilde{U}(V)^\dagger \mathfrak{M}_\sigma \widetilde{U}(V)$.
For $V \geq 0$ this weight satisfies $\varphi|_{V} \leq \varphi_V$  with agreement on $\mathfrak{M}_V \subset \mathfrak{M}|_V$.

We need the commutant HSMI. 
Since $\mathscr{H}$ is in standard form we get a half sided modular inclusion on the commutants.
More specifically we note that $\sigma' = \sigma(J \cdot J)$ is an nsf weight on $\mathcal{A}'$
and
\begin{equation}
U(-a) \mathcal{A}' U(a) \subset \mathcal{A}' \qquad a < 0
\end{equation}
satisfying $U(a) \eta_{\sigma'}(U(-a) \cdot U(a)) = \eta_{\sigma'}(\cdot)$ for $a<0$.
Hence we have a half sided translation in the opposite sense. 
We note that $\mathcal{A}_V' = U(V)^\dagger \mathcal{A}' U(V)$ forms the complement family
with a family of weights $\varphi_V'$ and restricted weights $\varphi|_{V'}$.
We also define $\mathfrak{N}_{V'} = U(V)^\dagger \mathfrak{N}_{\sigma'} U(V)$ and the semicyclic representation is defined via $\eta_{V'}(n') = U(-V) \eta_{\sigma'}( U(V) n' U(-V))$.

The modular data for the weight $\varphi_V$ will be labelled $J_V, \Delta_V$
which are unitarily related to: $\Delta_V = U(-V) \Delta_\sigma U(V)$ and $J_V = U(-V) J U(V)$.
The relative modular operators, support projections and cocycles 
will be labelled in essentially the same way as the state case. They satisfy the same properties as outlined in Lemma~\ref{lem:flow} for the state case.

In the next two sections we prove the two main results that allow us to prove the QNEC.
The first is the relative entropy sum rule, that is the weight version of Proposition~\ref{prop:sum}, while the second is finiteness
of $P$ in various purifications, the weight case of Proposition~\ref{prop:finite}. 

\subsection{Relative entropy sum rule, weight case}
\label{sec:sumrule}

We now consider relative entropies in the half sided modular inclusion setting
for some $\psi \in \mathscr{H}$
\begin{equation}
D(\psi|_V) \equiv D(\omega_\psi|_V|\varphi_V)\,,
\qquad
D(\psi|_{V'}) \equiv D(\omega_\psi|_{V'}| \varphi_{V'})
\end{equation}
where use of these expressions should be qualified with the requirement that the relative entropy is well defined.

For $V\geq 0$, compared to the relative entropy for the restricted weights $\varphi|_V$ we have the bound:
\begin{equation}
\widetilde{D}(\psi|_V) - D(\psi|_V) \geq 0\,, \qquad \widetilde{D}(\psi|_V) \equiv  D(\omega_\psi|_V|\varphi|_V) 
\end{equation}
with equality when $V=0$. In fact our proof of the QNEC for $D(\psi|_V)$ also applies to $\widetilde{D}(\psi|_V)$ (albeit with a different translation generator) so these will both be monotonic convex functions. There is no obvious bound between the two functions when $V< 0$. 

Note that a simple result that follows from $\varphi_{V_2} \subset \varphi_{V_1}$ when $V_1 \leq V_2$ and Lemma~\ref{lem:mono-weight} is the following. Suppose that $D(\psi|_V)$ is well defined and finite for some $V$ then it is well defined 
and monotonic non-increasing for all $V$. The same applies to the commutant.

Our main task for this section is then to prove the relative entropy sum rule:
\begin{proposition}
\label{prop:sum-weight}
Consider two cuts $V_1 < V_2$ and $\psi \in \mathcal{D}(\sqrt{P})$. Suppose that the four relative entropies
$D(\psi|_{V_1})$, $D(\psi|_{V_1'})$, $D(\psi|_{V_2})$ and $D(\psi|_{V_2'})$ are well defined and satisfy
\[
D(\psi|_{V_2}),D(\psi|_{V_1'}) < \infty \,, \qquad -\infty < D(\psi|_{V_2'}),D(\psi|_{V_1}).
\]
Then:
\begin{equation}
\label{eq:sumweight}
2\pi  (V_2-V_1) \| \sqrt{P} \left| \psi \right> \|^2 = \big( D(\psi|_{V_1}) - D(\psi|_{V_2}) \big) + \big( D(\psi|_{V_2'}) - D(\psi|_{V_1'}) \big)
\end{equation}
Furthermore for all $V \in \mathbb{R}$ the relative entropies
are well defined and finite: $-\infty < D(\psi|_{V}),D(\psi|_{V'}) < \infty$.
\end{proposition}

We start by proving \eqref{eq:sumweight} under some more restrictive assumptions. 

\begin{proposition*}{\bf\ref{prop:sum-weight}\,(a)} The sum rule \eqref{eq:sumweight} holds for all $\psi \in \mathscr{H}$ under the extra assumption that $D(\psi|_{V_1}),D(\psi|_{V_2'}) < \infty$ and
$-\infty < D(\psi|_{V_2}), D(\psi|_{V_1'})$. That is, we assume all relative entropies in the sum rule are well defined and finite and $\psi \in \mathcal{D}(\sqrt{P})$.
\end{proposition*}

\begin{proof}
We consider the cocycle relation with $a = V_2 - V_1 > 0$ derived using the weight analog of Lemma~\ref{lem:flow} and the HSMI relations Theorem~\ref{thm:hsmi-weight} (c). 
\begin{align}
&\left<\psi \right| (D\varphi_{V_1} :D\omega_\psi|_{V_1})_s^\dagger (D\varphi_{V_2} :D\omega_\psi|_{V_2})_s \left| \psi \right>\\
&= \left<\psi \right| (D\varphi_{V_1'} :D\omega_\psi|_{V_1'})_s^\dagger \exp( -i a( 1-e^{-2\pi s}) P) (D\varphi_{V_2'} :D\omega_\psi|_{V_2'})_s \left| \psi \right>
\end{align}
and we can take derivatives at $s=0$ as follows. Let $w_s, v_s, w_s', v_s'$ be the cocycles appearing above so that
the above identity reads $\omega_\psi((w_s)^\dagger v_s) = \omega_\psi( (w_s')^\dagger e^{ -i a_s P} v_s')$
where $a_s = a (1-e^{-2\pi s}) \in \mathbb{R}$. Then we can turn this
identity into:
\begin{align}
\label{eq:bothhands}
&\omega_\psi( (w_s)^\dagger-1) + \omega_\psi(v_s-1) - \omega_\psi((w_s')^\dagger-1) - \omega_\psi(v_s'-1) - \omega_\psi ( e^{ -i a_s P}-1)
\\ &= -\omega_\psi( ( (w_s)^\dagger-1)(v_s-1)) + \omega_\psi(((w_s')^\dagger-1) ( e^{ -i a_s P}-1)(v_s'-1))   \nonumber \\
\nonumber
&+ \omega_\psi(((w_s')^\dagger-1) ( e^{ -i a_s P}-1)) + \omega_\psi(((w_s')^\dagger-1) (v_s'-1))+ \omega_\psi(( e^{ -i a_s P}-1)(v_s'-1))
\nonumber
\end{align}
If we divide by $s$ and take the limit $s \rightarrow 0^+$ the right hand side of \eqref{eq:bothhands} vanishes due to Cauchy-Schwarz estimates such as:
\begin{equation}
|\omega_\psi( ( (w_s)^\dagger-1)(v_s-1))|/s \leq 
(2{\rm Re} \omega_\psi( 1-w_s)/s )^{1/2}(2{\rm Re} \omega_\psi( 1-v_s)/s )^{1/2} 
\end{equation}
where we have used that the cocycles square to a projection that fixes $\psi$.
Both of the square root terms on the right vanish as $s \rightarrow 0^+$  since the relative entropies are well defined and finite so the derivative of $\omega_\psi(v_s)$ exists
at $s=0$ and is purely imaginary. See Lemma~\ref{lem:calculational}(1). Similar arguments apply to all the terms on the right hand side, since all the relative entropies
and $P_\psi$'s are finite. Hence the right hand side of \eqref{eq:bothhands} vanishes in the limit. Lemma~\ref{lem:calculational}(1) then allows
us to calculate the limits on the left hand side of \eqref{eq:bothhands} which must add up to zero and this lands on the sum rule \eqref{eq:sumweight}.
\end{proof}

At this point we note that we can assume that $\varphi$ is indeed unbounded $\varphi(1) = \infty$ (the bounded relative entropy sum rule Proposition~\ref{prop:sum} was proven with only the input being the proof immediately above).

\begin{proposition*}{\bf\ref{prop:sum-weight}\,(b)} 
The sum rule \eqref{eq:sumweight} holds for all $\psi \in \mathcal{D}(\Delta_{\eta_{V_2},\psi;V_2}^\theta) \cap \mathcal{D}(\Delta_{\eta_{V_1'},\psi; V_1'}^\theta)$
for some $0< \theta\leq 1/2$ along with the rest of the assumptions in Proposition~\ref{prop:sum-weight}.
\end{proposition*}
\begin{proof}
We can proceed as in the proof for the state case, Proposition~\ref{prop:sum}. 
Monotonicity
of the relative modular operators (see for example \eqref{eq:resbound}) implies $\psi \in \mathcal{D}(\Delta_{\eta_{V_1},\psi;V_1}^\theta) \cap \mathcal{D}(\Delta_{\eta_{V_2'},\psi;V_2'}^\theta) $.
From Lemma~\ref{lem:calculational} (2) we see that all the relative entropies are $> -\infty$. Combined  with the extra assumptions in Proposition~\ref{prop:sum-weight}
so that  $D(\psi|_{V_2})$ and $D(\psi|_{V_1'})$ are finite
we need only treat the sum rule when one of $D(\psi|_{V_2'})$ or $D(\psi|_{V_1})$
equals $\infty$ which leads to a contradiction with the sum rule. So let us assume $D(\psi|_{V_1}) =+\infty$
aiming for a contradiction (the case $D(\psi|_{V_2'}) =+\infty$ can be treated similarly).

Define the functions:
\begin{align}
\label{eq:firsth}
h_{V_1}(z) &\equiv  \left< \psi \right|\left(1-  Z_{V_1}^{z}\Delta_{\eta_{V_1},\psi;V_1}^z \right)\left| \psi \right>\,, \quad h_{V_2'}(z) \equiv  \left< \psi \right|\left(1 - Z_{V_2'}^{z} \Delta_{\eta_{V_2'},\psi;V_2'}^{z} \right)\left| \psi \right>
\\
\| \psi \|^2 -f(z) &= \left< \psi \right| \Delta_{\eta_{V_2'},\psi;V_2'}^z \Delta_{\eta_{V_1},\psi;V_1}^z \left| \psi \right>
=  \left< \psi \right| \Delta_{\eta_{V_2},\psi;V_2}^z \exp(i a P(1-e^{-2\pi i z} )) \Delta_{\eta_{V_1'},\psi;V_1'}^z \left| \psi \right>
\label{eq:torhs5}
\end{align}
where the last equality can be established along ${\rm Re} z =0$ using Lemma~\ref{lem:flow} (generalized to a weight) and Theorem~\ref{thm:hsmi-weight} and then inside the strip $0 <{\rm Re} z < \theta$ using equality at the lower edge,
analyticity and continuity in the closure of this strip (using the domain condition in the assumptions for this Proposition) and the edge of the wedge theorem.
In \eqref{eq:firsth} we have included normalizations $Z_{V_1},Z_{V_2'}$ that we use to fix $h_{V_1}(\theta) = h_{V_2'}(\theta) =0$
so that for real $x$ the functions $h_{V_1}(x)/x, h_{V_2'}(x)/x$ are real positive and increasing as $x \rightarrow 0$
as established in Lemma~\ref{lem:calculational}(2) which also proves the limits exist and give:
\begin{equation}
\lim_{x \rightarrow 0^+} h_{V_1}(x)/x = \infty\,, \qquad -\infty <  \lim_{x \rightarrow 0^+} h_{V_2'}(x)/x \leq \infty 
\end{equation}
increasingly. Similar considerations imply that we can take the limit on $f(x)/x$ based on the right hand side
expression of \eqref{eq:torhs5} with:
\begin{equation}
-\infty < \lim_{x \rightarrow 0^+} f(x)/x < \infty
\end{equation}
where the domain condition is used as well as boundedness of $\| \exp(i a P(1-e^{-2\pi i x} )) \| \leq 1$ for $0\leq x\leq \theta$.

Based on the left hand side expression in \eqref{eq:torhs5} and a Cauchy-Schwarz estimate we have:
\begin{equation}
\label{eq:torhs12}
|h_{V_1}(x) +h_{V_2'}(x) -\tilde{f}(x)|^2 \leq (2h_{V_1}(x) -h_{V_1}(2x)) (2h_{V_2'}(x) - h_{V_2'}(2x))
\end{equation}
where $\tilde{f}(x) = Z_{V_1}^x  Z_{V_2'}^x f(x) - \left( Z_{V_1}^x  Z_{V_2'}^x-1\right) \| \psi\|^2$. Note that $\tilde{f}(x)/x$ has a well
defined finite limit as $x\rightarrow 0^+$.  We can drop the term $- (2h_{V_1}(x) -h_{V_1}(2x)) h_{V_2'}(2x) < 0$
on the right hand side of \eqref{eq:torhs12}, where the inequality uses the properties of $h_{V_1}, h_{V_2'}$ stated above. 
We get two inequalities by also dropping the terms given by switching $V_1 \leftrightarrow V_{2}'$:
\begin{equation}
(h_{V_1}(x) -h_{V_2'}(x))^2 -2{\rm Re}(\tilde{f}(x))(h_{V_1}(x) +h_{V_2'}(x)) \leq  \begin{cases}  -2h_{V_1}(2x) h_{V_2'}(x) \\ -2h_{V_2'}(2x) h_{V_1}(x) \end{cases}
\end{equation}
Adding these two together:
\begin{equation}
0 \leq (h_{V_1}(x) -h_{V_2'}(x))^2 \leq ( - h_{V_1}(2x) +2 {\rm Re}\tilde{f}(x) ) h_{V_2'}(x) + 
( - h_{V_2'}(2x) +2{\rm Re} \tilde{f}(x) ) h_{V_1}(x) 
\end{equation}
implying that at least one of $( - h_{V_1}(2x) +2 {\rm Re}\tilde{f}(x) )$ or
$( - h_{V_2'}(2x) +2 {\rm Re}\tilde{f}(x) )$ must be non-negative at any given $x$.
Notably if $h_{V_1}(x)/x$ and $h_{V_2'}(x)/x$ both diverged this would lead to a contradiction. 
This implies that $h_{V_2'}(x)/x$ must have a finite limit. 
We can return to the inequality \eqref{eq:torhs12} and drop both $h_{V_1}(2x), h_{V_2'}(2x)$ terms:
\begin{equation}
(h_{V_1}(x)/x +h_{V_2'}(x)/x -{\rm Re}\tilde{f}(x)/x)^2 \leq 4 (h_{V_1}(x)/x )( h_{V_2'}(x)/x)
\end{equation}
which leads to a contradiction since $h_{V_1}(x)/x$ diverges while all the other terms have finite limits.
\end{proof}

To finish the proof we need the following approximations for $\psi$:
\begin{lemma}
\label{lem:approx-sum}
Consider $V_1 < V_2$.
For all $\psi \in \mathcal{D}(\sqrt{P})$, then there exists a net of states $\psi_\iota \in \mathscr{H}$ with $\lim_\iota \| \psi - \psi_\iota \| = 0$
and with the following properties:
\begin{enumerate}
\item $\psi_\iota \in \mathcal{D}( \Delta_{\eta_{V_2},\psi_\iota;V_2}^{1/2}) \cap \mathcal{D}( \Delta_{\eta_{V_1'},\psi_\iota;V_1'}^{1/2}) $
\item $D(\psi_\iota|_{V_2}), D(\psi_\iota|_{V_1'}) < \infty$
\item $ \sup_\iota \| \sqrt{ P } \psi_\iota \| < \infty$
\end{enumerate}
\end{lemma}
\begin{proof}
See Appendix~\ref{app:approxsumrule}.
\end{proof}

\begin{proof}[Proof of Proposition~\ref{prop:sum-weight}]
 The final step is then to approximate all states by ones satisfying the assumptions in 
 Proposition~\ref{prop:sum-weight}(b). We use Lemma~\ref{lem:approx-sum} to approximate $\psi$ by the net $\psi_\iota$
 where the corollary to Proposition~\ref{prop:sum-weight}(b) is the fact that all relative entropies in the sum rule are well defined and finite. 
  
The sum rule for $\psi_\iota$ in \eqref{eq:sumweight} along with monotonicity Lemma~\ref{lem:mono-weight} and
the uniform bound on $\| \sqrt{P} \psi_\iota \|$ implies that:
\begin{equation}
0 \leq \liminf_\iota (D(\psi_\iota|_{V_2'}) - D(\psi_\iota|_{V_1'}) ) < \infty
\end{equation}
Using Lemma~\ref{lem:weight-semi} the difference is semicontinuous along $\iota$,
which applies because the relative entropies along the net are well defined and finite 
and by the assumptions of this Proposition $D(\psi|_{V_1'}) < \infty $ and $D(\psi|_{V_2'}) > -\infty$
hence:
\begin{equation}
0 \leq D(\psi|_{V_2'}) - D(\psi|_{V_1'})  < \infty
\end{equation}
But this implies that $D(\psi|_{V_2'}) < \infty$ and $D(\psi|_{V_1'}) > -\infty$ and so both these relative entropies are finite. 

A similar argument shows that:
\begin{equation}
0 \leq \liminf_\iota (D( \psi_\iota|_{V_1}) - D(\psi_\iota|_{V_2})) < \infty
\end{equation}
and again semicontinuity of the difference can be applied:
\begin{equation}
0 \leq D(\psi |_{V_1}) - D(\psi |_{V_2})  < \infty
\end{equation}
given the assumptions in the statement. 
So $-\infty < D(\psi|_{V_1}) < \infty$ and $-\infty < D(\psi|_{V_2}) < \infty$.
Hence we have verified all the relative entropies in the sum rule are finite and thus we can apply
the original proof  of Proposition~\ref{prop:sum-weight}(a) to give the sum rule under these more general circumstances.

For other cuts $V$ well definedness follows from Lemma~\ref{lem:mono-weight}. If we grow $V_2 \rightarrow V$ with $V > V_2$ then it is easy to check using monotonicity that we maintain the finiteness assumptions that let us derive the sum rule - hence $-\infty < D(\psi|_V), D(\psi|_{V'}) < \infty$ for $V \geq V_2$. Similarly if we shrink $V_2 \rightarrow V$ with $V_1 \leq V < V_2$ we maintain the sum rule and $-\infty < D(\psi|_V), D(\psi|_{V'}) < \infty$ for these $V$'s. Finally if we shrink or grow $V_1 \rightarrow V$ we also maintain the sum rule and thus finiteness of all relative entropies along the $V$ family. 
\end{proof}

\subsection{Finiteness, weight case}
\label{sec:finite-weight}

Our aim in this section is to prove Proposition~\ref{prop:finite} in the general weight case. 
The discussion will be somewhat similar to \cite{Ceyhan:2018zfg} with some slightly different definitions, some streamlining
of the original arguments, and the need to deal with various complications due to the unboundedness of the weight. 

In this section we work with a fixed algebra, denoted $\mathcal{A}$ satisfying the assumption
of a HSMI Theorem~\ref{thm:hsmi-weight}, with weight $\varphi$ and commutant weight $\varphi'$.
We denote the corresponding semicyclic representation as $\eta$ and $\eta'$. Of course all these
results extend to the family $\mathcal{A}_V$ in a natural way. 

One of the key steps is to find the analytic continuation of:
\begin{equation}
\label{eq:gpsi}
g_\psi (z,s) = \left< \psi_s \right| U(-b) \left| \psi_s \right> \,,\qquad U(-b) = e^{ i b P} \,, \quad b = \epsilon e^{2\pi z}\,, \quad s,z \in \mathbb{R}
\end{equation}
to a holomorphic function of two complex variables $(s,z)$ in various complex domains. 
\begin{equation}
\left| \psi_s \right> = u_s' \left| \psi \right> = (D \varphi' : D\omega_{\psi}|_{\mathcal{A}'})_s \left| \psi \right>
\end{equation}
This is one of our first tasks.  We start by describing the triangular tubular region $T$
into which this vector valued function potentially has a continuation. First define the \emph{two complex dimensional} domain of holomorphy based on a triangle: 
\begin{equation}
T = \{ (z,s) \in \mathbb{C}^2:  {\rm Im} z > 0 \,,\,\, {\rm Im} s > 0 \,,\,\,  {\rm Im} s < 1/2-{\rm Im}z \} 
\end{equation}
The three edges of this triangle each consist of a single complex strip $S = \{ w\in \mathbb{C} :0< {\rm Im }w< 1/2 \}$ and one real parameter:
\begin{equation}
\partial T = \bigcup_{i=1,2,3} \overline{f_i( S  \times \mathbb{R} )} \equiv  \bigcup_{i=1,2,3} \overline{\Sigma_i}
\end{equation}
where  $f_i$ map to the edges of $T$ via (where $r$ is real and $w$ is valued in the open complex strip $0< {\rm Im }w< 1/2$):
\begin{align}\label{eq:s123}
f_1,f_2,f_3 : (w,r) \rightarrow (z,s)  = (w, r)\,,\,\, (r,w)\,,\,\, (i/2-w+r,w)
\end{align}
with image $\Sigma_i \subset \overline{T}$.   See the following figure:

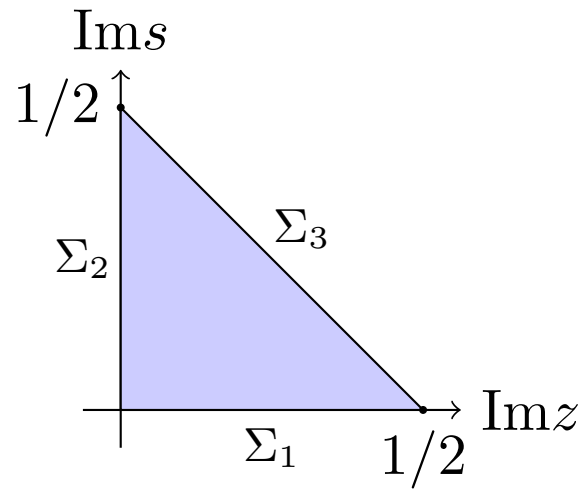
\begin{center}
\begin{tikzpicture}[scale=.5]
    \draw[->] (-.5,0) -- (4.5,0) node[right] {\({\rm Im} z\)};
    \draw[->] (0,-0.5) -- (0,4.5)  node[above] {\({\rm Im} s\)};

    \fill[blue!20] (0,0) -- (4,0) -- (0,4) -- cycle;
  
     \draw (4,0) -- (0,0)  -- (0,4);
      \draw (4,0) -- (0,4);

    \fill (4,0) circle (1.5pt) node[below] {\(1/2\)};
    \fill (0,4) circle (1.5pt) node[left] {\(1/2\)};

    \node at (-3,1.5) {\(T = \)};

    \node at (2.4,2.4) {\footnotesize \( \Sigma_3\)};
     \node at (2,-0.5) {\footnotesize \( \Sigma_1\)};
   \node at (-0.5,2) {\footnotesize \( \Sigma_2\)};

\end{tikzpicture} \hspace{1cm}

\end{center}

Define the set of holomorphic $\mathscr{H}$-valued functions inside a one or two dimensional complex sub-domain $O \subset S$
or $O \subset T$ as $\mathcal{H}(O)$. Define the set of weakly continuous $\mathscr{H}$-valued functions
over some subset $D \subset \overline{S}$ or $D \subset \overline{T}$ as $\mathcal{C}(D)$.
We firstly construct the following vector on this triangle $\overline{T}$:

\begin{lemma}
\label{lem:gamma}
Given some vector $\psi \in\mathscr{P}^\natural$, operators  $x' \in \mathcal{A}',y \in \mathcal{A}$
and some $\epsilon >0$,
then there exists a holomorphic $\mathscr{H}$-valued function on $\overline{T}$, with $\Gamma(x',y;\cdot , \cdot) \in \mathcal{H}(T)\cap  \mathcal{C}(\overline{T})$
with holomorphic edges $\Gamma(x',y; f_i( \cdot, r) ) \in \mathcal{H}(S)
\cap \mathcal{C}(\overline{\Sigma_i})$ for $i=1,2,3$, $r \in \mathbb{R}$ and with the following properties:
\begin{enumerate}[label=(\alph*)]
\item The value on the bottom edge is:
\begin{equation}
  \label{eq:thevector}
\big| \Gamma(x',y;z,s) \big> = \left[ \sigma_{\eta,\psi}'^s(x')^\dagger U(-\epsilon e^{2\pi z}) \sigma_{\eta,\psi}'^s(x') \right] y \left| \psi \right>
\,, \quad  (z,s) \in \overline{\Sigma_1}
\end{equation}
where:
\begin{equation}
\sigma_{\eta,\psi}'^{s}(x') \equiv  \Delta^{is}  x' \Delta_{\eta,\psi}^{-is}
\end{equation}

\item
At the top corner $(z,s) \in \overline{\Sigma_2} \cap \overline{\Sigma_3}$ the vector equals:
\begin{align}
    \label{eq:thevector-top}
& \left| \Gamma(x',y;z,s = i/2+t) \right> 
= U(-\epsilon e^{2\pi z}) y J \left| \Gamma(x',1;z,t) \right> 
\end{align}
\item For all $(z,s) \in \overline{T}$ we have the domain condition: $y^\dagger \big| \Gamma(x',y;z,s) \big>
\in \mathcal{D}(\Delta_\psi^{{\rm Im} s})$  such that:
\begin{equation}
    \label{eq:newsymmr}
\Delta_{\psi}^{-is} y^\dagger \big| \Gamma(x',y;z,s) \big>=  \pi' (x')^\dagger J \big| \Gamma(j(y),j(x');z^\star + s^\star+ i/2, - s^\star) \big> 
\end{equation}
where $\pi'$ is the support projection for $\psi,\mathcal{A}'$.
\item The norm of $\Gamma$ is bounded by $\| x' \|^2  \| y \| \| \psi \| $ for all $(z,s) \in \overline{T}$.

\end{enumerate}
\end{lemma}
\begin{remark}
When necessary we will label $\Gamma \rightarrow \Gamma_\psi$ to make clear the input state.
\end{remark}

\begin{proof}
We start with $\Gamma(z,s)$ defined on $\overline{\Sigma_1}$ which obviously satisfies $\Gamma(\cdot , \cdot) \in \mathcal{C}(\overline{\Sigma_1})$ and $\Gamma(\cdot,t) \in \mathcal{H}(S)$ for each $t \in \mathbb{R}$. Assuming $y^\dagger \in \mathfrak{N}_\varphi$, (b), (c)  and (d) follow the same procedure as in \cite{Ceyhan:2018zfg} where the condition on $y^\dagger$ allows us  to analytically continue in the same way as the bounded state case. We leave the details to Appendix~\ref{app:prop2}.

Then for general $y$ we consider an approximate identity $\pi_\iota$, a net of projection operators 
such that $\lim_\iota \pi_\iota = 1$, given by (b) of Lemma~\ref{lem:approxid}.
Then $ (\pi_\iota y)^\dagger \in \mathfrak{N}_\varphi$ so we can use
(a-d) along the net. Then define for general $y$:
\begin{equation}
\big| \Gamma(x',y;z,s) \big>\equiv w-\lim_\iota \big| \Gamma(x',\pi_\iota y;z,s ) \big> \qquad (z,s) \in \overline{T}
\end{equation}
Certainly the limit exists along $\overline{\Sigma_1} \cup (\overline{\Sigma_2} \cap \overline{\Sigma_3})$
with limiting value given by the explicit expressions in \eqref{eq:thevector} and \eqref{eq:thevector-top} thus confirming (a) and (b) and $ \Gamma(\cdot,\cdot) 
\in \mathcal{C}(\overline{\Sigma_1}), \mathcal{C}(\overline{\Sigma_2} \cap \overline{\Sigma_3})$.
Furthermore it is not hard to check that convergence is uniform on compact subsets
of $\overline{\Sigma_1}$ and $\overline{\Sigma_2} \cap \overline{\Sigma_3}$. 
  We can then extend convergence further to $T \cup \Sigma_2 \cup \Sigma_3$ using the Poisson integral kernel for a family of strips
that cross between the top corner $\overline{\Sigma_2} \cap \overline{\Sigma_3}$ and the bottom edge $\overline{\Sigma_1}$: $(z,s) = ((i/2-w) \alpha + r, w)$ 
where $w \in S$ and $0\leq \alpha \leq 1$ parameterizes which strip. 
In particular we can take the limit inside the Poisson integral by uniform convergence on compacts:
that is after dropping the integral contributions at large real values along the edge of the strip where the kernel decays exponentially.

This implies the limit exists inside $T$ and also inside $\Sigma_{1,2,3}$. 
That is everywhere in $\overline{T}$ where it satisfies (d) via the three line lemma. 

We can also argue for $(c)$ as follows. The domain condition can be established of the  limit $\lim_\iota$ by first acting with $\left( a \psi + \chi,\cdot \right)$, for $a \in \mathcal{A}$ and $\chi \in (1-\pi')\mathscr{H}$,
and then taking the limit. The resulting function in $\overline{T}$ is bounded by $\| x' \|^2 \| y \|^2  \| \psi \| \| a \psi \|$ (using the right hand side of \eqref{eq:newsymmr} along the $\iota$ net with $j(x') \rightarrow \pi_\iota j(x')$) 
and applying this for all $a \in \mathcal{A}$
we prove the stated domain condition since $\mathcal{A} \psi + (1-\pi') \mathscr{H}$ is a core for $\Delta_\psi^{1/2}$
and hence for $\Delta_\psi^{i s^\star}$.
This argument also
shows that the $\iota$ limit on the left hand side of \eqref{eq:newsymmr} commutes with the action of $\Delta_{\psi}^{-is}$.

 The right hand side of the symmetry relation \eqref{eq:newsymmr} is defined in the same way as a weak limit along $\iota$ with $j(x') \rightarrow \pi_\iota j(x')$. But then the relation as written does not hold along $\iota$.
Rather, since the symmetry relation fixes the bottom edge $\Sigma_1$ and the top corner $\overline{\Sigma_2} \cap \overline{\Sigma_3}$ and acts on the family of strips as $\alpha \rightarrow 1-\alpha$,
it must be compatible with the Poisson integrals and so we only
need to check that the symmetry relation applies at the edges. This is indeed the case.

We know, due to the Weierstrass convergence theorem\footnote{Convergence inside $T$ is uniform on compacts due to the Poisson integrals that define it. This allows application of Weierstrass, see \cite{schiff2013normal}, which guarantees convergence to an analytic function. Weierstrass is usually stated 
for sequences, but since the topology of uniform convergence on compact sets is metrizable,
Weierstrass is simply the statement that analytic functions in $T$ are closed with respect
to the topology of uniform convergence on compacts. Hence Weierstrass also applies to nets.}, that $ \Gamma(x',y;\cdot,\cdot)  \in \mathcal{H}(T)$. We however do not obviously retain continuity on $\overline{T}$, rather the limit fragments into various potentially discontinuous regions with decreasing real dimensionalities: $T,\Sigma_i, \overline{\Sigma_i} \cap \overline{\Sigma_j}$ with $ i,j =1,2,3$ and $i \neq j$. The functions at the top corner and bottom edge still determine $\Gamma$
inside $T$ via the Poisson integrals. 
Our final task is to show that this fragmentation does not occur and confirm continuity in $\overline{T}$.

Certainly on the three edges we maintain analyticity $\Gamma(f_i(\cdot,r))  \in \mathcal{H}(S)$
for all $r \in \mathbb{R}$ and $i=1,2,3$ via further applications of the Weierstrass convergence theorem. 
Weak continuity on $\overline{\Sigma_{2,3}}$ follows from the Poisson integral formulas at $\alpha=0,1$ and the fact that
the boundary data on the edge of these strips is bounded and weakly continuous (jointly so in both the boundary point and $r$). So $\Gamma(\cdot,\cdot) \in \mathcal{C}(\overline{\Sigma_i})$ for $i=1,2,3$. 

Finally to establish weak continuity in all of $\overline{T}$ we apply the Malgrange-Zerner theorem \cite{borchers1995use,epstein1960edge} in the form \cite{Druzkowski} to $\left< \xi \right. \left| \Gamma(\cdot,\cdot) \right>$, for all $\xi \in \mathscr{H}$, and restricted to pairs of strips $\overline{\Sigma_{i}} \cup \overline{\Sigma_{j}}$ for $i\neq j \in \{1,2,3\}$. We derive from this a holomorphic vector valued function which is also weakly continuous to the edges: $\Gamma^{ij}(\cdot,\cdot) \in \mathcal{H}(T) \cap \mathcal{C}(\overline{T} \backslash \overline{\Sigma_k})$ where $\Sigma_k$ is the third edge.
$\Gamma^{ij}$ restricts to $\Gamma$ on the edges $\Sigma_i \cup \Sigma_j \cup ( \overline{\Sigma_i} \cap \overline{\Sigma_j}) $. 
It follows from the edge of the wedge theorem (one dimensional, along the complex strips $( (i/2-w) \alpha + r, w)$)
that $\Gamma^{ij} = \Gamma$ in $T$ for all pairs $ij$
thus establishing continuity of $\Gamma$ on $\overline{T} \backslash \overline{\Sigma_k}$ and
hence in all $\overline{T}$. 
\end{proof}

\begin{figure}[ht]
\centering
\begin{tikzpicture}[scale=.5]

    \fill[blue!20] (0,0) -- (4,0) -- (4,-4) -- cycle;

    \fill[blue!20] (4,0) -- (0,4) -- (0,0) -- cycle;

    \fill (4,0) circle (1.5pt) node[above] {\(1/2\)};
    \fill (0,4) circle (1.5pt) node[left] {\(1/2\)};
    \draw[->] (-.5,0) -- (4.5,0) node[right] {\({\rm Im} z\)};
    \draw[->] (0,-.5) -- (0,4.5) node[above] {\({\rm Im} s\)};

    \node at (-3.5,0.5) {\(P = \)};

\end{tikzpicture}
\hspace{1cm}
\begin{tikzpicture}[scale=.5]

    \fill[blue!20] (0,0) -- (4,0) -- (4,-4) -- cycle;

    \fill[blue!20] (4,0) -- (0,4) -- (0,0) -- cycle;

    \fill[red!20] (0,-4) -- (4,-4) -- (0,0) -- cycle;

    \fill[red!20] (0,4) -- (4,4) -- (4,0) -- cycle;

    \fill (4,0) circle (1.5pt) node[above] {\(1/2\)};
    \fill (0,4) circle (1.5pt) node[left] {\(1/2\)};
    \draw[->] (-.5,0) -- (4.5,0) node[right] {\({\rm Im} z\)};
    \draw[->] (0,-.5) -- (0,4.5) node[above] {\({\rm Im} s\)};

    \node at (-3.5,0.5) {\(R = \)};

\end{tikzpicture}
\caption{The regions $P = {\rm int}(\overline{T \cup \iota_\star(T)})$ and $R = {\rm int}(\overline{Q \cup \iota_\star(Q)})$ obtained by extending the triangle $T$ through the reflections $\mu_Q$ and $\iota_\star$.}
\label{fig:regions}
\end{figure}
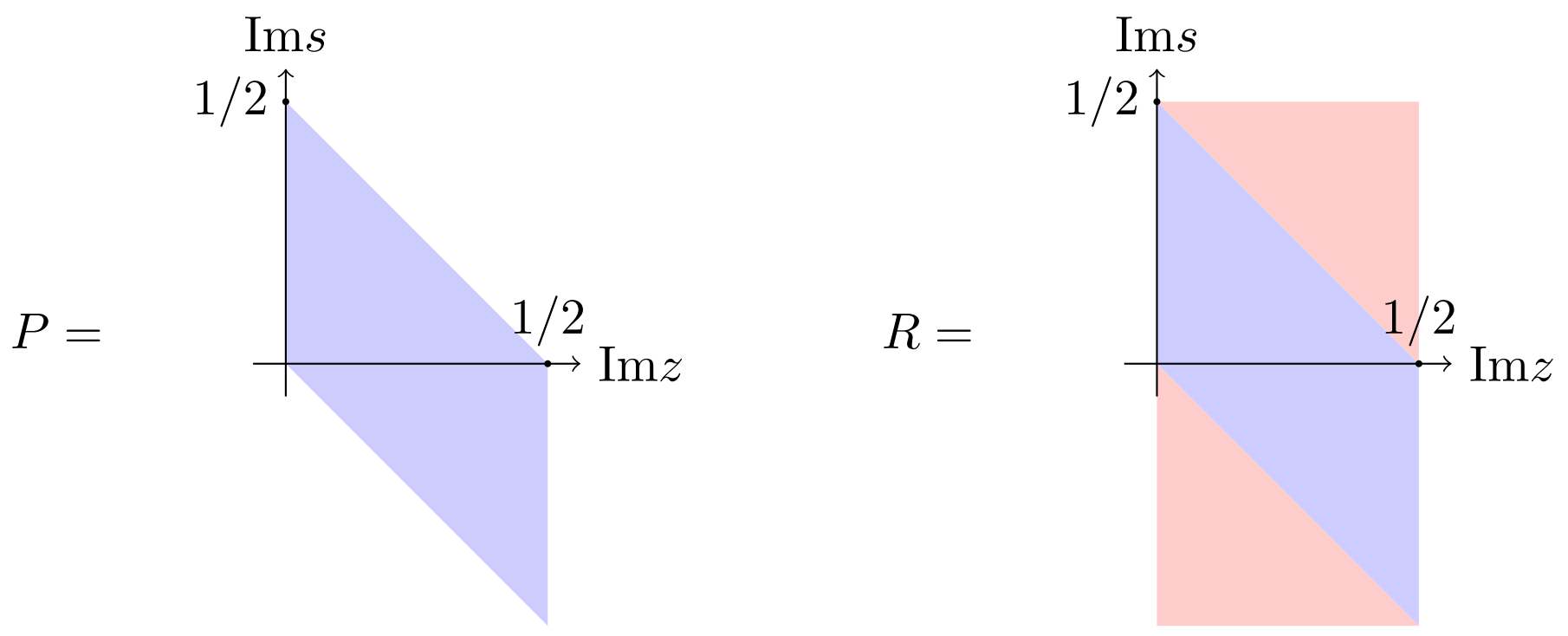

We can now summarize some symmetry properties of $\Gamma$. Define the following symmetry transformations:
\begin{align}
 \mu_Q  : (z,s) \rightarrow (z+s,-s+i/2) \qquad \iota_\star : (z,s) \rightarrow (i/2 
 +z^\star, s^\star)
\end{align}
These maps extend $T$ to the regions
\begin{align*}
Q &= {\rm int}\left(\overline{T \cup \mu_Q(T)}\right), \qquad
P = {\rm int}\left(\overline{T \cup \iota_\star(T)}\right), \\
R &= {\rm int}\left(\overline{Q \cup \iota_\star(Q)}\right).
\end{align*}
See Figure~\ref{fig:regions}.

Note that $\mu_Q$ sends $\overline{\Sigma_2} \rightarrow \overline{\Sigma_3}$. 
Then for $(z,s) \in \overline{\Sigma_3} \cap \overline{\Sigma_2}$ the symmetry relation \eqref{eq:newsymmr} for $\Gamma$ combined with \eqref{eq:thevector-top} 
can be extended to all of $(z,s) \in \overline{\Sigma_3}$ where:

\begin{equation}
    \label{eq:newreflection}
\Delta_{\psi}^{-is} y^\dagger \big| \Gamma(x',y;z,s) \big> = \pi' (x')^\dagger U(-\epsilon e^{2\pi (z+s)})
 x' \big| \Gamma(j(y),1; \mu_Q^{-1}(z,s)) \big>
\end{equation}

This relation will allow us to continue across $\Sigma_3$ into the larger $Q$
region, which contains the image of $\mu_Q$ (the upper pink region in Figure~\ref{fig:regions}).
 We can also invoke reality to allow us to extend to the image of $\iota_\star$
 which gives the entire rectangular region in Figure~\ref{fig:regions}.

\begin{lemma}
    \label{lem:g}
In the same context as Lemma~\ref{lem:gamma}, consider:
\begin{equation}
g_\psi(x',y;z,s) \equiv \big< \psi  \big| y^\dagger \big| \Gamma(x',y;z,s) \big> 
\end{equation}
which is a holomorphic function $(z,s) \in T$ uniformly bounded by $\| x' \|^2  \| y\|^2 \| \psi \|^2$ and continuous in the closure.
We have an analytic continuation of this function to larger regions as follows:
\begin{enumerate}
\item This function has analytic continuation to $(z,s) \in P$ satisfying:
\begin{equation}
\label{eq:reflect}
g_\psi(x',y;z,s) = (g_\psi(x',y; \iota_\star(z,s)))^\star
\end{equation}
 where it has the uniform bound $\| x'\|^2   \| y\|^2 \| \psi \|^2$ and $g_\psi$ is continuous on the closure.
\item If $x' \psi$ is entire analytic for translations $U(a)$, then we have an analytic continuation to $(z,s) \in {Q}$ with continuous closure satisfying:
\begin{equation}
\label{eq:comesfrom}
g_\psi(x',y;z,s) = \big< \psi \big| (x')^\dagger U({-\epsilon e^{2\pi (z+s)}})x' \big| \Gamma(j(y),1;\mu_Q^{-1}(z,s)) \big> 
\end{equation}
when $(z,s) \in \mu_Q(T)$. 
\item Combining 1 and 2, then we can include the reflection \eqref{eq:reflect} which is now holomorphic in $(z,s) \in {R}$ and continuous in the closure of 
the rectangular region. 
\end{enumerate}
\end{lemma}
\begin{proof}
Straightforward application of the edge of the wedge theorem, Lemma~\ref{lem:gamma}
and in particular for (2) we needed \eqref{eq:newreflection}.
\end{proof}

 For a vector $\psi \in \mathscr{H}$ we can always write $\theta' \psi = \widehat{\psi} \in \mathscr{P}^\natural$ with partial isometry $\theta' \in \mathcal{A}'$ so we can apply the previous Lemma~\ref{lem:g}:
\begin{proposition} 
\label{prop:finite-cocycle}
For a vector $\psi$ with $P_\psi < \infty$, then for all $s \in \mathbb{R}$,  $P_s \equiv \| \sqrt{P} u_s' \left| \psi \right> \|^2 < \infty$.
\end{proposition}
\begin{proof}
Without loss of generality we can take $\psi$ to be normalized. 
Using Lemma~\ref{lem:g} we set:
\begin{equation}
g_\psi(z,s) = g_{\widehat{\psi}}((\theta')^\dagger,1;z,s)\,, \quad (z,s) \in P
\end{equation}
and define:  $\gamma_\psi(s) = \left(1-g_\psi(i/4,s)\right)/\epsilon$.
We have the bound $|g_\psi(z,s)| \leq 1$ from (1) of Lemma~\ref{lem:g} and this implies that:
\begin{equation}
\label{eq:bound0}
{\rm Re} \gamma_\psi(s) \geq  (1 - | g_\psi(i/4,s)| )/\epsilon \geq 0
\end{equation}
such that $e^{-\gamma_\psi(s)}$ are uniformly bounded holomorphic functions in the strip $-1/4 < {\rm Im }s < 1/4$. Using the spectral resolution of $P$:
\begin{equation}
\gamma_\psi(s) = \int_0^\infty d  \left< \psi_s \right| E_\lambda(P) \left| \psi_s \right> \frac{( 1- e^{- \epsilon \lambda})}{\epsilon}
\end{equation}
Now $( 1- e^{- \epsilon \lambda})/\epsilon$ is monotonically decreasing as a function of $\epsilon > 0$ and for all $\lambda \geq 0$.
By the monotone convergence theorem we know that the limit $\epsilon \rightarrow 0$ exists although it could possibly be $\infty$ (it exists on the extended real line).  Thus the pointwise limit
\begin{equation}
h(s) = \lim_{\epsilon \rightarrow 0} e^{-\gamma_\psi(s)} \,, \qquad s \in \mathbb{R}
\end{equation}
exists and satisfies $0 \leq h(s) \leq 1$. Also $h(0) = e^{- P_\psi}$. 

The functions 
\begin{equation}
f_\epsilon(s) = e^{-\gamma_\psi(s)} \, \qquad \epsilon > 0
\end{equation}
are  uniformly bounded  by $1$ and holomorphic  on the open connected region:
\begin{equation} 
 -1/4 <  {\rm Im } s  < 1/4
\end{equation}
and the limit $\epsilon \rightarrow 0$ exists pointwise on a subset of this region $L$ with an accumulation point: the real line $L=\{ s :{\rm Im}\, s = 0\}$.
Then by the Vitali-Porter theorem\footnote{\label{foot:VP} A locally bounded sequence of analytic functions, $\{f_{\epsilon_n}(s), n \in \mathbb{N} \}$ (where $\epsilon_n \rightarrow 0$), on some open connected region $\Sigma$ is contained in a set of analytic functions that are compact with respect to the topology of uniform convergence on compact subsets of $\Sigma$. This is Montel's theorem. Vitali-Porter combines Montel with the identity theorem applied to $L \subset \Sigma$, a subset with a limit point in $\Sigma$. Assuming convergence along $L$ then there exists precisely one limit function in such a sequence of functions, $g(s)$, thus guaranteeing convergence everywhere in $\Sigma$. See \cite{schiff2013normal} for details. }
the $\epsilon$ limit of the function $f_\epsilon(s)$ exists uniformly on any compact subset of the strip and is equal to the holomorphic function:
\begin{equation}
h(s) =  \lim_{\epsilon \rightarrow 0} f_{\epsilon}(s) \,\qquad -1/4 <  {\rm Im } s  < 1/4
\end{equation}
where $|h(s)| \leq 1$. We also know that $f_{\epsilon}(s)$ cannot vanish for fixed $\epsilon >0$. The limit of a sequence of nowhere vanishing holomorphic functions $f_\epsilon(s)$ that converges uniformly to a holomorphic function $h(s)$ either vanishes everywhere or nowhere. The former possibility is not allowed since $h(0) = e^{ - P_\psi} \neq 0 $ by assumption, so $|h(s)|>0$ in the strip. 

Thus for all $s \in \mathbb{R}$:
\begin{equation}
P_s = - \ln h(s) < \infty
\end{equation}
which was what we wanted to establish.  We also note that $P_s \equiv - \ln h(s)$ extends this function to a holomorphic function in the strip $-1/4 < {\rm Im}\, s < 1/4$.
\end{proof}
\medskip

\begin{proposition}
\label{prop:finite-cone}
For all vector states with $P_\psi < \infty$, then $\widehat{P} \equiv \| \sqrt{P} \theta' \psi \|^2 < \infty$ and $\widetilde{P}  \equiv \| \sqrt{P} \theta \psi \|^2  < \infty$
where $\theta \in \mathcal{A}$ is the other partial isometry, which sends $\psi$ to $\theta \psi \in \mathscr{P}^\natural$.
\end{proposition}
\begin{proof} This is slightly more tricky. We essentially need a way to analytically continue from $P_s$ to $\widehat{P}$ where we can use our knowledge of the existence of $P_s$ to tell us about $\widehat{P}$. In order to do this we will start by taking $\psi \rightarrow \psi_\Lambda$ to be an entire state for the translation group generated by $P$ and approach a more general state via a limit of spectral projections:
\begin{equation}
\label{eq:entire}
\left| \psi \right> = \lim_{\Lambda} \left| \psi_\Lambda \right>\,, \qquad \left| \psi_\Lambda \right> = E_{[0,\Lambda]}(P) \left| \psi \right>/\sqrt{\left< \psi \right| E_{[0,\Lambda]}(P) \left| \psi \right>}
\end{equation}
We will drop the $\Lambda$ label from now on and leave it implicit in the discussion. Since $(\theta')^\dagger \widehat{\psi}$ is entire analytic in $P$ we can use part (2) of Lemma~\ref{lem:g}
to extend $g_\psi(z,s) \equiv g_{\widehat{\psi}}((\theta')^\dagger,1;z,s)$  to the larger region $R$  and also $\gamma_\psi(s) = (1 - g_\psi(i/4,s))/\epsilon$ now
has an extension to $-1/2 < {\rm Im}\, s < 1/2 $.

Now in the extended strip $-1/2 < {\rm Im} s < 1/2$ we can still show that the limit $\epsilon \rightarrow 0$ of $\gamma_\psi$ exists and is finite. This follows from another application of the Vitali-Porter theorem to $e^{-\gamma_\psi(s)}$, see Footnote~\ref{foot:VP}. Working inside the extended strip $-1/2< {\rm Im}\, s<1/2$, which is an open connected region that we call $\Sigma$, we will now establish that $e^{-\gamma_\psi(s)}$ is locally bounded in $\Sigma$. Start with the bound:
\begin{equation}
|g_\psi(i/4,s)|^2 \leq \left< \psi \right| e^{  2 \mathcal{N}_s \epsilon P} \left| \psi \right>
\leq e^{ 2 \mathcal{N}_s \Lambda \epsilon}  \qquad \mathcal{N}_s = - e^{2\pi {\rm Re} s} \cos( 2\pi {\rm Im} s) \geq 0
\end{equation}
for $1/4 \leq |{\rm Im}\, s| \leq 1/2$ and where $\Lambda$ was defined in \eqref{eq:entire}.  So for $0 < \epsilon < \epsilon_0$ we have:
\begin{equation}
{\rm Re} \gamma_\psi(s) \geq (1 - e^{\epsilon \Lambda \mathcal{N}_s})/\epsilon \geq  (1 - e^{\epsilon_0 \Lambda \mathcal{N}_s})/\epsilon_0
\end{equation}
In the extended strip we can combine this bound with \eqref{eq:bound0} to find:
\begin{equation}
\label{eq:defms}
{\rm Re} \gamma_\psi(s)  \geq  \min\{0, (1 - e^{\epsilon_0 \Lambda \mathcal{N}_s})/\epsilon_0 \}
\,, \qquad  -1/2 < {\rm Im}\, s < 1/2
\end{equation}
implying that $\exp(-\gamma_\psi)$ is locally bounded in the above strip.

We also already know that  the limit of $e^{-\gamma_\psi(s)}$ exists in the smaller strip
$L = \{ s: -1/4 < {\rm Im }s < 1/4 \}$. Thus we can apply Vitali-Porter to $e^{-\gamma_\psi(s)}$ 
and conclude that the limit $\epsilon \rightarrow 0$ exists uniformly on all compact subsets of the strip $-1/2< {\rm Im }s < 1/2$ to a holomorphic function. The limit function is also nowhere vanishing in the extended strip because we are taking a limit of nowhere vanishing holomorphic functions where we already know that the limit is non-zero at some points in the strip. In this way we have shown that 
\begin{equation}
P_s = \lim_{\epsilon \rightarrow 0} \gamma_\psi(s) < \infty \,\qquad -1/2 < {\rm Im}\, s < 1/2
\end{equation}

Set:
\begin{equation}
\label{eq:ghat}
g_{\widehat{\psi}}(z,s) = 
g_{\widehat{\psi}}(1,1;z,s)
 \,, \qquad (z,s) \in \overline{P}
\end{equation}
We would now like to compare $\gamma_\psi$ to the following holomorphic function:
\begin{equation}
\gamma_{\widehat{\psi}}(s) \equiv
 (1-g_{\widehat{\psi}}(\mu_Q^{-1}(i/4,s)))/\epsilon \,,
\end{equation}
for $  1/4<{\rm Im}\, s<3/4$.

Comparing the function $\gamma_{\widehat{\psi}}(s)$ to \eqref{eq:comesfrom} (for $x' = (\theta')^\dagger, y = 1, \psi \rightarrow \widehat{\psi}$ and $z = i/4$) there is an extra factor of $U(-\epsilon e^{2\pi (z+s)})$. Hence in the limit $\epsilon \rightarrow 0$ we can see that $\gamma_{\widehat{\psi}}$ will be related to  $\gamma_\psi$ which will allow us to deduce the existence of the former limit from the latter limit.
The function $\gamma_{\widehat{\psi}}$ has the same analyticity/continuity properties as $\gamma_\psi$ for non-entire states in $P$, except that the complex strip is shifted.

In more detail, we can compare $\gamma_\psi$ and $\gamma_{\widehat{\psi}}$ on the overlap region $ 1/4 < {\rm Im} s < 1/2$ using $\Gamma$ from Lemma~\ref{lem:gamma}
\begin{align}
&\left| \gamma_\psi(s) - \gamma_{\widehat{\psi}}(s) +\frac{\left< \psi \right| \left( e^{-\epsilon e^{2\pi s} P}-1 \right) \pi' e^{- \epsilon e^{ 2\pi s^\star} P } \left| \psi \right>}{\epsilon} \right| \\ \nonumber
& = \left| \left(  \theta'  \left(1 - e^{- \epsilon e^{2\pi s^\star } P} \right) \left| \psi \right>, \big| \Gamma_{\widehat{\psi}} \big>- \theta' e^{-\epsilon e^{2\pi s^\star} P} \big| \psi \big>  \right)  \right|/\epsilon\\
 & \qquad \qquad \leq \| \theta' e^{-\epsilon e^{2\pi s^\star} P} \big| \psi \big> - \big| \Gamma_{\widehat{\psi}} \big> \| \,\, \|  \left(1 - e^{- \epsilon e^{2\pi s^\star } P} \right) \left| \psi \right> \| /\epsilon
 \label{eq:rhs}
\end{align}
where $\big| \Gamma_{\widehat{\psi}} \big> \equiv \big|  \Gamma_{\widehat{\psi}}(1,1;\mu_Q^{-1}(i/4,s)) \big> $
and $(\theta')^\dagger \theta' = \pi'$ the support projection for $\psi,\mathcal{A}'$. 
Now we can compute/bound each term on the right hand side:
\begin{align}
\| \theta' e^{-\epsilon e^{2\pi s^\star} P} \big| \psi \big> - \big| \Gamma_{\widehat{\psi}} \big> \|^2 /\epsilon
&\leq \left(  \left< \psi \right| e^{-2\epsilon {\rm Re} e^{2\pi s} P } \left| \psi \right> -1\right)/\epsilon + 2 {\rm Re} \gamma_\psi(s)  \\   & \mathop{\rightarrow}_{\epsilon \rightarrow 0} 2 {\rm Re} \left( P_s - e^{2\pi s} P_\psi \right)
\end{align}
where we used $\big< \Gamma_{\widehat{\psi}} \big|  \Gamma_{\widehat{\psi}}  \big> \leq 1$ and the finiteness of the limit on $\gamma_\psi(s)$ in the extended strip. Also:
\begin{align}
\lim_{\epsilon \rightarrow 0} \|  \left(1 - e^{- \epsilon e^{2\pi s^\star } P} \right) \left| \psi \right> \|^2 /\epsilon = 0
\end{align}
and here we have used the fact that $\psi$ is entire in order to take the limit.

We conclude, from the vanishing of the right hand side of \eqref{eq:rhs}, that the following limit exists:
\begin{equation}
\lim_{\epsilon \rightarrow 0} \gamma_{\widehat{\psi}}(s) = P_s - e^{2\pi s} P_\psi \,, \quad 1/4 < {\rm Im} s < 1/2
\end{equation}
Note that ${\rm Re} \gamma_{\widehat{\psi}} \geq 0$ for $1/4 \leq {\rm Im} s \leq 3/4$ so we can again apply the Vitali-Porter theorem for $e^{-\gamma_{\widehat{\psi}}}$ to conclude that the limit of $\gamma_{\widehat{\psi}}$ exists and is finite on compact subsets of the strip: $1/4 < {\rm Im} s < 3/4$. Thus:
\begin{equation}
\widehat{P} 
= \lim_{\epsilon \rightarrow 0} \frac{1 - g_{\widehat{\psi}}(i/4,0) }{\epsilon} = \lim_{\epsilon \rightarrow 0} \gamma_{\widehat{\psi}}(i/2)< \infty
\end{equation}
where the relation to $\widehat{P}$ follows from the definition \eqref{eq:ghat}, (a) of Lemma~\ref{lem:gamma}, and monotonicity of the limit $\epsilon \rightarrow 0$.
We can prove $\widetilde{P}$ is finite in the same way except making the replacement $\psi \rightarrow J \psi$ from the outset. 

Now for general states we simply note that continuity of $P_\Lambda$, and semicontinuity of $\widehat{P}_\Lambda$ and $\widetilde{P}_\Lambda$
coupled with the sum rule  $2P_\Lambda = \widehat{P}_\Lambda + \widetilde{P}_\Lambda$ (using Lemma~\ref{lem:sum} and $\theta \theta' =  J J_\psi$) imply, after sending $\Lambda \rightarrow \infty$, that $\widehat{P}$ 
and $\widetilde{P}$ remain bounded. 
\end{proof}

\subsection{The QNEC for weights}
\label{sec:weightQNEC}

Note that the quantum expansion $\Theta(\psi|_V)$ is defined in exactly the same way as the state case in \eqref{eq:intro-theta}. 

We will need a weight version of Lemma~\ref{lem:calc} (the calculation of the quantum expansion in terms of modular flow), Lemma~\ref{lem:sum} (the $P$ sum rule for local products of partial isometries),
Lemma~\ref{lem:duality} (the $P$ sum rule in terms of quantum expansions)
and Lemma~\ref{lem:cont} (semicontinuity of the quantum expansion). The statements and proofs of these Lemmas are essentially identical to the non-weight case so they are not worth restating. The proofs do require the weight-case finiteness of Proposition~\ref{prop:finite-cocycle} and Proposition~\ref{prop:finite-cone} in place of Proposition~\ref{prop:finite},
and the weight versions of Lemma~\ref{lem:calc} and Lemma~\ref{lem:sum} are used
to prove the weight versions of Lemma~\ref{lem:duality} and Lemma~\ref{lem:cont}. 

Some care is needed with the weight version of Lemma~\ref{lem:deriv}:
\begin{lemma}
\label{lem:deriv-weight}
Suppose that $\psi \in \mathcal{D}(\sqrt{P})$ and that $D(\psi|_{V_1})$ is well defined and finite 
$-\infty < D(\psi|_{V_1}) < \infty$, then the following right derivative exists:
\begin{equation}
\label{eq:deriv-weight}
- \partial_V^+ D(\psi|_{V})  =2 \pi \Theta(\psi|_{V})
\end{equation}
for all $V \geq V_1$. 
\end{lemma}
\begin{proof}
Since the derivative is taken from the right,
without loss of generality we can replace $\psi \rightarrow \psi = \xi_{\omega_\psi|_{V_1}}$. 
The symmetry of states in the natural cone then implies that $ -\infty <  D(\psi|_{V_1}) = D(\psi|_{V_1'})< \infty$ and $D(\psi|_{V}) \leq  D(\psi|_{V_1}) < \infty$
and $D(\psi|_{V'}) \geq D(\psi|_{V_1'}) > -\infty$,
and Proposition~\ref{prop:finite-cone} means that we maintain $\psi \in \mathcal{D}(\sqrt{P})$. 
The sum rule Proposition~\ref{prop:sum-weight} applies and so all $D(\psi|_{\widetilde{V}})$ and
$D(\psi|_{\widetilde{V}'})$ are well defined and finite for all $\widetilde{V}$. 

The rest of the proof for the weight case of Lemma~\ref{lem:deriv} proceeds if we can show finiteness of all relative entropies for $\psi_s = \Delta_{\eta_{V_1},\psi;V_1}^{is} \Delta_{\psi;V_1}^{-is}\psi $ and also $\psi_s' = \Delta_{\eta_{V_2'},\psi;V_2'}^{is} \Delta_{\psi;V_2'}^{-is}\psi $.
From Proposition~\ref{prop:finite-cocycle} we maintain $\psi_s, \psi_s' \in \mathcal{D}(\sqrt{P})$.  Then \eqref{eq:focuslast} holds in the weight case because 
$\varphi_{W}(U(c) \cdot U(-c)) = \varphi_{W+c}(\cdot)$ for all $W,c$.
So the flowed state $\psi_s$ maintains finiteness at $V_1,V_1'$ and $V$. We again use Proposition~\ref{prop:sum-weight} and conclude all relative entropies (that is also at $V'$) are finite.
The weight version of Lemma~\ref{lem:duality} allows us to compute $\| \sqrt{ P} \psi_s \|^2$, $\| \sqrt{ P} \psi_s' \|^2$ in the sum rules and to derive the weight version of \eqref{eq:sbound1}. Similar reasoning applies to $\psi_s'$ and the weight version of \eqref{eq:sbound2}.

Combining we then get the weight version of \eqref{eq:to-deriv}. 
The weight version of semicontinuity Lemma~\ref{lem:cont} allows us to take the derivative.
\end{proof}

Combining we prove a limited version of the QNEC still involving the assumption that $\Theta(\psi|_V) < \infty$. That is:
\begin{theorem}
\label{thm:mainthmweight}
For a normal state $\rho \in (\mathcal{A}_{V_0})_\star^+$ satisfying $-\infty < D(\rho|_{V_0}) < \infty$ and  $\Theta(\rho|_{V_0}) < \infty$, then
for all cuts $V_0 \leq V < \infty$ we have:
\begin{enumerate}
\item $-\infty < D(\rho|_V) < \infty$ is a continuous, monotonically non-increasing, convex function of $V$.
\item $\Theta(\rho|_V)$ is a non-negative, monotonically non-increasing, continuous function of $V$.
\item The relation holds for all $V_2 > V_1 \geq V_0$:
\begin{equation}
D(\rho|_{V_1})- D(\rho|_{V_2})
= 2\pi \int_{V_1}^{V_2} d V\Theta(\rho|_V)
\end{equation}
\end{enumerate} 
\end{theorem}
\begin{proof}
Essentially the same as the proof of Theorem~\ref{thm:initialproof} with appropriate substitutions
of the weight versions of the Lemmas and Propositions. 
\end{proof}

And for states with well defined relative entropies we maintain convexity in $V$:
\begin{theorem}
\label{thm:weight-convexity}
For any normal state $\rho \in \mathcal{B}(\mathscr{H})^+_\star$ with well defined and finite relative entropy restricted to some $V_0$, then for all $V \in \mathbb{R}$, $V \rightarrow D(\rho|_V) \in [-\infty,\infty]$ is an extended real valued function, monotonically non-increasing and convex where $D(\rho|_V)$ is finite.
\end{theorem}
\begin{proof}
We have already established the well definedness of all $D(\rho|_V)$ and monotonicity,
under the hypothesis, thanks to Lemma~\ref{lem:mono-weight}.
For convexity we pick $V_3 > V_1$ where $D$ is finite for all $V \in[V_1,V_3]$.
We regularize $\psi = \xi_{\rho|_{V_1}}$ using Lemma~\ref{lem:weight-entropy-approximation} for $V_1 < V_2 < V_3$ so we have purifications $\psi_\iota$ satisfying the assumptions in Theorem~\ref{thm:mainthmweight} so that convexity 
is equivalent to:
\begin{equation}
\frac{D(\psi_\iota|_{V_1})- D(\psi_\iota|_{V_2})}{V_2-V_1} \geq \frac{D(\psi_\iota |_{V_1})- D(\psi_\iota|_{V_3})}{V_3-V_1}
\end{equation}
Taking the limit on the left using Lemma~\ref{lem:weight-entropy-approximation} and on the right
using Lemma~\ref{lem:weight-semi}:
\begin{equation}
         \frac{D(\psi|_{V_1})- D(\psi|_{V_2})}{V_2-V_1} = \liminf_\iota \frac{D(\psi_\iota|_{V_1})- D(\psi_\iota|_{V_2})}{V_2-V_1} \geq \frac{D(\psi |_{V_1})- D(\psi|_{V_3})}{V_3-V_1}
\end{equation} 
\end{proof}

\section{Discussion}
\label{sec:conclusions}

In this paper we have given a general proof of the QNEC in the algebraic setting of a half sided translation.
We worked with both a vacuum state, where the proofs streamline and improve upon \cite{Ceyhan:2018zfg}, and a vacuum weight.
Our proofs of the convexity of relative entropy as a function of the cut $V$ are assumption free, removing the need for a purification with finite averaged null energy.

The main open question pertains to the physical applicability of the QNEC for weights. 
These were motivated by a class of evaporating black holes or inflationary cosmologies where a weight with geometric modular flow was shown to exist \cite{Chen:2024rpx,Chen:2026zxc,Speranza:2025joj}. The question that remains is whether a half sided translation or modular inclusion also exists in these settings.
In this discussion section we give some non-rigorous physics arguments, and hope a more rigorous discussion will be made in the future.

We focus on the Schwarzschild black hole with the quantum fields living in the evaporating Hilbert space constructed by \cite{Chen:2026zxc}.
Most of the HSMI ingredients, Definition~\ref{def:hsmi-weight}, are physically natural.  1. Ref.~\cite[Appendix~B]{Chen:2026zxc} has proven, under a natural axiomatic framework for quantum fields in the black hole background, that there is an nsf weight $\sigma$ on the evaporating Hilbert space, whose modular flow implements the geometric time translations of the spacetime.
2. Positive geometric flows take future null cuts of the horizon, parameterized by $V$, into themselves.
3. As discussed in Appendix~\ref{app:hsmi-proof} and Lemma~\ref{lem:removecyc} we can replace the assumption of cyclicity for the subalgebra $\mathcal{B}$ with additivity $\bigvee_{V>0}\mathcal{A}_V = \mathcal{A}$ which is easy to prove using the same axioms in \cite[Appendix~B]{Chen:2026zxc}: local normality and the split property allow us to pass this question to the Hartle-Hawking Hilbert space where additivity is automatic by the Reeh-Schlieder theorem for the HH state.

The one remaining question is 4. semifiniteness for $\sigma$ restricted to any of the $\mathcal{A}_V$ with $V>0$.
In fact we only need $\sigma$ to be semifinite for some subalgebra associated to a geometric region outside the black hole - spacelike separated from the horizon ($\mathcal{A}_{\Sigma_{R,a}}$ for $a>0$ as defined in \cite[Appendix~B]{Chen:2026zxc})
since $\mathcal{A}_{\Sigma_{R,a}} \subset \mathcal{A}_V$ for some $V>0$ and so this passes semifiniteness
to $\mathcal{A}_V$. 

The weight is unbounded because it supports evaporation at all times. An (approximate) projective measurement $\pi_i
\in \mathfrak{N}_\sigma$ cuts this off, say via an approximate measurement of the initial time of evaporation. For example we could take $\pi_i$ to approximately represent a projective measurement of this time $T_i$. The measurement $\pi_i$ can be constructed far from the black hole in the neighborhood of a fixed time slice. 

Another way to say this is to consider the divergent behavior as arising from the thermality of the HH state
that extends far from the black hole. In particular the two disjoint Hilbert spaces in the black hole, HH and evaporating,  are analogous to the two sectors of a QFT in Minkowski space at finite temperature and the vacuum (zero temperature). In the vacuum, thermal physics is represented by the unbounded weight ${\rm Tr}_{\mathscr{H}_{\rm vac}}e^{ - \beta H}(\cdot)$ for the Hamiltonian of the QFT and bounded states can be constructed by measurements of operators $\pi_i$ that extend to spatial infinity and that project to the vacuum outside of a compact region which then contains a localized thermal state that evaporates to $\infty$ in the forward time direction. In the black hole case there is an additional localized factor (using the split property) near the black hole horizon that accounts for entanglement across the horizon. In \cite[Appendix~B]{Chen:2026zxc} local normality is used to relate the two sectors for this part.

Since spatial infinity is contained in all regions $\Sigma_{R,a}$, the $\pi_i$ measurements can in principle be designed to extract finite states $\sigma(\pi_i \cdot \pi_i)$ and supposing that a sequence of such measurements are available from some fixed $\Sigma_{R,a}$
with $so-\pi_i \rightarrow 1$ we would have that semifiniteness extends to all of $\mathcal{A}_V$.   The reason we have not yet managed to upgrade this to a theorem based on the axioms in \cite[Appendix~B]{Chen:2026zxc} is that the local near horizon factor and the part containing spatial infinity are coupled by the time flow of the black hole background and semifiniteness on the subfactor associated to the split depends on the details of this coupling.

As this example illustrates, the weight case is a genuinely non-equilibrium setting, since there are no normal states that are invariant under the time translation isometry
of the black hole. 
Other examples were illustrated in \cite{Chen:2024rpx,Speranza:2025joj}, for example similar non-equilibrium physics arises in inflationary cosmology.
We expect similar applications of our results to this setting, once semifiniteness on the subalgebras is established.

An even simpler setting where a weight HSMI arises and the semifinite condition is obviously true is in the original Hartle-Hawking state but now including the ADM mass mode motivated
by semiclassical gravity - this mode can be thought of as an idealization of a clock. The standard Hilbert space is:
\begin{equation}
\mathscr{H}_{HH} \otimes L^2(\mathbb{R})_L \otimes L^2(\mathbb{R})_R
\end{equation}
The algebras become $\mathcal{A}_V = \mathcal{A}_V^{HH} \otimes 1_L \otimes \mathcal{B}( L^2(\mathbb{R})_R)$. Consider the weight on $\mathcal{A}_V$
determined by:
\begin{equation}
\varphi(a \otimes b_R) = \omega_{\eta} (a) \, {\rm Tr} e^{- X_R} (b_R) 
\end{equation}
with $J = J_{HH} \otimes J_{LR}$ where  $J_{LR}$ is the antilinear operator that exchanges the $L$ and $R$ Hilbert spaces.

The half sided modular inclusion Definition~\ref{def:hsmi-weight} applies to $\mathcal{A}_V^{HH}$ in the state case for $\eta$ and then for $\mathcal{A}_V$ in the weight case.
The clock time $P_R$ is conjugate to $X_R$ and introduces a non-equilibrium character to the system so it looks formally similar to the black hole evaporation case.
 We would then approximately project onto the initial time of the clock:
\begin{equation}
\pi_i = \frac{1}{2\pi} \int_{-T_i}^{T_i} dp e^{ i p X_R } \left| \alpha_i \right> \left< \alpha_i \right |  e^{ - i p X_R}  \qquad \left| \alpha_i \right>
 =  \int d x e^{ - x^2/\alpha_i} \left| x \right>
\end{equation}
where  we pick $\alpha_i= T_i$ and send $T_i \rightarrow \infty$. So that:
\begin{equation}
\left< x_1 \right| \pi_i \left| x_2 \right> 
=  \frac{1}{2\pi} \exp( - (x_1^2+x_2^2)/\alpha_i)  \int_{-T_i}^{T_i}  dp e^{i p(x_1-x_2)}
\rightarrow \delta(x_1-x_2)
\end{equation}
and also
\begin{equation}
{\rm Tr} e^{- X_R}  \pi_i = \frac{1}{2\pi} \int_{-T_i}^{T_i} dp \int_{-\infty}^{\infty} d x e^{-x - 2x^2/\alpha_i}  < \infty
\end{equation}
While this gives a non-trivial example of a weighted QNEC $\mathcal{A}_V \in \mathcal{HS}(\mathcal{A},\varphi)$ for $V>0$, the QNEC in this case follows directly from
the state case QNEC. This can be traced to the fact that the corresponding translation generator $P = P_{HH} \otimes 1_L \otimes 1_R$ (not to be confused with $P_L,P_R$), derived from Theorem~\ref{thm:trtohm}, has a non-empty kernel
${\rm ker} P \geq \left| \eta \right> \left< \eta \right| \otimes 1_L \otimes 1_R$ (using the correspondence between closed subspaces and projections).

We can however use this structure to give an example of a different $\widehat{P} \in\mathcal{TR}(\mathcal{A},\varphi)$ which nevertheless leads to the same HSMI.
That is pick $\widehat{P} = P_{HH}  + \delta( e^{-2\pi P_R} + e^{2\pi P_L})$, for any $\delta > 0$ and where we have suppressed any $\otimes 1$'s. 
This works because of the Heisenberg algebra of the $P_{R,L}, X_{R,L}$ (establishing (c) in Theorem~\ref{thm:hsmi-weight}) and because a short analytic continuation argument establishes that:
\begin{equation}
    \label{eq:UR}
U_R(-a) \mathcal{D}(e^{-X_R/2}) \subset \mathcal{D}(e^{-X_R/2}) \,, 
\quad   e^{-X_R/2} U_R(-a)\xi =   U_R(a) e^{-X_R/2}\xi \,, \quad \forall \xi \in \mathcal{D}(e^{-X_R/2})  
\end{equation}
where $U_R(a) = \exp( -i a \delta e^{-2\pi P_R} )$ and $a>0$ which implies (a) in Theorem~\ref{thm:hsmi-weight} for the generator $\widehat{P}$.
Null translations $e^{i V \widehat{P}} \mathcal{A} e^{-i V \widehat{P}} $ 
also maintain the same family of nested algebras and now ${\rm ker} \widehat{P} = 0$. The translated weight $\varphi_V$ is different from the restriction of $\varphi$ since the subset in \eqref{eq:UR} is strict. 

The physical interpretation of these examples is less clear.
The centralizer of this weight is the crossed product algebra constructed in \cite{Witten:2021unn,Chandrasekaran:2022eqq}. Physically, restricting to the centralizer can be understood as gauging modular flow of the weight, which can then be interpreted as the gravitational Hamiltonian constraint for the HH Hilbert space
where the additional modes 
 $X_R,X_L$ represent the $\mathcal{O}(1)$, in $G_N$, fluctuations of the ADM energies outside the two asymptotic black hole regions.
 Since the HH state has issues in Minkowski space, the simplest setting where this description is without issue is the AdS black hole.
 
However there is a well known difficulty in preserving local subalgebras while gauging such large $\mathcal{O}(1)$ fluctuations to the relative $P_L - P_R$ time shift mode~\cite{Chen:2024rpx,Faulkner:2024gst}. Correspondingly it is expected that the resulting type-II$_\infty$ algebra cannot furnish a HSMI.\footnote{
While non-trivial HSMI implies type-III$_1$ algebras  in the state case, it is not clear if this is also true in the weight case \cite[Theorem~2.2 and the subsequent remarks]{araki2005extension}.
We thank Antony Speranza for extensive discussions on this and related questions.} 
Hence, in this setting, the QNEC is mostly of interest in a semiclassical limit where such time shift fluctuations are parametrically smaller in $G_N$. In this case we first ignore the constraint, and study the consequences of the QNEC for the gravitational theory by treating the constraint in perturbation theory.

We emphasize that it remains an open question whether an example of the genuine weight case exists (where ${\rm ker}P = 0$ for the canonical generator given by the inverse of the surjective map in Theorem~\ref{thm:trtohm}.)
The evaporating black hole Hilbert space \cite{Chen:2026zxc} is a promising place to study this question.

\section{Acknowledgements}

We are grateful for discussions with Antony Speranza, Xuchen Cao and Zhencheng Wang. This work is supported by the DOE
award number DE-SC0015655. 

LLMs (Claude Opus 5.5 and GPT-6 Astra) were used at various stages of this work. We acknowledge their help with minor editing, proof checking and some substantial contributions to the ideas in Appendix~\ref{app:entropy-differences}. In particular LLMs found a gap in the original formulation of Theorem~\ref{thm:hsmi-weight} which then led to the discovery that translations and modular inclusions are not bijective for weights. Also the LLMs were asked to combine the ideas in Appendix~\ref{app:approxsumrule} and Appendix~\ref{app:finiteDTh} in order to remove the assumption $\Theta(\psi|_{V_0}) < \infty$ in the formulation of the weight QNEC. The result was Appendix~\ref{app:entropy-differences} which was understood, checked and written up in the author's own words and formulation. Other than that Appendix the main proof ideas were found well before the AIpocalypse  (a rough draft was written at least a year ago) without the help of A.I. The author takes full responsibility for the work herein. 

\appendix

\section{Some proofs}
\label{app:proofs}

\subsection{Lemma~\ref{lem:sum}}
\label{app:prop1}

This result was first written down by the author in an unpublished note \cite{faulkner-note}.
Take $a>0$ and consider:
\begin{align}
D &=  \left< \psi \right| U({-a}) \left| \psi \right> +  \left< ww' \psi \right| U({-a})  \left| w w' \psi \right>  - \left< w \psi \right|  U({-a})  \left| w \psi \right>
- \left<w' \psi \right|   U({-a}) \left| w' \psi \right>  
\end{align}
where $U(a) = \exp( - i a P)$. 
Note that:
\begin{equation}
 \left< ww' \psi \right| U({-a})  \left| w w' \psi \right> = \left< \psi \right| \left(w^\dagger U({-a}) w U(a) \right) (w')^\dagger U({-a}) w' \left| \psi \right>
\end{equation}
where we have used the fact that $w^\dagger U({-a}) w U(a) \in \mathcal{A}$ for $a>0$ by the half-sided modular inclusion. We can rewrite this as a sum of four terms:
\begin{align}
 |D| & = |D_1 + D_2 + D_3+D_4| \leq |D_1| + |D_2| + |D_3| +|D_4|
\end{align}
where
\begin{align}
D_1
&= \left( w^\dagger (U(a)-1) w \left| \psi \right>, U(a)   (w')^\dagger (U({-a})-1) w' \left| \psi \right>  \right) \\
D_2 & = \left(  w^\dagger (U({a})-1) w\left| \psi \right>,  (U(a)-1) \left| \psi \right> \right) \\
D_3 & = \left( (U({-a})-1) \left| \psi \right>, (w')^\dagger (U({-a})-1) w' \left| \psi \right> \right) \\
D_4 & = - \left( (U({-a})-1)\left| \psi \right> ,(U({-a})-1) \left| \psi \right> \right)
\end{align}
We can bound all of these as:
\begin{equation}
|D_i| \leq \| ( U(a) - 1) \left| \phi \right> \| \,\, \| ( U(a) - 1) \left| \chi \right> \| \,, \quad  i = 1,2,3,4
\end{equation}
where $\phi,\chi \in \{ \psi, w \psi, w' \psi  \}$. We can compute the following limit
using the dominated convergence theorem for the Lebesgue integral in the spectral representation of $P$:
\begin{equation}
 \left< \phi \right| (1 - U(a) )/a \left| \phi \right>  = \int_0^\infty \left< \phi \right| dE_\lambda(P) \left| \phi \right>(1 - e^{ - i a \lambda} )/a
\end{equation}
where $| 1 - e^{-i a \lambda}|/a \leq \lambda$ and we know that $P_\phi$ is finite by assumption so we can take the limit inside the integral:
\begin{equation}
\lim_{a \rightarrow 0^+}  \left< \phi \right| (1 - U(a) )/a \left| \phi \right> = i P_\phi
\end{equation}
It follows that:
\begin{equation}
\lim_{ a \rightarrow 0^+} \frac{ \| ( U(a) - 1) \left| \phi \right> \|^2}{a} = \lim_{ a \rightarrow 0^+} 2{\rm Re} \left< \phi \right| (1 - U(a) )/a \left| \phi \right> = 0
\end{equation}
and
\begin{equation}
\lim_{a \rightarrow 0^+} |D_i|/a = 0 \, \implies  \lim_{a \rightarrow 0^+} D/a = 0
\end{equation}
and the latter limit can again be computed (via dominated convergence given finiteness of the $P$'s). One finds:
\begin{equation}
0 = \lim_{a \rightarrow 0^+} D/(ia) = P_{\psi} + P_{ww' \psi} - P_{w \psi} - P_{w' \psi}
\end{equation}
which completes the proof. 

\subsection{Strong resolvent continuity of modular operators}
\label{app:rescont}
\begin{lemma}
    \label{lem:rescont}
As a function of $\rho \in \mathcal{A}_\star^+$ (equipped with the norm topology) the relative modular operator  $\Delta_{\eta_\sigma,\xi_{\rho}} $ is continuous in the strong resolvent sense when acting on $\pi' \mathscr{H}$. That is for a net $\rho_\iota \rightarrow \rho$:
\begin{equation}
    \label{eq:thestatement}
so-\lim_\iota \frac{1}{\beta+ \Delta_{\eta_\sigma,\xi_{\rho_\iota}} }\pi' =\frac{1}{\beta+\Delta_{\eta_\sigma,\xi_\rho} } \pi'
\end{equation}
for all $\beta > 0$, where $j(\pi')$ is the support projection for $\rho$. 
\end{lemma}
\begin{proof}
See Propositions 4.9 and 4.10 of Ohya-Petz \cite{ohya-petz} where this was established if $\sigma$ is a state. It is not hard to follow the same
arguments using a slightly different core for the unbounded operator $\Delta_{\eta_\sigma,\xi_{\rho}}^{1/2}$.
We give some key steps below:

Set $\pi_\iota =$ the support projection for $\rho_\iota$
and define $\pi = j(\pi')$ and $\pi_\iota' = j(\pi_\iota)$. 
Then convergence of $\xi_{\rho_\iota}
\rightarrow \xi_\rho$ implies that $\lim_\iota (1-\pi_\iota) a' \xi_{\rho} = 0$ for all $a' \in \mathcal{A}'$
so that $\lim_\iota (1-\pi_\iota) \pi =0$. Set $\Delta_\iota \equiv \Delta_{\eta_\sigma,\xi_{\rho_\iota}}$ 
and $\Delta \equiv \Delta_{\eta_\sigma,\xi_{\rho}}$. Then for $x \in \mathfrak{N}_\sigma^\dagger$:
\begin{equation}
    \label{eq:deltahalf}
   \lim_\iota \| \Delta_\iota^{1/2} x \pi \xi_{\rho_\iota} - \Delta^{1/2} x \xi_\rho \|  =
\lim_\iota \| \eta_\sigma(\pi_\iota \pi x^\dagger) - \eta_\sigma(\pi x^\dagger) \| = 0
\end{equation}
using the resolvent identity $(R_z(\Delta_\iota^{1/2}) - R_z(\Delta^{1/2})) R_z(\Delta^{1/2})^{-1}
= R_z(\Delta_\iota^{1/2})(\Delta^{1/2} - \Delta_\iota^{1/2})$,  where $R_z(Q) = 1/(z+Q)$,  we can now use \eqref{eq:deltahalf} and $R_z(\Delta_\iota^{1/2})\Delta_\iota^{1/2} x\pi  (\xi_{\rho_\iota} - \xi_\rho) \rightarrow 0$ to show that:
\begin{equation}
    \label{eq:givesstrong}
    \lim_\iota \|(R_z(\Delta_\iota^{1/2}) - R_z(\Delta^{1/2})) R_z(\Delta^{1/2})^{-1} x \xi_\rho \| =0
\end{equation}
And for any $\chi \in \pi' \mathscr{H}$ we can approximate $R_z(\Delta^{1/2})\chi = \lim_n x_n \xi_\rho$
for $x_n \in \mathfrak{N}_\sigma^\dagger$ 
such that:
\begin{equation}
\lim_n R_z(\Delta^{1/2})^{-1} x_n \xi_\rho  = \chi
\end{equation}
since $\mathfrak{N}_\sigma^\dagger \xi_\rho$ is a core for $\Delta^{1/2}$ on $\pi' \mathscr{H}$.
Hence \eqref{eq:givesstrong} gives strong convergence of a net of bounded operators on a dense subset of $\pi' \mathscr{H}$. We can upgrade this to strong convergence by uniform boundedness. Picking $z = \pm i \sqrt{\beta}$ and taking the product gives the statement \eqref{eq:thestatement}. 
\end{proof}

\subsection{HSMI weight case}
\label{app:hsmi-proof}
Here we provide a proof of Theorem~\ref{thm:hsmi-weight}, Lemma~\ref{lem:onlyc} and Theorem~\ref{thm:trtohm} in that order.

\begin{proof}
(a) + (c) $\implies$ (b). Consider the modular operator for the restricted weight $ \sigma|_{\mathcal{A}_a}$
where $\mathcal{A}_a = U(a)^\dagger \mathcal{A} U(a)$ with $a>0$. 
Define also another $\mathcal{A}_a$ weight via $\sigma_a = \sigma( U(a)\cdot U(-a))$.
The condition (a) implies that:
\begin{equation}
\sigma|_a(n^\dagger n)  = \sigma_a(n^\dagger n) < \infty\qquad n \in U(-a) \mathfrak{N}_\sigma U(a)
\end{equation}
so that the restricted weight is normal semifinite. However generally
we only get $\sigma|_a \leq \sigma_a$ since the restricted weight could be finite on more
than $U(-a) \mathfrak{N}_\sigma U(a) \subset \mathfrak{N}_\sigma \cap \mathcal{A}_a$.  
Still (a) implies that $\eta_\sigma$ is semi-cyclic 
when restricted to $\mathcal{A}_a$. Hence the modular operators can be defined in the usual
way $S_{\sigma|_a} \eta_\sigma(n) = \eta_\sigma(n^\dagger)$ for $n \in \mathfrak{N}_\sigma^\dagger \cap \mathfrak{N}_\sigma \cap \mathcal{A}_a$ and $S_{\sigma|_a} = J_{\sigma|_a} \Delta^{1/2}_{\sigma|_a}$.

Covariance allows us to construct the standard form and semicyclic representation for $\mathcal{A}_a$ with weight $\sigma_a$ on $\mathscr{H}$ as $\eta_a(\cdot) = U(-a)\eta(U(a)\cdot U(-a))$.
Hence the modular data for $\sigma_a$ and $\mathcal{A}_a$ 
is given by $J_{\sigma_a} = U(-a) J U(a)$ and $\Delta_{\sigma_a} = U(-a) \Delta_\sigma U(a)$. 
Combining this with $(c)$ gives:
\begin{equation}
    \label{eq:leftac}
\Delta_{\sigma_a}^{-is} \Delta_{\sigma}^{is}  = U(a(e^{2\pi s} -1)) 
\end{equation}
But since one weight $\sigma|_a$ is an extension of $\sigma_a$ the same applies to the
semicyclic representation $\eta_a(U(-a)n U(a)) =  U(-a) \eta_\sigma(n)
= \eta(U(-a)n U(a))$, for $n \in \mathfrak{N}_\sigma$, so the  
Tomita operator satisfies:
$S_{\sigma_a} \subset S_{\sigma|_a} \subset S_\sigma$ and hence $\Delta_\sigma \leq \Delta_{\sigma|_a} \leq \Delta_{\sigma_a}$.
Hence the left hand side of \eqref{eq:leftac} has a bounded analytic extension to the strip $- 1/2 \leq {\rm Im} s \leq 0$ with norm bounded by $1$ and this implies, say setting $s =- i/4$,
that $\| e^{ - a P} \| \leq 1$ implying that $P \geq 0$. 

(a) + (b) $\implies$ (c). 
We start by observing:
\begin{equation}
(J \eta_\sigma(x), \eta_\sigma(y) ) =  ( \eta_\sigma(x^\dagger), J \eta_\sigma(y^\dagger) ) = ( \eta_\sigma(y^\dagger), J \eta_\sigma(x^\dagger) )
\end{equation}
for $x,y \in \mathfrak{N}_\sigma \cap \mathfrak{N}_\sigma^\dagger$. This can be established with modular theory but should
be thought of as the statement that $J x^\dagger J$ commutes with $y$. Apply this to:
\begin{equation}
\label{eq:sameman}
\left(J \eta_\sigma(x_1), \Delta_{\sigma}^{is} U(-a e^{2\pi s}) \Delta_{\sigma}^{-is} \eta_\sigma(x_2) \right)
= \left( \eta_\sigma(x_2^\dagger), \Delta_{\sigma}^{is} U(a e^{2\pi s}) \Delta_{\sigma}^{-is} J \eta_\sigma(x_1^\dagger) \right)
\end{equation}
for $x_1,x_2 \in \mathfrak{N}_\sigma \cap \mathfrak{N}_\sigma^\dagger $ with $a>0$ and $s \in \mathbb{R}$ so
(a) shows that the second entry on the left hand side is $\eta_\sigma(y)$ where
$y = \sigma_{-s}(U(-ae ^{2\pi s}) \sigma_s( x_2)  U(ae ^{2\pi s})) $. 

Positivity (b) allows us to analytically continue the first expression into the $s$ strip
with $0 \leq {\rm Im} s \leq 1/2$. Compute the function
at $s=i/2+ t$:
\begin{equation}
= \left(\eta_\sigma(x_1^\dagger), \Delta_{\sigma}^{it} U(a e^{2\pi t}) \Delta_{\sigma}^{-it}  J\eta_\sigma(x_2^\dagger) \right)
\end{equation}
We can also continue the second expression into the strip below the real axis to find at $s = -i/2 + t$:
\begin{equation}
= \left( J \eta_\sigma(x_2), \Delta_{\sigma}^{it} U(-a e^{2\pi t}) \Delta_{\sigma}^{-it}  \eta_\sigma(x_1) \right)
= \left( \eta_\sigma(x_1^\dagger), \Delta_{\sigma}^{it} U(a e^{2\pi t}) \Delta_{\sigma}^{-it} J \eta_\sigma(x_2^\dagger) \right)
\end{equation}
where we used the same manipulation as in \eqref{eq:sameman}. We thus find a periodic function in the complex $s$ plane
with the function at $t+i/2$ identical to that at $t-i/2$. Holomorphy and boundedness (which are easy enough to establish using (b))
imply that the function is independent of $s$. Hence:
\begin{align}
&\left(J \eta_\sigma(x_1), \Delta_{\sigma}^{is} U(-a e^{2\pi s}) \Delta_{\sigma}^{-is} \eta_\sigma(x_2) \right) =\left(J \eta_\sigma(x_1), U(-a)\eta_\sigma(x_2) \right) \\
 &\left( \eta_\sigma(x_2^\dagger), U(a) J \eta_\sigma(x_1^\dagger) \right) = \left(\eta_\sigma(x_1^\dagger), U(a) J\eta_\sigma(x_2^\dagger) \right)
= \left(\eta_\sigma(x_2^\dagger), JU(-a) \eta_\sigma(x_1^\dagger) \right)  
\end{align}
implying the result since the operators are bounded and we have agreement on a dense domain.  

In the bounded case we still need to show that (c) $\implies$ (a). But since the weight
is a state we get the GNS cyclic and separating vector $\eta \in \mathscr{H}$ and (a) is just invariance $U(a) \eta = \eta$. To show this we note that $P$ is a self-adjoint operator by Stone's theorem and we have the usual spectral theory for such operators. In particular we get a finite Borel measure on $\mathbb{R}$: $B \rightarrow \mu(B) = \left<\eta\right| E_B(P) \left| \eta \right>$ 
for Borel subsets $B \subset \mathbb{R}$. But (c) and invariance of $\eta$ under modular flow gives 
$\mu(B) = \mu( B \lambda)$ for any $\lambda >0$. The only finite Borel measure on $\mathbb{R}$ satisfying this is proportional to the delta function measure at $0$ so that $\eta \in {\rm ker} P$ and hence $P \eta =0$ or $U(a) \eta = \eta$ as required.   
\end{proof}

Next we give a proof of Lemma~\ref{lem:onlyc}.
\begin{proof}
Following the proof of Theorem~\ref{thm:hsmi-weight}, for $a>0$ we can construct
a translated weight $\sigma_a$ on $\mathcal{A}_a$ which is unitarily related to $\sigma$
so is nsf. Modular data can be computed with the translated semi-cyclic representation $\eta_a$
and the modular data satisfies 
$\Delta_{\sigma_a}^{-is} \Delta_{\sigma}^{is}  = U(a(e^{2\pi s} -1))$ from (c) so that
\begin{equation}
    \label{eq:samealgebras}
\Delta_{\sigma}^{-is} \mathcal{A}_a \Delta_{\sigma}^{is}
=  \Delta_{\sigma}^{-is} \Delta_{\sigma_a}^{is} \mathcal{A}_a \Delta_{\sigma_a}^{-is} \Delta_{\sigma}^{is} = U(-a(e^{2\pi s} -1)) \mathcal{A}_a U(a(e^{2\pi s} -1)) = \mathcal{A}_{a e^{2\pi s}} 
\subset \mathcal{A}_a
\end{equation}
for $s > 0$. The condition \eqref{eq:condN} gives:
\begin{equation}
U(-a)\mathfrak{N}_\sigma U(a) \subset \mathfrak{N}_\sigma \cap \mathcal{A}_a 
\end{equation}
and since the left hand side is finite for $\sigma$ and ultraweakly dense in $\mathcal{A}_a$ we see that $\sigma|_{\mathcal{A}_a}$ is also nsf. 

Finally note that strong continuity of $U(a)$ implies additivity $\bigvee_{a>0} \mathcal{A}_{a} = \mathcal{A}$ so we can apply Lemma~\ref{lem:removecyc} which removes the Araki-Zsid\'o assumption of semi-cyclicity of $\eta_\sigma$ restricted to the subalgebra $\mathcal{B} = \mathcal{A}_1$, in favour of the aforementioned additivity.
\end{proof}

\begin{lemma}
    \label{lem:removecyc}
Consider a von Neumann algebra $\mathcal{A}$ with an nsf weight $\sigma$ and a 
von Neumann subalgebra $\mathcal{B} \subset \mathcal{A}$ such that $\sigma|_{\mathcal{B}}$ remains semifinite.
Suppose also that $\sigma_{\sigma;\mathcal{A}}^s(\mathcal{B}) \subset \mathcal{B}$ for all $s\geq 0$, then:
$\bigvee_{s \in \mathbb{R}} \sigma_{\sigma;\mathcal{A}}^s(\mathcal{B}) = \mathcal{A}$ iff the semicyclic representation for $\sigma$ is cyclic on $\mathcal{B} \cap \mathfrak{N}_\sigma$. 
\end{lemma}
\begin{proof}
Define for $V>0$
\begin{equation}
\mathcal{A}_V = \sigma_{\sigma;\mathcal{A}}^s(\mathcal{B}) \qquad V = e^{2\pi s} 
\end{equation}
By assumption $\sigma$ is semifinite when restricted to any of these. 

Given \cite{araki2005extension} if we have cyclicity on the subalgebra then we get a strongly continuous translation group
$U(a)$ so that  
$\bigvee_{s \in \mathbb{R}} \sigma_{\sigma;\mathcal{A}}^s(\mathcal{B}) = 
\bigvee_{a>0} U(-a)\mathcal{A}U(a) = \mathcal{A}$ by strong continuity.  

In the other direction. Consider a Reeh-Schlieder like argument. Define the support projection for $a>0$:
\begin{equation}
\pi_\eta({\mathcal{A}'_a}) = \overline{ \eta( \mathfrak{N}_\sigma \cap \mathfrak{N}_\sigma^\dagger \cap
\mathcal{A}_a)}
\end{equation}
It must be that $\pi_\eta(\mathcal{A}'_b) \leq \pi_\eta(\mathcal{A}'_a)$ for $a < b$ and also:
\begin{equation}
    \label{eq:spi}
\sigma_{\sigma;\mathcal{A}}^s(\pi_\eta(\mathcal{A}'_b) )  = \pi_\eta(\mathcal{A}'_{b e^{2\pi s}}) 
\end{equation}
Then consider some $\xi \in 1-\pi_\eta(\mathcal{A}'_b)$ and write:
\begin{equation}
f(s) = \left< \xi \right| \Delta_{\sigma}^{-is} \left| \eta(n_b) \right> 
 = \left< \xi \right| \left. \eta(\sigma_{s;\mathcal{A}}(n_b)) \right> \qquad n_b \in \mathfrak{N}_\sigma \cap \mathfrak{N}_\sigma^\dagger \cap
\mathcal{A}_b
\end{equation} 
which now vanishes $f(s) = 0$ for $s\geq 0$ using $\sigma_{\sigma;\mathcal{A}}^s(\mathcal{B}) \subset \mathcal{B}$. We can analytically continue this to the strip with continuity in the closure implying that $f(s) =0$ for all $s$.
Since this applies to all $n_b \in \mathfrak{N}_\sigma \cap \mathfrak{N}_\sigma^\dagger \cap
\mathcal{A}_b$ we see that $\xi \in 1-\pi_\eta({\mathcal{A}'_a})$ for $0 < a < b$ and hence
$\pi_\eta({\mathcal{A}'_a}) \leq \pi_\eta({\mathcal{A}'_b})$ so equality $\pi_\eta({\mathcal{A}'_a}) = \pi$ for all $a>0$. 

Additivity implies that:
\begin{equation}
\pi \in \bigwedge_{a>0} \mathcal{A}_a' = \left( \bigvee_{a>0} \mathcal{A}_a \right)' = \mathcal{A}'
\end{equation}
But we also have that \eqref{eq:spi} implies that $[\Delta_\sigma^{is},\pi]=0$. 
That is $\pi$ is in the centralizer. 
Now $J \eta_\sigma(n) = \Delta_{\sigma}^{1/2} \eta_\sigma(n^\dagger) = \pi \Delta_{\sigma}^{1/2} \eta_\sigma(n^\dagger)  =
\pi J \eta_\sigma(n) $ where $n \in \mathfrak{N}_\sigma \cap \mathfrak{N}_\sigma^\dagger \cap
\mathcal{B}$. Hence $J \pi= \pi J \pi$. But taking the $\dagger$ implies $[\pi,J]=0$ and so 
$\pi = J \pi J \in \mathcal{A}$.  Hence $\pi \in Z(\mathcal{A})$. 
Also $\sigma(n^\dagger (1-\pi) n) = \| \eta_\sigma( (1-\pi) n) \|^2 = 0$ for $n \in  \mathfrak{N}_\sigma \cap \mathfrak{N}_\sigma^\dagger \cap \mathcal{B} $. So faithfulness implies
that $(1-\pi) n  =0$ so that semifiniteness of the restriction implies that $\pi=1$. 
\end{proof}

And finally a proof of Theorem~\ref{thm:trtohm}.
\begin{proof}
Apply Lemma~\ref{lem:onlyc} to prove that $e^{ i P} \mathcal{A} e^{- i P} \subset_\sigma \mathcal{A}$
is a HSMI based on $\sigma$. Conversely given some $\mathcal{B} \in \mathcal{HS}(\mathcal{A},\sigma)$ Araki-Zsido showed all (a,b,c) of Theorem~\ref{thm:hsmi-weight} for the corresponding $\widetilde{P}$ given
in \eqref{eq:promise} along with $e^{ i \widetilde{P}} \mathcal{A} e^{- i \widetilde{P}} = \mathcal{B}$ and $\sigma|_{\mathcal{A}_a} = \sigma( \widetilde{U}(a) \cdot \widetilde{U}(-a))$ for all $a\geq0$, so that the map defined in this theorem is surjective. For $\mathcal{B}  = e^{ i P} \mathcal{A} e^{- i P}$ we just need
to show that the extra condition $\sigma|_{\mathcal{A}_a} = \sigma( U(a) \cdot U(-a))$ implies that $P = \widetilde{P}$ proving invertibility on this set. For this we simply
note that the modular operators of these weights $\sigma|_{\mathcal{A}_a} = \sigma( U(a) \cdot U(-a)) \equiv \sigma_a $ on $\mathcal{A}_a$ each determine the two different translations so they must agree:
\begin{equation}
U(a(e^{2\pi s} -1))  = \Delta_{\sigma_a}^{-is} \Delta_{\sigma}^{is}  
= \Delta_{\sigma|_{\mathcal{A}_a}}^{-is} \Delta_{\sigma}^{is}   = \widetilde{U}(a(e^{2\pi s} -1))  
\end{equation}
\end{proof}

\subsection{Lemma~\ref{lem:gamma}, bounded case}

\label{app:prop2}

We prove Lemma~\ref{lem:gamma} for $y^\dagger \in \mathfrak{N}_\varphi$; the general case
follows in Section~\ref{sec:finite-weight} by approximating $y$ along $\pi_\iota y$. The argument
is that of \cite{Ceyhan:2018zfg} with the modular flow of the weight in place of that of the state. 
The proof in \cite{Ceyhan:2018zfg} is itself a generalization of standard
Tomita-Takesaki like arguments (see \cite{Witten:2018lha} for clear discussion).

\begin{proof}[Proof of Lemma~\ref{lem:gamma} for $y^\dagger \in \mathfrak{N}_\varphi$.]
On $\overline{\Sigma_1}$ we have $s = r \in \mathbb{R}$, and combining
$\sigma_{\eta,\psi}'^{r}(x') = \Delta^{ir} x' \Delta_{\eta,\psi}^{-ir}$ with
$\Delta^{-ir} U(a) \Delta^{ir} = U(a e^{2\pi r})$, from Theorem~\ref{thm:hsmi-weight},
\begin{equation}
\label{eq:Gammaw}
\big| \Gamma(x',y;z,s) \big> = \Delta_{\eta,\psi}^{is} (x')^\dagger U\!\left(- \epsilon e^{2\pi (z+s)}\right)
x' \Delta_{\eta,\psi}^{-is} y \left| \psi \right> \,, \qquad (z,s) \in \overline{\Sigma_1}
\end{equation}
This holds on the bottom edge only: for ${\rm Im}\, s = \theta >0$ the outer flow
$\Delta_{\eta,\psi}^{is} = \Delta_{\eta,\psi}^{-\theta} \Delta_{\eta,\psi}^{it}$ is unbounded, and
the vector to its right is not known to lie in its domain.

For $n' \in \mathfrak{N}_{\varphi'} \cap \mathfrak{N}_{\varphi'}^\dagger \subset \mathcal{A}'$ set
\begin{equation}
\label{eq:Gn}
G_{n'}(z,s) \equiv \left( \big| \eta'(n') \big>, \Delta_{\eta,\psi}^{is} (x')^\dagger
U\!\left(- \epsilon e^{2\pi (z+s)}\right) x' \Delta_{\eta,\psi}^{-is} y \left| \psi \right>\right)
\end{equation}
Here $\Delta_{\eta,\psi}^{-\theta}$ acts on $\eta'(n')$, which lies in
its domain by the commutant version of \eqref{eq:relmod-weight}, so
\eqref{eq:Gn} is defined on $\overline{T}$, holomorphic on $T$, continuous on $\overline{T}$, and
holomorphic on each $\Sigma_i$. Such $\eta'(n')$ are dense by semi-cyclicity.

We can use Tomita-Takesaki theory to estimate each modular factor for  $\theta \in [0,1/2]$:
\begin{equation}
\label{eq:crude}
\big| G_{n'}(z,s) \big| \leq \| x'\|^2\, \| \eta'(n') \|^{1-2\theta}
\big\| (n')^\dagger \left| \psi \right> \big\|^{2\theta}
\big\| y \left| \psi \right> \big\|^{1-2\theta} \varphi( y y^\dagger)^{\theta}
\end{equation}
finite on $\overline{T}$, although not uniform in $\eta'(n')$.

On $\overline{\Sigma_1}$ we have
$\| \Gamma \| \leq \| x' \|^2 \| y \| \| \psi \|$. At the top corner,
with $(z,s) \in \overline{\Sigma_2} \cap \overline{\Sigma_3}$, then
several lines of algebra (essentially the same as in \cite{Ceyhan:2018zfg})
give:
\begin{equation}
\label{eq:topcorner}
G_{n'}(z,s=i/2+t)  = \left( \big| \eta'(n') \big>, U(- \epsilon e^{2\pi z})y J \big| \Gamma(x',1;z,t) \big>\right)
\end{equation}
 With \eqref{eq:crude} controlling the interior we can apply 
Phragm\'en-Lindel\"of to the strips foliating $T$ giving:
\begin{equation}
\label{eq:unifbound}
\big| G_{n'}(z,s) \big| \leq \| \eta'(n') \| \,\| x' \|^2 \| y \| \| \psi \| \,, \qquad (z,s) \in \overline{T}
\end{equation}
which is (d). The Riesz representation theorem then defines $\big| \Gamma(x',y;z,s) \big>$ on $\overline{T}$ with
$\Gamma(x',y;\cdot,\cdot) \in \mathcal{H}(T) \cap \mathcal{C}(\overline{T})$ and
$\Gamma(x',y;f_i(\cdot,r)) \in \mathcal{H}(S) \cap \mathcal{C}(\overline{\Sigma_i})$
where the vector at the top corner can be computed using \eqref{eq:topcorner} giving (b).
The vector is an analytic continuation of \eqref{eq:Gammaw} from the bottom strip so that (a) is satisfied. 

For (c) note that at the top corner of $T$ the vector $\Gamma(x',y;z,i/2+t) =  X \psi$ for
some $X \in \mathcal{A}$. It is then possible to use an interpolation argument to show 
the domain
condition $y^\dagger \Gamma(x',y;z,s) \in \mathcal{D}(\Delta_{\psi}^{{\rm Im} (s)})$ and moreover that
$\Delta_\psi^{-is} y^\dagger \Gamma(x',y;z,s)$ has an analytic continuation 
to $T$ appropriately continuous on the closure. The edge of the wedge theorem then means we only need to check the symmetry relation along $\Sigma_1$ which is a short calculation. 
\end{proof}

\section{Approximation schemes}
\label{app:approximation}

\subsection{Approximate states with finite \texorpdfstring{$D$ and $\Theta$}{relative entropy and quantum expansion}}
\label{app:finiteDTh}

In this Appendix we aim to prove Theorem~\ref{thm:fullapprox}.
We break this down:
\begin{lemma}
\label{lem:partway}
Given some $\rho \in \mathcal{A}_\star^+$ there exists a sequence of operators $a_n' \in \mathcal{A}'$ such that $a_n' \eta \in \mathscr{P}^\natural$ and:
\begin{equation}
\| \rho - \rho_n \| \rightarrow 0 \,,\quad \lim_n D(\rho_n) = D(\rho)\,, \quad \rho_n \equiv  \omega_{a_n' \eta}|_{\mathcal{A}}\,, \quad \rho_n(1) = \rho(1)
\end{equation}
Furthermore if $\Theta(\rho) < \infty$, then we can pick $a_n' \eta \in \mathcal{D}(\sqrt{P})$ with
\begin{equation}
\lim_n \Theta(\rho_n) = \Theta(\rho)
\end{equation}
\end{lemma}
\begin{proof}
Let $\xi_\rho \in \mathscr{P}^\natural$ be the vector purification of $\rho$. We apply a Gaussian smearing:
\begin{equation}
\psi_k = f_k(\ln \Delta_{\eta,\xi_\rho}) \xi_{\rho}\,, \qquad \rho_k = \omega_{\psi_k}|_{\mathcal{A}} \,, \qquad f_k(x) = \exp( - x^2/k)
\end{equation}
which have the property that $\| \psi_k - \xi_\rho \| \rightarrow 0$ and $\lim_k D(\rho_k) = D(\rho)$ and $\psi_k = \tilde{a}_k' \eta$ 
for some $\tilde{a}_k' \in \mathcal{A}'$ as established in Lemma 5 and Theorem 5 of \cite{Faulkner:2020iou}, see also \cite[Lemma~9]{Faulkner:2020recoveryII}. There is a partial isometry $\theta_k'$ such that $\theta_k' \psi_k \in \mathscr{P}^\natural$
and so we may set $a_k' = \theta_k' \tilde{a}_k'$. 

Now suppose that $\Theta(\rho) < \infty$, then we use the following form:
\begin{equation}
\psi_k = \int ds \tilde{f}_k(s)\Delta_{\eta,\xi_\rho}^{-is} \Delta_{\xi_{\rho}}^{is}  \xi_{\rho}
\qquad \tilde{f}_k(s) =(k/4\pi)^{1/2} \exp( - k s^2/4)
\end{equation}
which allows us to estimate using Lemma~\ref{lem:duality}
\begin{equation}
\| \sqrt{P} \psi_k \| \leq \int ds |\tilde{f}_k(s)| \| \sqrt{P} u_s \xi_{\rho} \|
=\sqrt{ \Theta(\rho)} \int ds  \tilde{f}_k(s)\sqrt{1 + e^{-2\pi s}}
\end{equation}
As $k \rightarrow \infty$ the right hand side converges to $\sqrt{2 \Theta(\rho)}$ hence:
\begin{equation}
\limsup_k \| \sqrt{P} \psi_k \|^2  \leq 2 \Theta(\rho)
\end{equation}
Lemma~\ref{lem:calc} combined with lower semicontinuity of the expectation value of $P$ then gives:
\begin{equation}
2\Theta(\rho) = \| \sqrt{P} \xi_\rho \|^2 \leq \liminf_k \| \sqrt{P} \psi_k \|^2  \leq  \limsup_k \| \sqrt{P} \psi_k \|^2  \leq 2 \Theta(\rho)
\end{equation}
implying equality through the chain. By Lemma~\ref{lem:cont}, $\Theta(\rho_k)$ and $\Theta(\psi_k|_{\mathcal{A}'})$ are both lower semicontinuous
to $\Theta(\rho)$ so the sum rule \eqref{eq:sumT} gives:
\begin{equation}
2 \Theta(\rho) = \lim_k \| \sqrt{P} \psi_k \|^2 \geq \liminf_k \Theta(\rho_k) + \liminf_k \Theta(\omega_{\psi_k}|_{\mathcal{A}'})
\geq 2 \Theta(\rho)
\end{equation}
and hence $\liminf_k  \Theta(\rho_k)  = \Theta(\rho)$. Hence we can pick a subsequence 
$\rho_n = \rho_{k_n}$ where $\lim_n\Theta(\rho_n) = \Theta(\rho)$. Rescaling $a_n'\rightarrow Z_n a_n'$ with $Z_n> 0$ and $\lim_n Z_n = 1$ does not affect any of the limits for both $D(\rho_n)$ and $\Theta(\rho_n)$ while also keeping $a_n' \eta \in \mathcal{D}(\sqrt{P})$, so we can choose these such that $\rho_n(1) = \rho(1)$ hence verifying
all the claims in the Lemma. 
\end{proof}

To finish the proof of Theorem~\ref{thm:fullapprox} we need to allow
for the possibility that $\Theta(\rho) = \infty$.
We will need some properties of the Araki-Masuda \cite{araki1982positive} $L_p$ norms based on the cyclic and separating $\eta$ and for $2 \leq p \leq \infty$ (these are also known to determine the sandwiched relative entropies):
Define the $L_p$ space as the subset of $\mathscr{H}$ for which the following norm is well defined and finite:
\begin{equation}
\label{Lpnorm}
\| \psi \|_{L_p(\mathcal{A}')} = \sup_{\chi \in \mathscr{H} : \| \chi \| \leq  1}  \| \Delta_{\eta,\chi; \mathcal{A}}^{1/p-1/2} \psi \|
\end{equation}
where if $\psi \in L_\infty(\mathcal{A}')$ then there exists some $a' \in \mathcal{A}'$,
$\psi = a' \eta$ and $\| a' \eta \|_{L_\infty(\mathcal{A}')}  = \| a' \|$. For $p=2$ then $L_2(\mathcal{A}')= \mathscr{H}$.
We will need:

\begin{lemma}
\label{lem:Lp}
Suppose that $\psi \in L_p(\mathcal{A}')$ with $2 \leq p \leq \infty$, then $e^{ -i \epsilon e^{-i \theta} P} \psi \in L_p(\mathcal{A}')$ for $ \epsilon > 0$ and $0 \leq \theta \leq 2\pi/p$:
\begin{equation}
\| e^{ -i \epsilon e^{-i \theta} P} \psi \|_{L_p(\mathcal{A}')} \leq \| \psi \|_{L_p(\mathcal{A}')}
\end{equation}
Also for $p > 2$ and $\| \psi \| = 1$ the following quantity is monotonically non-decreasing in $p$ and bounds the relative entropy
\begin{equation}
    \label{eq:p2}
\frac{2p}{p-2} \ln  \| \psi \|_{L_p(\mathcal{A}')} \geq D(\psi|_{\mathcal{A}})
\end{equation}
Furthermore, if $\psi \in L_q(\mathcal{A}')$ for some $q>2$, then as $p$ approaches $2$:
\begin{equation}
\lim_{p \rightarrow 2^+} \frac{2p}{p-2} \ln  \| \psi \|_{L_p(\mathcal{A}')} = D(\psi|_{\mathcal{A}})
\end{equation}
\end{lemma}
\begin{proof}
The first property follows from an interpolation argument that establishes that the following map
$T_{z}(\mathcal{A}' \subset \mathcal{B}') \equiv \Delta_{\mathcal{B}}^{-z} \Delta_{\mathcal{A}}^{z}: L_{p}(\mathcal{A}') \rightarrow L_{p}(\mathcal{B}')$ is a contraction for $0\leq {\rm Re} z \leq 1/p$:

We start by defining $T_{z}(\mathcal{A}' \subset \mathcal{B}')$ as the bounded (since $\Delta_{\mathcal{B}} \geq \Delta_{\mathcal{A}}$) extension of the product
of modular operators acting on $\mathscr{H}$.

We then check the contractive property at $z=1/p$ as follows. Act on $\psi = \Delta_{\eta;\mathcal{A}}^{-1/p} a' \eta$ so that:
\begin{equation}
T_{1/p}(\mathcal{A}' \subset \mathcal{B}') \psi = 
 \Delta_{\eta;\mathcal{B}}^{-1/p}  a' \eta = J_{\mathcal{B}'}^p J_{\mathcal{A}'}^p \psi
\end{equation}
where $J_{\mathcal{A}'}^p  = J_\mathcal{A} \Delta_{\mathcal{A}}^{1/p-1/2}$ is an antilinear isometry
acting on $L_p(\mathcal{A}')$ (see \cite{araki1982positive}) and similarly for $\mathcal{B}'$. 
And since $L_p(\mathcal{A}') \subset L_p(\mathcal{B}')$ contractively, \cite{Berta:2016vnw}, we find
that $T_{1/p}(\mathcal{A}' \subset \mathcal{B}')$ is a contraction at $z=1/p$ from density.

 For the other values of $z$  consider:
\begin{align}
\left| \Gamma(z) \right> = \Delta_{\eta,\chi;\mathcal{B}}^{1/p-1/2} \Delta_{\eta;\mathcal{B}}^{-z} \Delta_{\eta;\mathcal{A}}^{z}
\psi
\end{align}
for $\| \chi \| \leq 1$.
A standard argument defines this vector and proves that it is 
 analytic inside the strip $0 \leq {\rm Re} z \leq 1/p$.
 The argument involves checking this statement weakly on $(b' \eta, \cdot)$ and then estimating
 the norm at the edges:
 \begin{align}
 \| \left| \Gamma(1/p+is) \right>  \| = 
 \| \Delta_{\eta,\chi;\mathcal{B}}^{1/p-1/2} \Delta_{\eta;\mathcal{B}}^{-is}  J_{\mathcal{B}'}^p J_{\mathcal{A}'}^p \Delta_{\eta;\mathcal{A}}^{is }
\psi \| \leq \| \psi \|_{L_p(\mathcal{A}')} \\
 \| \left| \Gamma(is) \right>  \| = 
 \| \Delta_{\eta,\chi;\mathcal{B}}^{1/p-1/2} \Delta_{\eta;\mathcal{B}}^{-is}  \Delta_{\eta;\mathcal{A}}^{is }
\psi \| \leq \| \psi \|_{L_p(\mathcal{A}')}
 \end{align}
 and then using the Phragm\'en-Lindel\"of theorem to extend this bound into the interior of the strip.
 The fact that $\mathcal{B}' \eta$ is a core for $\Delta_{\eta,\chi;\mathcal{B}}^{1/p-1/2} $ establishes
 that $\Delta_{\eta;\mathcal{B}}^{-z} \Delta_{\eta;\mathcal{A}}^{z}
\psi \in \mathcal{D}(\Delta_{\eta,\chi;\mathcal{B}}^{1/p-1/2})$. 
The explicit bound also gives:
\begin{equation}
\| T_{z}(\mathcal{A}' \subset \mathcal{B}') \psi \|_{L_p(\mathcal{B}')}
= \sup_{\chi \in \mathscr{H} : \|\chi \| \leq 1} \| \left| \Gamma(z) \right>  \| \leq \| \psi \|_{L_p(\mathcal{A}')} 
\end{equation}

We apply this for $z=\theta/(2\pi)$ here
and $\mathcal{B} = e^{ i \epsilon P} \mathcal{A} e^{- i \epsilon P} \subset \mathcal{A}$
so that:
\begin{equation}
\| e^{ -i \epsilon e^{-i\theta} P }\psi \|_{L_p(\mathcal{A}')}
=  \| e^{ i\epsilon(1- e^{-i\theta})  P }\psi \|_{L_p(\mathcal{B}')} 
= \| T_{\theta/(2\pi)}(\mathcal{A}' \subset \mathcal{B}')\psi \|_{L_p(\mathcal{B}')}  
\leq\| \psi \|_{L_p(\mathcal{A}')}  
\end{equation}

The other $L_p$ norm properties are standard and can be found in \cite{Berta:2016vnw} and \cite{Jencova:2018tqz}.
\end{proof}

\begin{proof}[Proof of Theorem~\ref{thm:fullapprox}] Without loss of generality
    we may take $\rho(1)=1$.
By Lemma~\ref{lem:partway}, only the case $\Theta(\rho) = \infty$ remains.
Using the states in Lemma~\ref{lem:partway} determined by a sequence $a_n' \in \mathcal{A}'$, we construct:
\begin{equation}
\psi_n = e^{-i \epsilon_n e^{-i \pi/p_n} P} a_n' \eta\,, \quad p_n> 2, \epsilon_n >0\,
\end{equation}

We aim to find a sequence $p_n \rightarrow 2$  and $\epsilon_n \rightarrow 0$ such that
$\rho_n = \omega_{\psi_n}|_{\mathcal{A}}$
satisfies the claims in this Theorem. 
For any such sequence:
\begin{equation}
 \| \psi_n - \xi_\rho \|  \leq \| a_n' \eta - \xi_\rho \| +  \| (e^{-i \epsilon_n e^{-i \pi/p_n} P} -1)a_n'  \eta \|  
 \leq 3\| a_n' \eta - \xi_\rho \| +\| (e^{-i \epsilon_n e^{-i \pi/p_n} P} -1)\xi_\rho \| 
\end{equation}
using first the triangle inequality and then the norm bound on $\| e^{-i \epsilon_n e^{-i \pi/p_n} P} \| \leq 1$.
The first term vanishes by Lemma~\ref{lem:partway}. The second vanishes using the spectral
representation of $P$ and dominated convergence.

Using the spectral representation of $P$ and the fact that ${\rm Re}( -i \epsilon_n e^{-i \pi/p_n} ) < 0$ then it
 is clear that $\psi_n \in \mathcal{D}(\sqrt{P})$ so that $\Theta(\rho_n) < \infty$.

Also Lemma~\ref{lem:Lp} gives: 
\begin{equation}
\label{eq:quickref}
s_n  + \frac{1}{\rho_n(1)} D(\rho_n)  \leq \frac{2p_n}{p_n-2} \ln  \| \psi_n \|_{L_{p_n}(\mathcal{A}')}
\leq \frac{2p_n}{p_n-2} \ln  \| a_n' \eta \|_{L_{p_n}(\mathcal{A}')} 
\end{equation}
where $s_n =  \frac{2 \ln \rho_n(1)}{p_n-2} $. In the first inequality we applied \eqref{eq:p2} to the unit vector $\psi_n/\| \psi_n\|$.
The
right hand side of \eqref{eq:quickref} is bounded by $2 \ln \| a'_n \|  < \infty$
so that $D(\rho_n) < \infty$.

Since $\| \psi_n - \xi_\rho \| \rightarrow 0$ lower semicontinuity gives
\begin{equation}
\label{eq:bothlower}
\liminf_n D(\rho_n) \geq D(\rho) \qquad \liminf_n \Theta(\rho_n) \geq \Theta(\rho)
\end{equation}
Since we have assumed $\Theta(\rho) = \infty$ this implies the second limit in \eqref{eq:toprovelim}.
If $D(\rho) = \infty$ then \eqref{eq:bothlower} already proves the first limit in \eqref{eq:toprovelim} and we may pick $p_n \rightarrow 2, \epsilon_n \rightarrow 0$ arbitrarily, skipping
the rest of this proof. 

If $D(\rho) < \infty$ then we fix $p_n,\epsilon_n$ as follows. 
At fixed $n$, allowing $p$ to vary independently in the right hand side of \eqref{eq:quickref}, we have:
\begin{equation}
 \lim_{p \rightarrow 2^+} \frac{2p}{p-2} \ln  \| a_n' \eta \|_{L_{p}(\mathcal{A}')} = D(\omega_{a_n' \eta}|_{\mathcal{A}}) 
\end{equation}
using the normalization condition $\| a_n' \eta \| = 1$ in Lemma~\ref{lem:partway}. And since
$D(\omega_{a_n' \eta}|_{\mathcal{A}})  \rightarrow D(\rho)$ there is a sequence $p_n \rightarrow 2$ for which:
\begin{equation}
\lim_n \frac{2p_n}{p_n-2} \ln  \| a_n' \eta \|_{L_{p_n}(\mathcal{A}')} =  D(\rho)
\end{equation}
Then we may choose $\epsilon_n \rightarrow 0$ so that $\rho_n(1) \rightarrow 1$ fast enough
to make $s_n \rightarrow 0$.  
Lower semicontinuity of relative entropy and \eqref{eq:quickref} gives:
\begin{equation}
D(\rho) \leq \liminf_n D(\rho_n) \leq \limsup_n D(\rho_n) \leq \lim_n \frac{2p_n}{p_n-2} \rho_n(1) \ln  \| a_n' \eta \|_{L_{p_n}(\mathcal{A}')} = D(\rho)
\end{equation}
implying the rest of \eqref{eq:toprovelim}. 
\end{proof}

\subsection{Approximate identity}
\label{app:approxid}

In this appendix we remind the readers about the existence of approximate
identities for normal semifinite weights. We will need two different types throughout the paper:

\begin{lemma}
\label{lem:approxid}
Given an unbounded nsf weight $\varphi$ there exists an approximate identity:
a net of positive operators $x_\iota \in \mathfrak{M}_\varphi$ such that $so-\lim_\iota x_\iota =1$ and $\| x_\iota \| \leq 1$.
We can additionally choose the net to satisfy one of the following conditions:
\begin{itemize}
\item[(a)] $x_\iota$ is monotonically increasing along the net.
\item[(b)] each operator is a projection $x_\iota^2 = x_\iota$.
\end{itemize}
\end{lemma}
\begin{proof}
(a) We can pick the directed set defining the net simply to be 
\begin{equation}
\iota = \{ x : x \in \mathfrak{M}_\varphi^+, \| x \| < 1 \}
\end{equation}
with $x_\iota$ as the identity map. This is a directed set, under the usual order on positive operators,
since for all $a,b \in \iota$ we can find a larger element:
\begin{equation}
c = R[ R^{-1}[a]+R^{-1}[b]] \qquad R[x] = \frac{x}{1+x}
\end{equation}
This is larger since $c = R[R^{-1}[a]+R^{-1}[b]] \geq R[ R^{-1}[a]] = a$
due to the fact that $R$ is an operator monotone function. It is also larger than $b$ by an identical argument. 
Also since $\| a \|, \| b\| < 1$
the operators $R^{-1}(a), R^{-1}(b)$ are bounded operators so that $ \| c\| < 1$. 
Finally $R^{-1}(a) \leq a ( 1 - \| a\|)^{-1}$ so that $R^{-1}(a),R^{-1}(b) \in \mathfrak{M}_\varphi^+$
and since $c \leq R^{-1}[a]+R^{-1}[b]$ it must be that $c \in  \mathfrak{M}_\varphi^+$.  

We now check this net converges to $1$. The semifinite condition on $\varphi$ guarantees the existence of some net converging to $so-\lim_\alpha n_\alpha = 1$ and we can take this net to be positive operators with norm $< 1$ by rescaling $n'_\alpha = \ell(n_\alpha) \varphi(n_\alpha)/(1+\varphi(n_\alpha))$
 where $\ell(x) = \min(x,1)$ and this maintains $n_\alpha' \in \mathfrak{M}_\varphi$ and convergence $so-\lim_\alpha n_\alpha'  = 1$ due to the unboundedness and lower semicontinuity of $\varphi$. Hence for any $n_\alpha'$ there is a larger element in $x_\iota$ showing that $so-\lim_\iota x_\iota = 1$.

(b) We can now sacrifice monotonicity and simply use the spectral projections
of $x_\iota$ constructed above $\pi_\iota = E_{[1/2,1]}(x_\iota)$. 
Then $\pi_\iota \leq 2 x_\iota$ so that $\pi_\iota \in \mathfrak{M}_\varphi^+$
and $(1- \pi_\iota) \leq 2 ( 1- x_\iota)$ so that $so-\lim_\iota \pi_\iota = 1$. 
\end{proof}

\subsection{Approximations for the relative entropy sum rule
}
\label{app:approxsumrule}

In this Appendix we prove Lemma~\ref{lem:approx-sum}. 
The main difficulty is finding states that satisfy the domain condition 
$\psi_\iota \in \mathcal{D}( \Delta_{\eta_{V_2},\psi_\iota;V_2}^{1/2}) \cap \mathcal{D}( \Delta_{\eta_{V_1'},\psi_\iota;V_1'}^{1/2})$. In fact this condition is equivalent to a condition on supports of the induced states:
\begin{equation}
    \label{eq:requiredfor}
\pi(\psi_\iota|_{V_2}) \in \mathfrak{N}_{V_2}\,,\quad \pi(\psi_\iota|_{V_1'}) \in \mathfrak{N}_{V_1'}  
\end{equation}
as can be seen from the definitions of the Tomita operators:
\[
\Delta_{\eta_{V_2},\psi_\iota;V_2}^{1/2} \psi_\iota = (\theta')^\dagger J_{V_2} \eta(\pi(\psi_\iota|_{V_2})).
\]
Here $\theta' \in \mathcal{A}_{V_2}'$ is the partial isometry taking $\psi_\iota$ to the natural cone. The same applies
for $V_1'$. We will achieve this by acting on $\psi$ with an approximate
identity for the two algebras with supports satisfying the above conditions. 
We need these to also maintain nice finite $P$ expectation and relative entropies. 

For now we ignore the support condition requirement and construct an approximate identity that is well behaved with respect to $P$ and $\ln \Delta_\varphi$. We will state these requirements
for a single algebra $\mathcal{A}$:
\begin{lemma}
\label{lem:approxunit0}
There exists an approximate identity $x_\iota$, for the  definition domain $\mathfrak{M}_\varphi \subset \mathcal{A}$ of $\varphi$ with $\| x_\iota \| \leq 1, x_\iota \geq 0$ and 
such that the following limits exist as bounded operators in $\mathcal{A}$:
\begin{equation}
    \label{eq:toestab}
so-\lim_{s \rightarrow 0} \frac{ \sigma_{\varphi}^{s}(x_\iota) -x_\iota }{ i s} \equiv [x_{\iota}, \ln \Delta_{\varphi}]  \qquad
so-\lim_{a \rightarrow 0^+} \frac{ U(a)^\dagger x_\iota U(a) - x_\iota}{ia} \equiv [P,x_{\iota}]
\end{equation}
where the right-hand side of both equations are uniformly bounded operators along the net. 
Additionally we can take $x_\iota$ to be entire analytic under modular flow $\mathbb{C} \ni z \rightarrow \sigma_{\varphi}^{z}(x_\iota) \in \mathfrak{M}_\varphi$
maintaining $so-\lim_\iota  \sigma_{\varphi}^{z}(x_\iota) = 1$ for all $z\in \mathbb{C}$. 
\end{lemma}
\begin{proof}
Consider an increasing approximate identity given in (a) of Lemma~\ref{lem:approxid} that we will
re-label as $y_\iota \in \mathfrak{M}_\varphi^+$ here.
We then smear the modular flow and positive null translates against a smooth function:
\begin{equation}
x_\iota = \int_{-\infty}^{\infty } d s \int_{0}^\infty d a g(a) e^{-s^2} \sigma_{\varphi}^s(U(a)^\dagger y_\iota U(a))
\end{equation}
where we take $g(a) \geq 0$ and $ \int_{0}^\infty d a  g(a)  =1/\sqrt{\pi}$ so the smearing over $s,a$ is normalized. We also take $g(a)$ to be uniformly differentiable, compactly supported in $(0,\infty)$ and  with the estimate:
\begin{equation}
    \label{eq:gprime}
 \int_{0}^\infty da  | g'(a) | < \infty
\end{equation}
Obviously a function $g$ satisfying these properties can be found. 

We now establish $x_\iota$ is entire under modular flow:
\begin{equation}
    \label{eq:normbound}
\| \sigma^z_{\varphi}(x_\iota) \| \leq  \int_{-\infty}^{\infty} d s | e^{-(s-z)^2} |\int_0^\infty da g(a)
\leq e^{({\rm Im} z)^2}
\end{equation}
which also gives the requirement $\| x_\iota \| \leq 1$. 
Also it is clear $\sigma_{\varphi}^z(x_\iota) \in \mathfrak{M}_\varphi$ since the definition domain is invariant under modular flow and positive translations give the bound $\varphi( x_\iota ) \leq \varphi(y_\iota)$ (see \eqref{eq:boundtranslate}).

We now show strong convergence to the identity. 
For all $\psi \in \mathscr{H}$ and $z \in \mathbb{C}$:
\begin{equation}
    \label{eq:tointegral}
\| (\sigma_{\varphi}^z(x_\iota)- 1) \left| \psi \right> \|
\leq\int_{-\infty}^{\infty } d s | e^{-(s-z)^2} |\int_{0}^\infty d a g(a)  \| ( \sigma_{\varphi}^s(U(a)^\dagger y_\iota U(a))-1) \left| \psi \right> \|
\end{equation}
Consider $(1-y_\iota)^2 \leq (1-y_\iota)$ so that:
\begin{equation}
    \label{eq:twosides}
\| ( \sigma_{\varphi}^s(U(a)^\dagger y_\iota U(a))-1) \left| \psi \right> \|^2
\leq \left< \psi \right| \sigma_{\varphi}^s(U(a)^\dagger (1-y_\iota) U(a)) \left| \psi \right>
\end{equation}
and define $f_\iota(a,s)$ as the right hand side of the above equation \eqref{eq:twosides}
then this is a decreasing
net over $\iota$ to $f_\iota(a,s) \rightarrow 0$. Since $f_\iota(a,s)$ are continuous in $a,s$ we get, via Dini's theorem, uniform convergence on compact subsets
of $\mathbb{R}_{\geq 0} \times \mathbb{R}$. This implies the same
mode of convergence for the left hand side of \eqref{eq:twosides}.
Since we demanded that $g(a)$ is continuous and compactly supported then we get convergence of the integral \eqref{eq:tointegral} to zero from the uniform
convergence condition and estimates of the gaussian integral at large $s$.
That is $\sigma_{\varphi}^z(x_\iota) \rightarrow 1$ in the strong operator topology. 
The above argument was given in Pedersen and Takesaki \cite{pedersen1973radon} and Zsid\'o's paper \cite{zsido2022equality}.

The Cauchy integral formula establishes the first limit on \eqref{eq:toestab} and the uniform bound on 
$\| [ \ln \Delta_\varphi, x_\iota] \|$. For the $P$ version of this bound consider:
\begin{equation}
i [x_\iota, P] \equiv so-\lim_{b \rightarrow 0} (x_\iota - U(b)^\dagger x_\iota U(b))/b
= \int_{-\infty}^{\infty} d s e^{- s^2 - 2 \pi s}  \int_{0}^\infty d a  g'(a)  \sigma_{\varphi}^s( U(a)^\dagger y_\iota U(a))
\end{equation}
where we can use dominated convergence to pass the limit inside the integral since we chose $g(a)$ to be uniformly differentiable. Thus establishing the existence of the limit in the second equation of \eqref{eq:toestab}.
The bound in \eqref{eq:gprime} leads to the estimate:
\begin{equation}
\| [x_\iota, P]\| \leq 
 \int_{-\infty}^{\infty} d s e^{-s^2 - 2\pi s} \int_{0}^\infty da  | g'(a) | \|  \sigma_{\varphi}^s( U(a)^\dagger y_\iota U(a)) \|  \leq \sqrt{\pi} e^{\pi^2} \int_{0}^\infty da  | g'(a) | 
\end{equation}
which is uniform along $\iota$. 
\end{proof}

This approximate identity does not, by construction, satisfy the support
condition required for \eqref{eq:requiredfor}. In order to fix this we
note that we can take certain functions of $x_\iota$ to fix this.  We need
to make sure this does not mess up the good behavior with respect to $P$ and $\ln \Delta_\varphi$. 
Consider:
\begin{lemma}
    \label{lem:f}
Consider any smooth bounded positive function $f : \mathbb{R}^+ \rightarrow \mathbb{R}^+$,
compactly supported with $f(t) = 0$ for $t \leq 1/2$. There exists some $C > 0$ 
such that for any positive $x \in \mathfrak{M}_\varphi^+$
\begin{enumerate}
    \item[(a)] the support of the new operator satisfies $\pi(f(x)) \in \mathfrak{M}_\varphi$  
    \item[(b)] and for any unbounded self-adjoint operator $Q$ with $[ x, Q] \in \mathcal{A}$, then $[f(x), Q] \in \mathcal{A}$ with $\| [f(x), Q]\| \leq C \| [ x, Q] \|$.
\end{enumerate}
where we define the commutators above as $[x,Q] = so-\lim_{b \rightarrow 0^+} i(e^{ i b Q} x e^{ - i b Q} - x)/b$ and similarly for $ x \rightarrow f(x)$. 
\end{lemma}
\begin{proof}
The conditions on $f$ imply that 
\begin{equation}
\| f'(t) \|_{L_\infty},\| \widetilde{f}(k) \|_{L_1}, \| \widetilde{f}(k)k \|_{L_1} < \infty
\end{equation}
where $\widetilde{f}$ is the Fourier transform. 

(a) The support of $f(x)$ is bounded by $E_{[1/2, \| x\| ]}(x)$, and:
\begin{equation}
\varphi( E_{[1/2, \| x \| ]}(x) ) \leq 2 \varphi( E_{[1/2, \| x \| ]}(x) x) 
\leq 2 \varphi( x) < \infty
\end{equation}
But since $f(x)(1- E_{[1/2, \| x \| ]}(x)) = f(x) E_{[0,1/2)}(x) = 0$
it must be that the support satisfies $\pi(f(x))  \leq E_{[1/2, \| x \| ]}(x)$
implying that $\pi(f(x)) \in \mathfrak{M}_\varphi$. 

(b) We will use the following formula (properly justified as a Bochner integral)
\begin{align}
 [\exp(i x),Q] &\equiv so-\lim_{b \rightarrow 0^+}  [\exp(i x),(e^{ibQ}-1)/ib]
\\ &= so-\lim_{b \rightarrow 0^+} \, i \int_0^1 d u \exp(i u x) [x, (e^{ibQ}-1)/ib] \exp(i (1-u) x)
    \label{eq:sameidea}
\end{align}
and we can pass the limit inside the integral with dominated convergence using the fact that $so-\lim_{b \rightarrow 0^+}  [x, (e^{ibQ}-1)/ib] = [x,Q]$
is a bounded operator so $ [x, (e^{ibQ}-1)/ib] $ is uniformly bounded in norm (i.e. because of the uniform boundedness principle).
We write:
\begin{equation}
 [\exp(i x),Q] = i\int_0^1 d u \exp(i u x) [x, Q] \exp(i (1-u) x)
\end{equation}

We can use the Fourier transform to represent $f(x)$ as a Bochner integral over
$e^{i k x}$. We can then pass the commutator of $Q$ inside using the same 
idea as in \eqref{eq:sameidea}:
\begin{equation}
 [f(x),Q] 
 = i\int d k \int_0^1 d u k \tilde{f}(k) \exp(i u k x) [x, Q] \exp(i (1-u) k x)
\end{equation}
The norm of the left hand side is bounded since $\| \widetilde{f}(k)k \|_{L_1} < \infty$
so that
\begin{equation}
\| [f(x),Q] \|
 \leq  \| [x, Q] \|  \|\tilde{f}(k)k \|_{L_1}
\end{equation}
and $[f(x),Q] \in \mathcal{A}$.
Also $C = \|\tilde{f}(k)k \|_{L_1}$ in the statement of the lemma.
\end{proof}

Combining these we have our final form of the approximate identity:
\begin{lemma}
\label{lem:approxunit}
There exists an approximate identity $x_\iota$, for the  definition domain $\mathfrak{M}_\varphi \subset \mathcal{A}$ of $\varphi$,
with non-trivial support satisfying $\pi(x_\iota) \in \mathfrak{M}_\varphi$ and satisfying the limits \eqref{eq:toestab} and uniform bound along the net given in Lemma~\ref{lem:approxunit0} and $\| x_\iota \| \leq 1, x_\iota \geq 0$. Additionally for any $\psi \in \mathscr{H}$ we have:
\begin{align}
    \label{eq:prem}
 \lim_{\theta \rightarrow 0^+} \frac{1}{\theta} [ \Delta_{\eta,\psi}^\theta,x_\iota] x_\iota  \left| \psi \right>
= [ \ln \Delta_{\varphi},x_\iota]   x_\iota \left| \psi \right>
\end{align}
\end{lemma}
\begin{proof}

We take the approximate identity of Lemma~\ref{lem:approxunit0} which we relabel
as $y_\iota$ and apply the
function $f$ of Lemma~\ref{lem:f} so that:
\begin{equation}
x_\iota = f(y_\iota)
\end{equation}
If we additionally require that $f(t) \leq 1$ and $f(1) = 1$ then $\lim_\iota x_\iota = f(1) = 1$
and also $\| x_\iota \| \leq \sup_t f(t) = 1$. 
We are left to check the final condition \eqref{eq:prem}. Recall that $x_\iota \in \mathfrak{M}_\varphi^+
\subset \mathfrak{N}_\varphi^+$
implies that $x_\iota \psi \in \mathcal{D}(\Delta_{\eta,\psi}^{1/2})$ 
and since $x_\iota^2 \leq \| x_\iota \| x_\iota \leq x_\iota$ we have $x_\iota^2 \in \mathfrak{M}_\varphi^+
\subset \mathfrak{N}_\varphi^+$ so also
$x_\iota^2 \psi \in \mathcal{D}(\Delta_{\eta,\psi}^{1/2})$ and the premise of \eqref{eq:prem}  makes sense. The limit can be computed using the Fourier transform representation
of $f(y_\iota)$:
\begin{align}
 & \lim_{\theta \rightarrow 0^+} \frac{1}{\theta} [x_\iota, \Delta_{\eta,\psi}^\theta] x_\iota  \left| \psi \right>
\\ &= \lim_{\theta \rightarrow 0^+} 
  i\int d k \int_0^1 d u k \tilde{f}(k) \exp(i u k y_\iota)\frac{1}{\theta} [y_\iota, \Delta_{\eta,\psi}^\theta] x_\iota \exp(i (1-u)k y_\iota) \left| \psi \right>
  \label{eq:thecase}
\end{align}
where we used the fact that $x_\iota$ commutes with functions of $y_\iota$ and so
the right hand side satisfies
 the domain condition on $\Delta_{\eta,\psi}^\theta$ for $0 \leq \theta \leq 1/2$.
Now given that $y_\iota$ is by construction entire analytic for modular flow (see
Lemma~\ref{lem:approxunit0} which applies to the net $y_\iota$) we can write:
\begin{align}
[y_\iota,\Delta_{\eta,\psi}^{\theta}]  &= [y_\iota,  \Delta_{\eta,\psi}^{\theta}] \Delta_{\eta,\psi}^{-\theta}  \Delta_{\eta,\psi}^{\theta}
 = (y_\iota - \sigma^{i \theta}_{\varphi}(y_\iota)) \Delta_{\eta,\psi}^{\theta}
\end{align}
where equality is guaranteed as long as both sides act on
vectors in their respective domains which is the case for \eqref{eq:thecase}.
Now we can take the limit of the integrand in \eqref{eq:thecase}
\begin{align}
\label{eq:thetaconv}
so-\lim_{\theta \rightarrow 0}\frac{1}{\theta} (y_\iota - \sigma^{i\theta}_{\varphi}(y_\iota))&= -i \partial_s \sigma^{s}_{\varphi}(y_\iota)|_{s=0}\\
\lim_{\theta \rightarrow 0} \Delta_{\eta,\psi}^{\theta}x_\iota \exp(i (1-u)k y_\iota) \left| \psi \right> &= x_\iota \exp(i (1-u)k y_\iota) \left| \psi \right>
\end{align}
which combines to:
\begin{align}
\lim_{\theta \rightarrow 0}\frac{1}{\theta} (y_\iota - \sigma^{i\theta}_{\varphi}(y_\iota))\Delta_{\eta,\psi}^{\theta}x_\iota \exp(i (1-u)k y_\iota) \left| \psi \right> &= -i \partial_s \sigma^{s}_{\varphi}(y_\iota)|_{s=0}\, x_\iota \exp(i (1-u)k y_\iota) \left| \psi \right>
\end{align}
Now convergence in \eqref{eq:thetaconv} implies, by the uniform boundedness principle, that $\|(y_\iota - \sigma^{i\theta}_{\varphi}(y_\iota))/ \theta \|$ is uniformly bounded for all $\theta$. 
Also the three line lemma gives the estimate:
\begin{equation}
\| \Delta_{\eta,\psi}^{\theta}x_\iota \exp(i (1-u)k y_\iota) \left| \psi \right>\|
\leq \varphi(x_\iota^2)^{\theta} \| \psi \|^{1-2\theta} 
\end{equation}
Combined we get a dominating function for the Bochner integral over $k,u$ and we can take the $\lim_\theta$ inside \eqref{eq:thecase} to conclude that:
\begin{align}
 \lim_{\theta \rightarrow 0^+} \frac{1}{\theta} [x_\iota,\Delta_{\eta,\psi}^\theta] x_\iota  \left| \psi \right>
= [x_\iota, \ln \Delta_{\varphi}] x_\iota \left| \psi \right>
\end{align}
where the existence of $[x_\iota, \ln \Delta_{\varphi}]
= [f(y_\iota), \ln \Delta_{\varphi}] $ is implied by Lemma~\ref{lem:approxunit0} and
Lemma~\ref{lem:f}. 

\end{proof}

The states $x_\iota \psi$ are well behaved for both null energy and relative entropy. 
For null energy:

\begin{lemma}
\label{lem:Pbound}
Given $\psi \in \mathcal{D}(\sqrt{P})$ and $x_\iota$ satisfying the properties of Lemma~\ref{lem:approxunit}, then $\| \sqrt{P} x_\iota \psi \|$ is uniformly bounded along $\iota$.
\end{lemma}
\begin{proof}
We have:
\begin{align}
\| [x_\iota, (P+1)^{1/2} ] \| &= \frac{1}{\pi} \left\| \int_0^\infty d \lambda \lambda^{1/2} \frac{1}{\lambda + P+ 1} [x_\iota,P]   \frac{1}{\lambda + P + 1} \right\| \\
& \leq \frac{1}{\pi} \left\|  [x_\iota,P] \right\| \int_0^\infty d \lambda \left\|\frac{\lambda^{1/2}}{(\lambda + P+1)^2}  \right\|
\leq \frac{1}{2} \left\|  [x_\iota,P] \right\| 
\end{align}
Then:
\begin{align}
\| \sqrt{P} x_\iota \left| \psi \right> \|^2 &=  \left< \psi\right| x_\iota (P+1) x_\iota \left| \psi \right> - \left< \psi \right| x_\iota^2 \left| \psi \right> 
\leq \| \sqrt{P+ 1} x_\iota \left| \psi \right> \|^2\\
&\leq \left( \frac{1}{2} \left\|  [x_\iota,P] \right\| + \sqrt{ \| \sqrt{P} \left| \psi \right> \|^2 + 1  } \right)^2  
\end{align}
picking $\| \psi \| = 1$ and we used $\| x_\iota \| \leq 1$. Thanks to Lemma~\ref{lem:approxunit}
this bound is then uniform as promised.
\end{proof}

And for relative entropy (we will not directly utilize this result for the relative entropy sum rule, but it will be useful
in Appendix~\ref{app:entropy-differences} for approximating relative entropy differences):
\begin{lemma}
    \label{lem:convrel}
Consider any $\psi \in \mathscr{H}$. Suppose that $D(\psi|\varphi)$ is well defined and finite, for any approximate identity $x_\iota$ such that $0 \leq x_\iota \leq 1$ and $\sup_\iota \|[ x_\iota, \ln \Delta_{\eta,\psi}]\| <\infty$
and the support projection has $\varphi( \pi(x_\iota)) < \infty$, 
then $D(x_\iota \psi |\varphi)$ is well defined and finite with:
\begin{equation}
    \label{eq:rellimit}
\lim_\iota D(x_\iota \psi |\varphi) = D(\psi|\varphi) 
\end{equation}
\end{lemma}
In order to prove this we will need some preliminary results on quadratic forms. Here we will treat $Q = -\ln \Delta_{\eta,\psi}$ as a quadratic form with domain $\mathcal{D}(|Q|^{1/2}) = \mathcal{D}(Q_+^{1/2}) \cap \mathcal{D}(Q_-^{1/2})
\subset \pi'(\psi) \mathscr{H}$ where $Q = Q_+ - Q_-$ is given by the spectral decomposition into positive and negative parts. Given this another way to characterize well defined and finite $D(\psi|\varphi)$ is $\iff$ $\psi \in \mathcal{D}(|Q|^{1/2})$.  For a general $Q$ we have the estimate:
\begin{lemma}
    \label{lem:Qgraph}
Let $Q$ be an unbounded, densely defined self-adjoint operator on $\mathscr{H}$ and $x$ a bounded operator 
with $\| [x,Q] \| < \infty$ (where the commutator is defined as in Lemma~\ref{lem:f}). 
Then $x$ is bounded on the graph norm of $|Q|^{1/2}$:
\begin{equation}
    \label{eq:Qhalf}
\| x \chi \|^2_{|Q|^{1/2}} 
\leq (2 \|x \|+\| [x,Q]\|) \| x \| \| \chi \|^2_{|Q|^{1/2}} \,, \qquad
\| \chi \|^2_{|Q|^{1/2}} \equiv\|  \chi \|^2 + \| |Q|^{1/2} \chi \|^2 
\end{equation}
for all $\chi \in \mathscr{H}$. 
\end{lemma}
\begin{proof}
Consider $\chi \in \mathcal{D}(Q)$ then $x \chi \in \mathcal{D}(Q)$ by the constraint on the commutator: 
\begin{align}
\| (Q^2 + 1)^{1/2} x \chi \|^2 &= \| Q x \chi \|^2 + \| x \chi \|^2
= \| x Q \chi + [Q,x] \chi \|^2 + \| x \chi \|^2 \\
& \leq ( \| x \| \| Q \chi \| + \| [Q,x] \| \| \chi \|)^2 + \| x \|^2 \| \chi \|^2 \\
& \leq \left( \| [ Q, x] \| \| \chi \| + \| x \|  \|(Q^2+1)^{1/2} \chi \| \right)^2 \\
& \leq  \left( (\| x \| + \| [ Q, x] \| )  \|(Q^2+1)^{1/2} \chi \| \right)^2  
\end{align}
where in the second inequality we used the elementary bound $(A+B)^2 + C^2 \leq ( B + (A^2+C^2)^{1/2})^2$ for $A,B\geq 0$. 
Now since $(Q^2+1)^{-1/2+is} \chi \in \mathcal{D}(Q)$ for $\chi \in \mathcal{D}(Q)$ we can interpolate this bound
as follows. Define:
\begin{align}
f(s) = \| (Q^2 + 1)^{-is} x (Q^2 + 1)^{is} \chi \|
\end{align}
Then the inequality we just proved gives $f(i/2+t) \leq (\| x \| + \| [ Q, x] \| )  \|\chi \|$.
Also $f(t) \leq \| x\| \| \chi \|$ and a three line estimate based on $\sup_{\xi \in \mathscr{H}_{\leq 1} }
\ln | \left< \xi \right| \left. (Q^2 + 1)^{-is} x (Q^2 + 1)^{+ is} \chi \right>|$ gives:
\begin{equation}
   \| (Q^2 + 1)^{1/4} x (Q^2 + 1)^{-1/4} \chi \| = f(i/4) \leq (\| x \| + \| [ Q, x] \| )^{1/2} \| x\|^{1/2} \|\chi \| 
\end{equation}
Considering $\zeta \in \mathcal{D}(|Q|^{3/2})$ then $\chi = (Q^2+1)^{1/4} \zeta$ is in $\mathcal{D}(Q)$
and
\begin{equation}
    \label{eq:abovebound}
   \| (Q^2 + 1)^{1/4} x \zeta \| = f(i/4) \leq (\| x \| + \| [ Q, x] \| )^{1/2}  \| x\|^{1/2} \|(Q^2 + 1)^{1/4}  \zeta  \| 
\end{equation}
Given any $\zeta \in \mathcal{D}(|Q|^{1/2})$ we can approximate it by one in $\zeta_n \in \mathcal{D}(|Q|^{3/2})$
using spectral projections for $|Q|$, $\zeta_n = E_{[0,n]}(|Q|) \zeta$. 
Dominated convergence controls the right hand side of \eqref{eq:abovebound}, $(Q^2+1)^{1/4}\zeta_n \rightarrow (Q^2+1)^{1/4} \zeta$ 
by the domain condition on $\zeta$.
Then $\| (Q^2+1)^{1/4} x( \zeta_n - \zeta_m)\| \rightarrow 0$ as a Cauchy
sequence by the above bound \eqref{eq:abovebound} and convergence of the right hand side. 
Hence by closedness of $(Q^2+1)^{1/4}$ we can take the limit $\zeta_n \rightarrow \zeta$
$\| (Q^2 + 1)^{1/4} x \zeta_n \| \rightarrow \| (Q^2 + 1)^{1/4} x \zeta \|$
and maintain the bound in \eqref{eq:abovebound}.
We can then extend this to all $\zeta \in \mathscr{H}$
where we interpret the right hand side as $=\infty$ if $\zeta \notin \mathcal{D}(|Q|^{1/2})$. 
Considering $|Q| \leq (Q^2+1)^{1/2} \leq |Q| +1$ gives \eqref{eq:Qhalf}. 
\end{proof}
\begin{proof}[Proof of Lemma~\ref{lem:convrel}]
    Without loss of generality we can take $\| \psi \| = 1$.
    Using $Q = - \ln \Delta_{\eta,\psi}$ and working on $\pi'(\psi) \mathscr{H}$ which is preserved by $x_\iota
    \in \mathcal{A}$,
    we note that $x_\iota \psi \in \mathcal{D}(|Q|^{1/2})$ from  Lemma~\ref{lem:Qgraph}. Now
    for any $\chi \in \mathcal{D}(Q)$ apply Cauchy-Schwarz:
    \begin{equation}
    \| |Q|^{1/2} (1-x_\iota) \chi \|^2 \leq \| (1-x_\iota) \chi \| \| Q (1- x_\iota) \chi \|
    \leq \| (1-x_\iota) \chi \| ( \| Q \chi \| + \| [Q,x_\iota] \| \| \chi \|)
    \end{equation}
    and using the uniform bound on $\| [Q,x_\iota]\|$ we can take the limit
    $\lim_\iota \| |Q|^{1/2} (1-x_\iota) \chi \| =0 $.  Now consider the Banach space $\mathcal{D}(|Q|^{1/2})$ with
    the norm given by the graph norm $\| \cdot \|_{|Q|^{1/2}}$. This is complete by the closedness of $|Q|^{1/2}$.
    Also $\mathcal{D}(Q) \subset \mathcal{D}(|Q|^{1/2})$ is dense in $\| \cdot \|_{|Q|^{1/2}}$
    since $\psi_n = E_{[0,n]}(|Q|) \psi$ satisfies $\lim_{n\rightarrow \infty} \| |Q|^{1/2} E_{[0,n]}(|Q|) \psi \|
     = \| |Q|^{1/2} \psi \|$ by monotone convergence. Also $ (1-x_\iota)$ thought of as an operator $\mathcal{D}(|Q|^{1/2}) \rightarrow \mathcal{D}(|Q|^{1/2})$
    equipped with the norm $\| \cdot \|_{|Q|^{1/2}}$, is uniformly bounded along $\iota$ from Lemma~\ref{lem:Qgraph}:
    \begin{equation}
    \| (1-x_\iota) \|_{\mathcal{D}(|Q|^{1/2}) \rightarrow \mathcal{D}(|Q|^{1/2})}
    \leq (2 \|1-x_\iota \| + \| [ x_\iota, Q]\|)^{1/2} \| 1-x_\iota\|^{1/2}
    \end{equation}
    Hence we can apply uniform boundedness
    for this Banach space, and this implies that $\lim_\iota |Q|^{1/2} (1-x_\iota)\left|\psi\right> =0$ for all $\psi \in \mathcal{D}(|Q|^{1/2})$. Since $x_\iota \psi \in \mathcal{D}(|Q|^{1/2})$ then $\left< \psi \right| x_\iota Q x_\iota \left| \psi \right>$ is well defined and finite. Bounds give:
    \begin{equation}
    \left| \left< \psi\right| x_\iota Q x_\iota \left| \psi \right> - D(\psi|\varphi) \right|
        \leq \left( \| x_\iota\psi \|_{|Q|^{1/2} }+\| \psi \|_{|Q|^{1/2}} \right) \| |Q|^{1/2}(1- x_\iota) \psi \|
        \rightarrow_\iota 0
    \end{equation}
The following relation can be derived using a cocycle relation along with the fact that each term is well defined and finite:
\begin{equation}
\left< \psi\right| x_\iota Q x_\iota \left| \psi \right>  = D(x_\iota \psi|\varphi) - D(x_\iota  \psi|\psi;\mathcal{A}')
\end{equation}
where the first term on the right is well defined and $> -\infty$ by the support condition on $\pi(x_\iota)$
and Lemma~\ref{lem:calculational} (2). The domination condition  $\omega_{x_\iota \psi }|_{\mathcal{A}'}
\leq \omega_\psi|_{\mathcal{A}'}$ gives $-\ln \Delta_{\eta,x_\iota \psi} \leq Q$
so that $D(x_\iota \psi|\varphi) \leq \left< \psi\right| x_\iota Q x_\iota \left| \psi \right> $ implying finite relative entropy. 
Unnormalized relative entropy is still bounded from below
as $D(x_\iota \psi|\psi; \mathcal{A}') \geq \| x_\iota \|^2 \ln \| x_\iota \psi \|^2 $ and together with the domination condition these
imply that $\lim_\iota D(x_\iota \psi|\psi; \mathcal{A}') = 0$.  Hence \eqref{eq:rellimit} follows.

\end{proof}

\begin{proof}[Proof of Lemma~\ref{lem:approx-sum}.]
We will build up the desired states over several steps. The construction uses two approximate identities from Lemma~\ref{lem:approxunit}, $x_\iota(V_2)$ for $\mathcal{A}_{V_2}$ and $x_\iota(V_1')$ for $\mathcal{A}_{V_1}'$, each with its own directed set. We index all nets in this proof by the product of these two directed sets (with the product order), regarding a net built from only one of the approximate identities as a net on the product that ignores the other coordinate; this does not change any limits. So a single index $\iota$ is used throughout, and all factors in the final states $\psi_\iota$ are evaluated at the same $\iota$.

Firstly consider a net of operators $x_\iota(V_2) \rightarrow 1$ as constructed in Lemma~\ref{lem:approxunit} but now for the algebra $\mathcal{A}_{V_2}$ and the appropriate weight,
that is $x_\iota(V_2) \in \mathfrak{M}_{V_2}$.

Our first net of states is designed to have well behaved relative entropy for $V_2$.
\begin{equation}
    \label{eq:seestep}
 \left|\xi_\iota\right> \equiv g_{\iota}(\ln \Delta_{\eta_{V_2}, x_\iota(V_2) \psi;V_2}) x_\iota(V_2) \left| \psi \right>
\end{equation}
where we choose the function $g_\iota$ as follows.
We start with a one parameter family of gaussian functions approaching $1$ as $\Lambda \rightarrow \infty$
\begin{equation}
g_\Lambda(t) = e^{ - t^2/\Lambda}
\end{equation}
Hence 
\begin{equation}
\lim_{\Lambda \rightarrow \infty} g_{\Lambda}(\ln \Delta_{\eta_{V_2}, x_\iota(V_2) \psi;V_2}) x_\iota(V_2) \left| \psi \right>= x_\iota(V_2) \left| \psi \right> 
\end{equation}
due to the dominated convergence theorem in the spectral representation of the relative
modular operator. So that for all $\iota$ we can choose $\Lambda_\iota \geq 1$ large enough that:
\begin{equation}
\| \left|\xi_\iota \right>- x_\iota(V_2) \left| \psi \right> \| \leq 1/ \varphi(x_\iota(V_2))
\end{equation}
for all $\Lambda \geq \Lambda_\iota$. We then choose $g_\iota = g_{\Lambda_\iota}$. 
We claim the $\xi_\iota$ states have the following properties:
\begin{itemize}
\item[(i)] $\lim_{\iota} \| \xi_\iota - \psi \| = 0$
\item[(ii)] $\| \sqrt{P} \xi_\iota\|$ is uniformly bounded along the net.
\item[(iii)] $\xi_\iota = \eta_{\varphi_{V_2'}}(A_\iota')$  for some $A_\iota' \in \mathfrak{N}_{V_2'}$. 
\end{itemize}
While (i) has been established by our choice of $\Lambda_\iota$, 
we will return towards the end of the proof to conditions (ii) and (iii). 
Note that condition (iii) implies that $\omega_{\xi_\iota}|_{\mathcal{A}_{V_2}}(\cdot) \leq
\| A_\iota' \|^2 \varphi_{V_2}(\cdot)$ which makes the relative entropy on $V_2$, $D( \xi_\iota|_{V_2}) <\infty$ (see \cite{Longo:2022lod}).

Our next step is to replace $A_\iota'$ with something suitably analytic under modular flow
for $V_2$. We consider:
\begin{equation}
(A_\iota')_\Sigma = \int ds \sqrt{\Sigma/\pi} e^{-s^2 \Sigma} \sigma_{\varphi';V_2'}^s (A_\iota') 
\end{equation}
As shown in \cite[eq.~(2.27) and the proof of Proposition~2.8]{araki2005extension} these operators have the property that $(A_\iota')_\Sigma \in \mathfrak{N}_{V_2'}$ and $so-\lim_{\Sigma \rightarrow \infty} (A_\iota')_\Sigma  = A_\iota'$
and $\lim_{\Sigma \rightarrow \infty} \eta_{V_2'}((A_\iota')_\Sigma) = \eta_{V_2'}(A_\iota') = \xi_\iota$. It is the latter property we need to maintain
and so for all $\iota$ we can choose $\Sigma_\iota \geq 1$ large enough that:
\begin{equation}
\| \eta_{V_2'}((A_\iota')_{\Sigma}) - \xi_\iota \| \leq 1/ \varphi_{V_2}(x_\iota(V_2)) \qquad \forall \Sigma \geq \Sigma_\iota
\end{equation}
Given this our second net of states will be $\zeta_\iota  \equiv \eta_{V_2'}((A_\iota')_{\Sigma_\iota})$ and these have the following properties:
\begin{itemize}
\item[(iv)] $\lim_{\iota} \| \zeta_\iota - \psi \| = 0$
\item[(v)] $\| \sqrt{P} \zeta_\iota\|$ is uniformly bounded along the net.
\item[(vi)] $\zeta_\iota = \eta_{V_2'}(B_\iota')$  for some $B_\iota' \in \mathfrak{N}_{V_2'}$
that is entire analytic for $z \rightarrow \sigma_{\varphi';V_2'}^z(B_\iota')$. 
\end{itemize}
Now (iv) and (vi) have been established, but we again leave the details of (v) to the end
of the proof. 

For our final states we need an approximate identity $x_\iota(V_1')$ as in Lemma~\ref{lem:approxunit} but now for $V_1'$. Combining with the previous approximate identity which commutes with $x_\iota(V_1')$, our final states are:
\begin{equation}
\psi_\iota = x_\iota(V_2)  x_\iota(V_1') \zeta_\iota
\end{equation} 
Combining (iv) and the approximate identity property of $x_\iota(V_2),  x_\iota(V_1')$
we get $\lim_{\iota} \| \psi_{\iota} - \psi \| = 0$. Lemma~\ref{lem:approx-sum} (1) follows from the support
condition on $x_\iota(V_2),  x_\iota(V_1')$ and the discussion around \eqref{eq:requiredfor}. 
Combining (v) with Lemma~\ref{lem:Pbound} we have an $\iota$ independent estimate for $\| \sqrt{P}\psi_\iota \|$. Hence we have established Lemma~\ref{lem:approx-sum} (3), the uniform $P$-expectation estimate. 

We are left with establishing Lemma~\ref{lem:approx-sum} (2); that is finiteness conditions on the relative entropy
for $V_2$ and $V_1'$ (the domain conditions of (1) establish both these relative entropies
are $> -\infty$). 

For (2) we can estimate the relative entropy as follows. 
Define:
\begin{equation}
\chi_\iota = \eta_{V_2'}( x_{\iota}(V_1') B_\iota')
\end{equation}
where $B_\iota'$ was defined in (vi) and in fact $x_{\iota}(V_1') B_\iota' \in\mathfrak{N}_{V_2'}$. 
Note that $\psi_\iota = x_\iota(V_2) \chi_\iota$ so that the induced states
satisfy $\omega_{\psi_{\iota}}|_{V_2'} \leq \| x_\iota(V_2) \|^2
\omega_{\chi_{\iota}}|_{V_2'}$ giving the bound
on the relative modular operators:
\begin{equation}
\Delta_{\varphi_{V_2}, \psi_\iota;V_2}^{-1}  = \Delta_{\varphi_{V_2}, x_\iota(V_2) \chi_\iota;V_2}^{-1} \leq \Delta_{\varphi_{V_2}, \chi_\iota;V_2}^{-1} \| x_\iota(V_2) \|^{2}
\leq \Delta_{\varphi_{V_2}, \chi_\iota;V_2}^{-1}  
\end{equation}
and using operator monotonicity of the $\ln$ we can turn this into a bound on the relative entropy:
\begin{equation}
    \label{eq:rhsinf}
-\infty < D(\psi_\iota | \varphi_{V_2} ; V_2) \leq -\left< \chi_\iota \right| x_\iota(V_2) \ln \Delta_{\varphi_{V_2},\chi_\iota;V_2} x_\iota(V_2) \left| \chi_\iota \right>
\end{equation}
implying that the right hand side is well defined through the spectral representation in
the same way that we defined relative entropy with respect to a weight in \eqref{eq:defplusminus} (in particular there are no $\infty -  \infty$ ambiguities that might occur when the spectral representation is on the entire real line). The right hand side of \eqref{eq:rhsinf} could still possibly be infinite. We rule that out now.
Note that:
\begin{equation}
F(z) = \left< \chi_\iota \right| \Delta_{\varphi_{V_2},\chi_\iota;V_2}^{z}  x_\iota(V_2)^2 \left| \chi_\iota \right>
\end{equation}
has an analytic continuation in the neighborhood of $z=0$: for positive ${\rm Re} z \geq 0$
we act to the right with the relative modular operator and use
the fact that $x_\iota(V_2)^2 \in \mathfrak{N}_{V_2}$ and for negative ${\rm Re}z \leq 0$
we use the bound on the linear functional:
\begin{equation}
    \label{eq:tolev}
\omega_{\chi_\iota}(\cdot)|_{V_2} \leq \| x_\iota(V_1') B_\iota' \|^2 \varphi_{V_2}(\cdot)|_{V_2} 
\end{equation}
which implies that \cite{takesaki2003theory,Longo:2022lod}:
\begin{equation}
    \Delta_{\chi_\iota;V_2}   \leq \| x_\iota(V_1') B_\iota' \|^2   \Delta_{\varphi_{V_2},\chi_\iota;V_2}   \quad \implies \quad \| \Delta_{\varphi_{V_2},\chi_\iota;V_2}^{-1/2} \chi_\iota  \| < \infty
\end{equation}
from operator monotonicity of $t \rightarrow -1/t$.
Thus $F(z)$ actually has an analytic continuation into the strip $-1/2 \leq {\rm Re }z \leq 1/2$.
Combining this with \eqref{eq:prem} applied to $x_\iota(V_2)$, we have:
\begin{align} \nonumber
\frac{1}{\theta}\left< \chi_\iota \right| x_\iota(V_2) (1- \Delta_{\varphi_{V_2},\chi_\iota;V_2}^\theta ) x_\iota(V_2) \left| \chi_\iota \right> & =
 \frac{1}{\theta} \left< \chi_\iota \right|  (1- \Delta_{\varphi_{V_2},\chi_\iota;V_2}^\theta)  x_\iota(V_2)^2 \left| \chi_\iota \right>  \\
&  + \frac{1}{\theta} \left< \chi_\iota \right| [ \Delta_{\varphi_{V_2},\chi_\iota;V_2}^\theta,x_\iota(V_2)] x_\iota(V_2) \left| \chi_\iota \right>  \label{eq:lhsfin}
\end{align}
so that both terms on the right have finite limits $\theta \rightarrow 0^+$ and so the left hand side of \eqref{eq:lhsfin} has a finite
limit and that limit must give the right hand side of \eqref{eq:rhsinf}:
\begin{equation}
D(\psi_\iota | \varphi_{V_2} ; V_2) \leq -\left< \chi_\iota \right| x_\iota(V_2) \ln \Delta_{\varphi_{V_2},\chi_\iota;V_2} x_\iota(V_2) \left| \chi_\iota \right> < \infty
\end{equation}
by standard methods (essentially the same method as the proof of Lemma~\ref{lem:calculational}(2)). 

To prove that the $V_1'$ relative entropy is finite 
we use the analyticity property of $B_\iota' \in \mathcal{A}_{V_2}'$ allowing us to write:
\begin{equation}
\eta_{V_2'}(B'_\iota) =  \Delta_{V_2'}^{-1/2}\Delta_{V_2'}^{1/2} \eta_{V_2'}(B'_\iota)
= \eta_{V_2}(A_\iota) \qquad A_\iota = \sigma^{i/2}_{\varphi_{V_2};V_2}( J_{V_2} (B'_\iota)^\dagger J_{V_2})
\end{equation}
where the second equality uses $(B_\iota')^\dagger \in \mathfrak{N}_{V_2'}$. To see this, let $e_\kappa'$ be an approximate identity for $\mathfrak{M}_{V_2'}$ as in Lemma~\ref{lem:approxid} and set $B_\kappa' = (e_\kappa' A_\iota')_{\Sigma_\iota}$. Since $\mathfrak{N}_{V_2'}$ is a left ideal, both $e_\kappa' A_\iota'$ and its adjoint $(A_\iota')^\dagger e_\kappa'$ lie in $\mathfrak{N}_{V_2'}$, so $B_\kappa' \in \mathfrak{N}_{V_2'} \cap \mathfrak{N}_{V_2'}^\dagger$ and the Tomita relation $J_{V_2} \Delta_{V_2'}^{1/2} \eta_{V_2'}(B_\kappa') = \eta_{V_2'}((B_\kappa')^\dagger)$ gives
\begin{align} \nonumber
\varphi_{V_2'}(B_\kappa' (B_\kappa')^\dagger) = \big\| \Delta_{V_2'}^{1/2}\, e^{-(\ln \Delta_{V_2'})^2/4\Sigma_\iota}\, e_\kappa' \, \eta_{V_2'}(A_\iota') \big\|^2 \leq e^{\Sigma_\iota/2} \, \| \eta_{V_2'}(A_\iota') \|^2
\end{align}
uniformly in $\kappa$. Since $B_\kappa' (B_\kappa')^\dagger \rightarrow B_\iota' (B_\iota')^\dagger$ weakly, lower semicontinuity of $\varphi_{V_2'}$ gives $\varphi_{V_2'}(B_\iota' (B_\iota')^\dagger) < \infty$. So $A_\iota \in \mathfrak{N}_{V_2} $
and thus:
\begin{equation}
\psi_\iota = x_\iota(V_1') \chi_\iota' \qquad \chi_\iota' \equiv \eta_{V_2}( x_\iota(V_2) A_\iota)
\end{equation}
So we can use an essentially identical argument for the commutant relative entropy, now mostly leveraging properties of $x_\iota(V_1') \in \mathfrak{N}_{V_1'}$ and replacing
the bound \eqref{eq:tolev} with:
\begin{equation}
\omega_{\chi'_\iota}(\cdot)|_{V_1'} \leq \| x_\iota(V_2) A_\iota \|^2 \varphi_{V_2'}(\cdot|_{V_1'}) 
\leq \| x_\iota(V_2) A_\iota \|^2 \varphi_{V_1'}(\cdot) 
\end{equation}
where we used \eqref{eq:weused} (for the commutant algebras).
Thus $-\infty < D(\psi_\iota | \varphi_{V_1'} ; V_1') < \infty$. Thus we have established property (2) and we are done.

We of course still need to prove (ii),(iii) and (v) above. 
We start with:

(iii) We can write the $\xi_\iota$ states as follows:
\begin{equation}
\xi_\iota = \int d s \tilde{g}_\iota(s)  \Delta_{\varphi_{V_2},x_\iota(V_2) \psi;V_2}^{is} x_{\iota}(V_2) \psi
= \int d s \tilde{g}_\iota(s-i/2)  \Delta_{\varphi_{V_2},x_\iota(V_2) \psi;V_2}^{is+1/2} x_{\iota}(V_2) \psi
\end{equation}
where $\tilde{g}_\iota(s) =\sqrt{\Lambda_\iota/4\pi} \exp( - s^2 \Lambda_\iota/4)$ is the Fourier transform of $g_\iota$.
In the second equality
we shifted the integration contour - this is justified because of the domain condition on $ \Delta_{\varphi_{V_2},x_\iota(V_2) \psi;V_2}^{1/2} x_\iota(V_2) \psi$ which is satisfied because the support of $x_\iota(V_2)$ is in the definition
domain of $\varphi_{V_2}$ (see \eqref{eq:requiredfor}). Then:
\begin{align}
\xi_\iota 
&= \int d s \tilde{g}_\iota(s-i/2)  \Delta_{\varphi_{V_2},x_\iota(V_2) \psi;V_2}^{is} (\theta')^\dagger  J_{V_2}  \eta( \pi(x_\iota(V_2)) ) 
 \\ &=  \int d s \tilde{g}_\iota(s-i/2)  \Delta_{\varphi_{V_2},x_\iota(V_2) \psi;V_2}^{is} (\theta')^\dagger  \Delta_{\varphi_{V_2};V_2}^{-is} \eta'(  \sigma_{\varphi_{V_2'};V_2'}^s(J_{V_2}\pi(x_\iota(V_2))J_{V_2}) ) 
  \\ &  =  \int d s \tilde{g}_\iota(s-i/2)  \eta'(  \Delta_{\varphi_{V_2},x_\iota(V_2) \psi;V_2}^{is} (\theta')^\dagger \Delta_{\varphi_{V_2};V_2}^{-is}  )  =  \eta'(A_\iota')
\end{align}
where the partial isometry $\theta' \in \mathcal{A}_{V_2}'$ takes $x_{\iota}(V_2) \psi$ to the canonical cone $\mathscr{P}^\natural(V_2)$ and in particular satisfies
$\theta' (\theta')^\dagger \leq J_{V_2} \pi(x_\iota(V_2))J_{V_2}$ since the $\mathcal{A}_{V_2}$
support of $x_\iota(V_2)\psi$ 
is less than or equal to the support of $x_\iota(V_2)$, see also
\eqref{eq:thetasupport}. Going from the second line to the third 
we can pull the cocycle into $\eta'$ and we used the aforementioned property
of $(\theta')^\dagger$ to absorb $J_{V_2} \pi(x_\iota(V_2))J_{V_2}$. The last equality on the last line pulls the integral
inside $\eta'$ and defines:
\begin{equation}
A_\iota' = \int d s \tilde{g}_\iota(s - i/2)  \Delta_{\varphi_{V_2},x_\iota(V_2) \psi;V_2}^{is} (\theta')^\dagger \Delta_{\varphi_{V_2};V_2}^{-is} 
\end{equation}
and this is justified by a standard estimate that establishes $A_\iota' \in \mathfrak{N}_{V_2'}$.  
This is the final form of $A_\iota'$ required in (iii).

(ii)  We need an estimate for:
\begin{align}
\| \sqrt{P} \xi_\iota\| = &\| \sqrt{P} g_\iota(\ln \Delta_{\varphi_{V_2}, x_\iota(V_2) \psi;V_2}) x_\iota(V_2) \left| \psi \right> \|
\leq  \int d s |\tilde{g}_\iota(s)| \| \sqrt{P} u_{-s} x_\iota(V_2) \left| \psi \right> \| \\
 &  = \int d s |\tilde{g}_\iota(s)| \left( e^{2\pi s} \Theta( x_\iota(V_2) \psi|_{V_2} ) + \Theta( x_\iota(V_2) \psi|_{V_2'} ) \right)^{1/2} \\
 & \leq \| \sqrt{P} x_\iota(V_2) \left| \psi \right> \|  \int d s |\tilde{g}_\iota(s)| (1+ e^{2\pi s})^{1/2} \label{eq:integralin}
\end{align} 
where
\begin{align} \nonumber
u_s = \Delta_{\varphi_{V_2};V_2}^{-is} \Delta_{x_\iota(V_2) \psi,\varphi_{V_2};V_2}^{is} \in \mathcal{A}_{V_2}\,, \qquad
\Delta_{\varphi_{V_2}, x_\iota(V_2) \psi;V_2}^{is}\, x_\iota(V_2) \psi = u_{-s}\, x_\iota(V_2) \psi
\end{align}
(the cocycle of Section~\ref{sec:states})
and we used the weight version of Lemma~\ref{lem:duality}.
The $s$ integral in \eqref{eq:integralin} is uniformly bounded for $\Lambda_\iota \geq 1$ and converges to $\sqrt{2}$ as $\Lambda_\iota \rightarrow \infty$. We also have Lemma~\ref{lem:Pbound} which concludes
the uniform estimate of $\| \sqrt{P} \xi_\iota\| $.  

(v) We have:
\begin{align}
\| \sqrt{P} \zeta_\iota  \| &= \| \int d s \sqrt{\Sigma_\iota/\pi} e^{-s^2 \Sigma_\iota} \sqrt{P} \Delta_{V_2'}^{is} \eta_{V_2'}(A_\iota') \| \\
&\leq \|\sqrt{P} \xi_\iota\| \int d s \sqrt{\Sigma_\iota/\pi} e^{-s^2 \Sigma_\iota} e^{\pi s}  = \|\sqrt{P} \xi_\iota\| e^{\pi^2/(4\Sigma_\iota)}
\end{align}
which can be estimated in an $\iota$ independent manner using (ii). 

\end{proof}

\subsection{Approximations for relative entropy differences}
\label{app:entropy-differences}

In this Appendix we prove an approximation result for relative entropy
differences with respect to a weight. This allows us to remove any assumptions about
finite $\Theta$ when proving convexity of $V \rightarrow D(\psi|_V)$.

We use the
HST structure for weights, Definition~\ref{def:tr-weight}, with positive translation generator $P \in \mathcal{TR}(\mathcal{A},\varphi_0)$, algebras $\mathcal{A}_V$, weights
$\varphi_V$, and semicyclic representations $\eta_V$. Our approach uses a combination of Lemma~\ref{lem:partway} from the state case,
and Appendix~\ref{app:approxsumrule} although here we do not track/constrain relative entropies on the commutant algebras. 

\begin{lemma}
\label{lem:weight-entropy-approximation}
Consider some $\psi \in \mathscr{H}$ with two cuts $V_1<V_2$, such that the relative
entropies $D(\psi|_{V_1})$ and $D(\psi|_{V_2})$ are well defined and finite.
Then there exists a net of states $\psi_\iota \in \mathcal{D}(\sqrt{P})$
and $\lim_\iota \|\psi_\iota-\psi \|=0$ 
such that the two relative entropies of each $\psi_\iota$ are well defined and finite and satisfy:
\begin{align}
\label{eq:weight-entropy-approximation}
 \lim_\iota D(\psi_\iota|_{V_1})=D(\psi|_{V_1}),\qquad
 \lim_\iota D(\psi_\iota|_{V_2})=D(\psi|_{V_2})
\end{align}
\end{lemma}

We need some preliminary results. 
Firstly we need to maintain semicontinuity which is usually absent for weights, but we can
retain it in the following situation:
\begin{lemma}\label{lem:support-lsc}
Suppose that $\psi_\iota\to\psi$ in $\mathscr{H}$ and that there is a fixed
projection $\pi\in\mathcal{A}$ with $\varphi(\pi)<\infty$ and
$\pi\psi_\iota=\psi_\iota$. Then the relative entropies are well defined, $>-\infty$, and:
\begin{equation}\label{eq:finite-support-entropy-lsc}
 D(\psi|\varphi)\leq\liminf_\iota D(\psi_\iota|\varphi).
\end{equation}
The value $+\infty$ is allowed.
\end{lemma}

\begin{proof}
Note that the limit gives $\pi\psi=\psi$. The support criterion
\eqref{eq:requiredfor} gives
\begin{equation}\label{eq:finite-support-half-power}
 \norm{\Delta_{\eta,\psi_\iota}^{1/2}\psi_\iota}^{2}
 =\varphi(\pi(\psi_\iota))\leq\varphi(\pi).
\end{equation}
Consequently, for $\Lambda \geq e$, the negative entropy tail satisfies
\begin{equation}\label{eq:uniformlarge}
 \begin{split}
 &\left<\psi_\iota\right|
   (\ln\Delta_{\eta,\psi_\iota})
   E_{[\Lambda,\infty)}(\Delta_{\eta,\psi_\iota})
  \left| \psi_\iota\right>\\
 &\hspace{25mm}\leq  \varphi(\pi) (\ln \Lambda)/ \Lambda .
 \end{split}
\end{equation}
Here we used $\ln\lambda\leq \lambda (\ln \Lambda)/\Lambda $ for $\lambda\geq \Lambda$.
Strong resolvent continuity, Lemma~\ref{lem:rescont},
gives convergence of the bounded truncations in the spectral representation \eqref{eq:specpm} of $D^-(\psi_\iota|\varphi)$.
The uniform bound \eqref{eq:uniformlarge} then proves
convergence of $D^-(\psi_\iota|\varphi)$. Lower semicontinuity of
$D^+(\psi_\iota|\varphi)$, established in Lemma~\ref{lem:calculational} (5), proves
\eqref{eq:finite-support-entropy-lsc}.
\end{proof}

We next give the weight version of Lemma~\ref{lem:Lp} pertaining to $L_p$ spaces. 
We are following the Araki-Masuda \cite{araki1982positive} approach, which was generalized to weights in \cite{masuda1983lp} but not in 
the form that we will work with here. 
Indeed we do not need the full machinery of Araki-Masuda, so it will suffice to introduce the natural
generalization to weights of the original definitions in \cite{araki1982positive}, and prove a limited number of results using this definition.  
Consider $2\leq p\leq\infty$ and define:
\begin{equation}
L_p(\mathcal{A}')^{\rm pre}= \left\{ \psi \in \bigcap_{\chi \in \mathscr{H}}\mathcal{D}( \Delta_{\eta,\chi;\mathcal{A}}^{1/p-1/2}), \| \psi \|_{L_p(\mathcal{A}')} < \infty \right\}
\end{equation}
which is a subspace of $\mathscr{H}$, we used:
\begin{equation}\label{eq:weighted-Lp-definition}
 \|\psi\|_{L_p(\mathcal{A}')}
 =\sup_{\chi\in\mathscr{H}:\,\norm{\chi}\leq1}
   \|\Delta_{\eta,\chi;\mathcal{A}}^{1/p-1/2}\psi\|.
\end{equation}
where here $\eta$ is the semicyclic map for $\varphi$. To define the actual $L_p(\mathcal{A}')$ spaces for $2 \leq p < \infty$ we must
pass to a completion. As opposed to the state case this procedure
will result in an extension from $\mathscr{H}$ for $p>2$\footnote{In the state case a bound can be derived  $\| \psi \|_{L_p}\geq \| \psi \| \| \eta \|^{2/p-1}$, allowing for an unnormalized vector $\eta$, which then clearly does not give a constraint in the unbounded case. For this reason we also no longer have monotonicity of the norms in $p$. Monotonicity usually arises from a combination of convexity of $1/2-1/p \rightarrow \| \psi \|_{L_p(\mathcal{A}')}$
and this aforementioned bound.}. 
Here we will only be interested in $\psi \in \mathscr{H} \cap L_p(\mathcal{A}') = L_p(\mathcal{A}')^{\rm pre}$ (where equality here uses closedness of the relative modular operators) so the $L_p$ norm can be written as above.  
As per this definition it is clear that $L_2(\mathcal{A}') = \mathscr{H}$ and slightly less obvious is 
$\eta'(\mathfrak{N}_{\varphi'}) = L_\infty(\mathcal{A}') \cap \mathscr{H}$:

\begin{lemma}
As per the above definitions $L_\infty(\mathcal{A}') \cap \mathscr{H} = \eta'(\mathfrak{N}_{\varphi'})$. 
\end{lemma}
\begin{proof}
We clearly have $\supset$. Given some $\psi \in \mathscr{H} \cap L_\infty(\mathcal{A}')$
then define the following operator:
\begin{equation}
    \label{eq:Tdef}
T \chi = \theta' \Delta_{\eta,\xi_\chi}^{-1/2} \psi \qquad \chi \in \mathscr{H}
\end{equation}
This is linear because for all $n' \in \mathfrak{N}_{\varphi'}$
\begin{equation} 
\theta' \Delta_{\eta, \xi_\chi}^{-1/2} \eta'(n') = j((n')^\dagger)(\chi_1+\chi_2) = \left(\theta_1' \Delta_{\eta,\xi_{\chi_1}}^{-1/2}  + \theta_2' \Delta_{\eta,\xi_{\chi_2}}^{-1/2}\right) \eta'(n') 
\end{equation}
where $\chi_1+\chi_2 = \chi$ with corresponding polar decompositions $\theta_1' \xi_{\chi_1} + \theta_2' \xi_{\chi_2}
 = \theta' \xi_{\chi}$. Extending this relation to $\eta'(n')  \rightarrow \psi$ is possible by using the domain
 condition on $\psi$ to construct a sequence in the common core $\eta'(\mathfrak{N}_{\varphi'})$ of all three relative modular operators and which converges $\eta'(n'_i) \rightarrow \psi$ in the graph norm of $\Delta_{\eta,\rho'}^{-1/2}$
 where $\rho' = \omega_{\xi_\chi}|_{\mathcal{A}'} + \omega_{\xi_{\chi_1}}|_{\mathcal{A}'} + \omega_{\xi_{\chi_2}}|_{\mathcal{A}'}$, and hence converges
 in the graph norm of the individual operators.  Hence $T(\chi_1+\chi_2) = T\chi_1+T\chi_2$. The norm of this operator is bounded:
 \begin{equation}
 \| T \chi \| \leq \| \psi \|_{L_{\infty}(\mathcal{A}')} \| \chi \| 
 \end{equation}
 Clearly $[T,u']=0$ for all $u' \in \mathcal{U}(\mathcal{A}')$ and hence $T \in \mathcal{A}$.
 For $\chi = \xi_{J \psi}$  we can invert \eqref{eq:Tdef} to find
 that $T \xi_{J \psi} \in \mathcal{D}(\Delta_{\eta,\xi_{J\psi}}^{1/2})$ implying that
 $\varphi(T j(\pi_{\psi}(\mathcal{A'})) T^\dagger) = \| \Delta_{\eta,\xi_{J\psi}}^{1/2} T \xi_{J\psi} \|^2$ so that 
 $Tj(\pi_{\psi}(\mathcal{A}')) \in \mathfrak{N}_\varphi^\dagger$
 and:
 \begin{equation}
 \psi = \Delta_{\eta,\xi_{J\psi}}^{1/2}  T \xi_{J\psi} = \eta'(\pi_{\psi}(\mathcal{A}') j(T^\dagger) )\in \eta'(\mathfrak{N}_{\varphi'})
 \end{equation}
Hence $\mathscr{H} \cap L_\infty(\mathcal{A}') = \eta'(\mathfrak{N}_{\varphi'})$. 
\end{proof}

The only result we need for these $L_p$ spaces for weights is the following combination
of the contractive and convexity properties proven in the state case in Lemma~\ref{lem:Lp}. While they 
are most likely also true individually for the weight versions, we only need a combination of the two bounds which has
a much shorter proof:
\begin{lemma}\label{lem:weighted-Lp-contraction}
Suppose that $\psi\in L_\infty(\mathcal{A}')\cap\mathscr{H}$, then
\begin{equation}\label{eq:weighted-sector-contraction}
 \|e^{-i \epsilon e^{-i\theta}P}\psi \|_{L_p(\mathcal{A}')}
 \leq \| \psi \|^{1-2/p}_{L_\infty(\mathcal{A}')} \| \psi \|^{2/p}\,,
 \qquad \epsilon>0,\qquad 0\leq\theta\leq\frac{2\pi}{p}.
\end{equation}
\end{lemma}
\begin{proof} 
We first check the $p=\infty$ endpoint. By
the commutant version of \eqref{eq:boundtranslate} and for $a\geq0$,
\begin{equation}
 \varphi'(U(a)(\cdot)|_{\mathcal{A}'}U(-a))\leq\varphi'(\cdot)
\end{equation}
hence $U(a) \mathfrak{N}_{\varphi'} U(-a) \subset \mathfrak{N}_{\varphi'}$. 
Suppose that $\psi = \eta'(n')$, for $n' \in \mathfrak{N}_{\varphi'}$, then $U(a) \psi = \eta'(U(a) n' U(-a)) \in L_\infty(\mathcal{A}') \cap \mathscr{H}$ so that
$\| U(a) \psi \|_{L_\infty(\mathcal{A}')} = \| n' \| = \| \psi \|_{L_\infty(\mathcal{A}')}$.
That is $U(a)$ acts as an isometry in  $L_\infty$ norm. 

Fix $0\leq\beta\leq2\pi$ and consider the analytic family
\begin{equation}
 T_z=U(\epsilon e^{-i \beta z}) = e^{ - i \epsilon e^{-i \beta z}P},\qquad 0\leq{\rm Re} z\leq 1/2.
\end{equation}
On the edge $z= i s$ its argument is positive real, and hence
$T_{i s}$ is an $L_\infty$ contraction. On the other edge ${\rm Re} z= 1/2$, and throughout
the strip, the argument of $U(a)$ is in the lower half-plane, so that by positivity of $P$, 
$\|T_z\|\leq 1$ as an operator on $\mathscr{H}$.
A three line estimate gives: 
\begin{align}
   \left. \|\Delta_{\eta,\chi;\mathcal{A}}^{z-1/2} T_z \psi\| \right|_{z=1/p}
   &\leq \sup_s \|\Delta_{\eta,\chi;\mathcal{A}}^{-1/2} T_{is} \psi\|^{1-2/p}  \sup_s \| T_{1/2+is} \psi\|^{2/p}  \\
   &\leq \| \psi \|^{1-2/p}_{L_\infty(\mathcal{A}')} \| \psi \|^{2/p}
\end{align}
after taking the supremum over $\chi$ we get:
\begin{equation}\label{eq:weighted-interpolation-conclusion}
 \|T_{1/p}\psi \|_{L_p(\mathcal{A}')}
 \leq \| \psi \|^{1-2/p}_{L_\infty(\mathcal{A}')} \| \psi \|^{2/p}
\end{equation}
Putting $\beta = p\theta$ proves
\eqref{eq:weighted-sector-contraction}. 
\end{proof}

We can put all these ingredients together to prove the original goal of this Appendix:

\begin{proof}[Proof of Lemma~\ref{lem:weight-entropy-approximation}]
Anticipating some future constraints on building an approximation, we start with $y_\iota \in \mathfrak{N}_{V_2}$ as given by
Lemma~\ref{lem:approxunit0}, and we pick functions $f,g$ as in Lemma~\ref{lem:f} with $f,g \leq 1$,
$f(1)=g(1)=1$ and such that $gf = f$. Set
$x_\iota =f (y_\iota)$ and $z_\iota = g(y_\iota)$ such that $z_\iota x_\iota = x_\iota$. 
These have the same properties as $x_\iota$ in Lemma~\ref{lem:approxunit}.

We start with $\psi_\iota = x_\iota \psi$. Since $[x_\iota, \ln \Delta_{\eta,\psi;V_2}]= [x_\iota, \ln \Delta_{V_2} ]$
we have the required uniform bound to apply Lemma~\ref{lem:convrel},
$\lim_\iota D(\psi_\iota|_{V_2}) = D(\psi|_{V_2})$. Also we can take the $s$ derivative of the following equation:
\begin{equation}
[\Delta_{V_1}^{is} \Delta_{V_2}^{-is},x_\iota] = [ e^{ -i (V_2-V_1) P(1-e^{-2\pi s})}, x_\iota]
\end{equation}
gives:
\begin{equation}
[\ln \Delta_{V_1}, x_\iota]- [\ln \Delta_{V_2}, x_\iota] = -2\pi (V_2-V_1)[P,x_\iota]
\end{equation}
so that $\sup_\iota \| [\ln \Delta_{V_1}, x_\iota]\| < \infty$ and $\varphi_{V_1}(\pi(x_\iota))
= \varphi_{V_2}(\pi(x_\iota)) < \infty$ so we can apply Lemma~\ref{lem:convrel} also to $V_1$
so that $\lim_\iota D(\psi_\iota|_{V_1}) = D(\psi|_{V_1})$. 

This approximation is however not forced to have finite $P$ expectation. Instead to do this we fix $\iota$
and introduce two new parameters worth of smearings. We reveal the final form:
\begin{equation}
\psi_{\iota,k,\epsilon} = z_\iota e^{ - \epsilon P} \chi_{\iota,k}\,, \quad
\chi_{\iota,k} = g_k( \ln \Delta_{\eta_{V_1}, x_\iota \psi; V_1}) x_\iota \psi
\end{equation}
where $g_k(t) = \exp(-t^2/k)$ are gaussians that become flat as $k \rightarrow \infty$. In order
of appearance from right to left, $g_k$ is designed to place $x_\iota \psi$
into $L_\infty(\mathcal{A}') \cap \mathscr{H}$ such that when we hit it
with $e^{-\epsilon P}$ we can constrain its $L_p$ norms (similar to Lemma~\ref{lem:Lp}) which will be used in an estimate for the spectral
representation of a relative modular operator.
 Furthermore $g_k$ gives a continuous approximation to relative entropy as $k \rightarrow \infty$. 
The $P$ factor gives finite $P$ expectation while
the $z_\iota$ puts back the finite support condition so that the
relative entropy with respect to the weight is bounded from below. It is this $z_\iota$ factor 
that prevents us from running the simpler argument in the proof of Theorem~\ref{thm:fullapprox} for the state case. 
Note that the good behavior of $z_\iota$ under $P$ commutators implies that $\psi_{\iota,k,\epsilon} \in \mathcal{D}(\sqrt{P})$ via Lemma~\ref{lem:Pbound} applied to $e^{ - \epsilon P} \chi_{\iota,k}$ and $z_\iota$.

We claim that the relative entropy of $\psi_{\iota,k,\epsilon}$ is well defined and finite
and:
\begin{equation}
    \label{eq:remainsestablish}
\limsup_{\epsilon \rightarrow 0} D(\psi_{\iota,k,\epsilon}|_{V_1}) 
\leq - \left< \chi_{\iota,k} \right| z_\iota \ln \Delta_{\eta_{V_1}, x_\iota \psi; V_1} z_\iota \left| \chi_{\iota,k} \right>
\end{equation}
we will establish this later. For now we see the consequence. Firstly note that the right hand side is
indeed well defined and finite: set $Q = -\ln \Delta_{\eta_{V_1}, x_\iota \psi; V_1}$ the gaussian
implies that $g_k(Q) x_\iota \psi \in \mathcal{D}(Q)$ and also since $[z_\iota,Q] = -[z_\iota, \ln \Delta_{V_1}]$ is a bounded operator we can apply Lemma~\ref{lem:Qgraph} to see that $z_\iota \chi_{\iota,k} $ is in the form domain of $|Q|^{1/2}$ and hence the claim. Furthermore we can use the bound in that Lemma 
to establish the following limit:
\begin{align}
\| |Q|^{1/2} (z_\iota \chi_{\iota,k} - x_\iota \psi) \| &= \| |Q|^{1/2} z_\iota (\chi_{\iota,k} - x_\iota \psi) \| \\
&\leq C(   \| \chi_{\iota,k} - x_\iota \psi \|^2 + \| |Q|^{1/2} ( \chi_{\iota,k} - x_\iota \psi) \|^2 )^{1/2} \\
& = C(   \| (g_k(Q) - 1) x_\iota \psi \|^2 + \| |Q|^{1/2} ((g_k(Q) - 1)x_\iota \psi) \|^2 )^{1/2}
\end{align}
where $C = \sup_\iota (2 +\| [z_\iota, Q] \|)^{1/2}$ is finite. We utilized $z_\iota x_\iota = x_\iota$ here.  The right hand side vanishes as $k \rightarrow \infty$ by
dominated convergence using the fact that $\| |Q|^{1/2}x_\iota \psi \| < \infty$ because the $x_\iota \psi$ relative entropy is well defined and finite as established in Lemma~\ref{lem:convrel}. Applying Cauchy-Schwarz gives:
\begin{equation}
\lim_{k \rightarrow \infty}  \left< \chi_{\iota,k} \right| z_\iota Q z_\iota \left| \chi_{\iota,k} \right>  
=  \left< \psi \right| x_\iota Q x_\iota \left| \psi \right> = D(x_\iota \psi|_{V_1})
\end{equation}
Hence it is always possible to find some $k_\epsilon$ (varying slowly enough with $\epsilon$) such that
\begin{equation}
\limsup_{\epsilon \rightarrow 0} D(\psi_{\iota,k_\epsilon,\epsilon}|_{V_1}) 
\leq  D(x_\iota \psi|_{V_1})
\end{equation}
But at fixed $\iota$ the uniform support condition $\varphi_{V_1}(\pi(z_\iota)) \leq \varphi_{V_2}(\pi(z_\iota)) < \infty$
implies that we can apply lower semicontinuity via Lemma~\ref{lem:support-lsc} 
along with the fact that $\lim_\epsilon \| \psi_{\iota,k_\epsilon,\epsilon} - x_\iota \psi \| =0$
to show the other bound $\liminf_{\epsilon \rightarrow 0} D(\psi_{\iota,k_\epsilon,\epsilon}|_{V_1}) 
\geq D(x_\iota \psi|_{V_1})$ and hence:
\begin{equation}
\lim_{\epsilon \rightarrow 0} D(\psi_{\iota,k_\epsilon,\epsilon}|_{V_1}) 
= D(x_\iota \psi|_{V_1})
\end{equation}
We can also apply lower semicontinuity to the $V_2$ relative entropy (again due to the support condition $\varphi_{V_2}(\pi(z_\iota)) < \infty$ and Lemma~\ref{lem:support-lsc}) and also to the relative entropy difference established in
Lemma~\ref{lem:weight-semi}, giving:
\begin{align}
\limsup_{\epsilon \rightarrow 0} \left( D(\psi_{\iota,k_\epsilon,\epsilon}|_{V_1})   -
D(\psi_{\iota,k_\epsilon,\epsilon}|_{V_2} ) \right) 
& \leq D(x_\iota \psi|_{V_1}) - D(x_\iota \psi|_{V_2})  \\
&\leq \liminf_{\epsilon \rightarrow 0} \left( D(\psi_{\iota,k_\epsilon,\epsilon}|_{V_1})   -
D(\psi_{\iota,k_\epsilon,\epsilon}|_{V_2} ) \right)  
\end{align}
and hence:
\begin{equation}
\lim_{\epsilon \rightarrow 0} D(\psi_{\iota,k_\epsilon,\epsilon}|_{V_2}) 
= D(x_\iota \psi|_{V_2})
\end{equation}
Convergence of the $D(x_\iota \psi|_{V_2,V_1})$ relative entropies along the net allows us to find
an $\epsilon_\iota < 1/ \varphi(x_\iota)$ which converges to zero (assuming unbounded $\varphi$)
such that $\psi_{\iota, k_{\epsilon_\iota},\epsilon_\iota}$ satisfies the properties
in the statement of this Lemma.

It remains to establish \eqref{eq:remainsestablish} and the well definedness and finiteness of the left hand side.
The support condition on $z_\iota$ establishes that $D(\psi_{\iota,\epsilon,k}|_{V_1}) > -\infty$ is well defined. 
Stripping off the $z_\iota$ factor we set $\zeta_\epsilon = e^{-P \epsilon} g_k(Q) x_\iota \psi
= e^{-P \epsilon} \chi_{\iota,k}$
so that $\psi_{\iota,k,\epsilon} = z_\iota \zeta_\epsilon$.
Using $\omega_{z_\iota \zeta_\epsilon}|_{V_1'} \leq \omega_{\zeta_\epsilon}|_{V_1'}$ on the relative
modular operator we get an upper bound:
\begin{equation}
    \label{eq:estabexist}
D(\psi_{\iota,k,\epsilon}|_{V_1}) \leq  \left< \zeta_\epsilon \right| z_\iota Q_\epsilon z_\iota \left| \zeta_\epsilon \right> 
\end{equation}
where 
$Q_\epsilon = -\ln \Delta_{\eta_{V_1}, \zeta_\epsilon; V_1}$. The right hand side can be defined through the spectral representation of $Q_\epsilon$ and the bound applies (and is useful) if we can show that the integral is well defined (and finite):
\begin{equation}
\label{eq:spectral-integral}
  \left< \zeta_\epsilon \right| z_\iota Q_\epsilon z_\iota \left| \zeta_\epsilon \right>  = \int_{-\infty}^{\infty} \lambda \left< \zeta_\epsilon \right| z_\iota d E_\lambda(Q_\epsilon) z_\iota \left| \zeta_\epsilon \right>
\end{equation}
we will show that this integral is well defined by finding bounds at large $| \lambda |$. 
These bounds will be uniform in $\epsilon$, which will also help us take the limit. 

The negative spectral tail is controlled by $z_\iota \in \mathfrak{N}_{V_1}$ with:
\begin{equation}
\| e^{-Q_\epsilon/2} z_\iota \zeta_\epsilon \|^2 
= \varphi_{V_1}(z_\iota \pi_{\zeta_\epsilon}(\mathcal{A}_{V_1}) z_\iota)
\leq \varphi_{V_1}(z_\iota^2)  < \infty
\end{equation}
So that for $R>1$ using $|\lambda| \leq R e^{|\lambda| - R}$ on the support $\lambda \leq -R$ gives:
\begin{equation}
 \int_{-\infty}^{-R} |\lambda| \left< \zeta_\epsilon \right| z_\iota d E_\lambda(Q_\epsilon) z_\iota \left| \zeta_\epsilon \right> \leq R e^{-R} \varphi_{V_1}(z_\iota^2)
\end{equation}

And large $\lambda >0$ is controlled by an $L_p$ norm estimate. Recall that $\chi_{\iota,k} = \eta_{V_1'}(A')$
for some $A' \in \mathfrak{N}_{V_1'}$ because of the gaussian smearing, see for example the discussion around \eqref{eq:seestep}. Hence $\chi_{\iota,k} \in L_\infty(\mathcal{A}_{V_1}') \cap \mathscr{H}$.
We use Lemma~\ref{lem:weighted-Lp-contraction} for $p=4$ and $\theta=\pi/2$, to show that $\zeta_\epsilon=
e^{-\epsilon P} \chi_{\iota,k} \in L_4(\mathcal{A}_{V_1}') \cap \mathscr{H}$ along with the explicit estimate:
\begin{equation}
\|\zeta_\epsilon \|_{L_4(\mathcal{A}_{V_1}')}^2 \leq \| \chi_{\iota,k} \|
\|\chi_{\iota,k} \|_{L_\infty(\mathcal{A}_{V_1}')}
\end{equation}
Combining this with an estimate from the variational form for the $L_p$ norms \eqref{eq:weighted-Lp-definition} with $\chi = \zeta_\epsilon/ \| \zeta_\epsilon\| $:
\begin{equation}
\| e^{Q_\epsilon/4} \zeta_\epsilon \|
= \| \Delta_{\eta_{V_1}, \zeta_\epsilon;V_1}^{-1/4} \zeta_\epsilon \| \leq \| \zeta_\epsilon\|^{1/2} \|\zeta_\epsilon \|_{L_4(\mathcal{A}_{V_1}')} \leq \| \chi_{\iota,k} \| \| A' \|^{1/2} 
\end{equation}

We need to insert $z_\iota$ into this bound. We will leverage the fact that $C \equiv \| [Q_\epsilon,z_\iota] \| = \|[\ln \Delta_{V_1},z_\iota]\| < \infty$. Consider: 
\begin{align}
\left\|  |Q_\epsilon|^{1/2} E_{[R,\infty)}(Q_\epsilon) z_\iota E_{(-\infty,0)}(Q_\epsilon)\right\|^2 & \leq
R \left\| E_{[R,\infty)}(Q_\epsilon) z_\iota E_{(-\infty,0)}(Q_\epsilon)\right\|^2  \\ &
\qquad  + 
\int_R^\infty d  r \left\|  E_{[r,\infty)}(Q_\epsilon) z_\iota E_{(-\infty,0)}(Q_\epsilon)\right\|^2
\end{align}
established writing the spectral representation of $Q_\epsilon  E_{[R,\infty)}(Q_\epsilon) $ and bounding
the integral.  Also consider the following integral relation:
\begin{align} \nonumber
&E_{[R,\infty)}(Q_\epsilon)( z_\iota - e^{-Q_\epsilon/2} z_\iota e^{Q_\epsilon/2}) E_{(-\infty,0)}(Q_\epsilon)
\\ & \qquad \qquad = \frac{1}{2}\int_0^1 dt E_{[R,\infty)}(Q_\epsilon) e^{-tQ_\epsilon/2} [Q_\epsilon,z_\iota] e^{tQ_\epsilon/2} E_{(-\infty,0)}(Q_\epsilon) 
\end{align}
allowing us to bound:
\begin{align}
\| E_{[R,\infty)}(Q_\epsilon) z_\iota E_{(-\infty,0)}(Q_\epsilon) \|&\leq 
\| [ z_\iota, \ln \Delta_{V_1}] \| \frac{2(1 - e^{-R/2})}{R} + \| z_\iota \| e^{-R/2} \\
&\leq 2(C +  e^{-1})/R
\end{align}
so that
\begin{equation}
\left\|  |Q_\epsilon|^{1/2} E_{[R,\infty)}(Q_\epsilon) z_\iota E_{(-\infty,0)}(Q_\epsilon)\right\|^2
\leq 8(C +  e^{-1} )^2 /R
\end{equation}
To bound the large $\lambda$ tail of the integral \eqref{eq:spectral-integral} we must consider:
\begin{align}
\| |Q_\epsilon|^{1/2} E_{[R,\infty)}(Q_\epsilon) z_\iota \zeta_\epsilon \| \nonumber
&\leq \| |Q_\epsilon|^{1/2} E_{[R,\infty)}(Q_\epsilon) z_\iota E_{(-\infty,0)}(Q_\epsilon) \zeta_\epsilon \| \\
& \qquad + \| |Q_\epsilon|^{1/2} E_{[R,\infty)}(Q_\epsilon) z_\iota E_{[0,\infty)}(Q_\epsilon) \zeta_\epsilon \| 
\end{align}
where we have already established an $\epsilon$ independent bound, that vanishes as $R\rightarrow \infty$ for the first.
For the second term we use:
\begin{align}
\| |Q_\epsilon|^{1/2} E_{[R,\infty)}(Q_\epsilon) z_\iota E_{[0,\infty)}(Q_\epsilon) \zeta_\epsilon \|
& \leq \frac{1}{\sqrt{R}} \| Q_\epsilon E_{[R,\infty)}(Q_\epsilon) z_\iota E_{[0,\infty)}(Q_\epsilon) \zeta_\epsilon \| \\
& \leq  \frac{1}{\sqrt{R}}\left( \| [z_\iota, \ln \Delta_{V_1}]\|\| \zeta_\epsilon \| + \| z_\iota \| \| Q_\epsilon E_{[0,\infty)}(Q_\epsilon) \zeta_\epsilon \| \right)\\
& \leq  \frac{1}{\sqrt{R}}\left( C + 4  \| A' \|^{1/2} \right) \| \chi_{\iota,k} \|
\end{align}
Hence the large $\lambda$ integral is bounded by:
\begin{equation}
 \int_{R}^{\infty} \lambda \left< \zeta_\epsilon \right| z_\iota d E_\lambda(Q_\epsilon) z_\iota \left| \zeta_\epsilon \right> \leq \frac{\| \chi_{\iota,k} \|^2}{R } \left( (1+2 \sqrt{2}) C + 4 \| A' \|^{1/2}  +2\sqrt{2} e^{-1}\right)^2
\end{equation}
These bounds establish that the right hand side of \eqref{eq:estabexist} is well defined and finite.  Also the fact
that these bounds were uniform in $\epsilon$ allows us to apply strong resolvent convergence \eqref{eq:thestatement} of the modular
operator $\Delta_{\eta_{V_1},\zeta_\epsilon;V_1}$ as $\lim_{\epsilon \rightarrow 0} \zeta_\epsilon = \chi_{\iota,k}$ to show the existence of the limit on the right hand side of \eqref{eq:estabexist}.
Hence:
\begin{equation}
\limsup_{\epsilon \rightarrow 0} D(\psi_{\iota,k,\epsilon}|_{V_1}) \leq  - \left< \chi_{\iota,k} \right| z_\iota 
\ln \Delta_{\eta_{V_1},\chi_{\iota,k};V_1} z_\iota \left| \chi_{\iota,k}\right>
\end{equation}

One final step shows that:
\begin{equation}
    \label{eq:combineswith}
\chi_{\iota,k} = g_k(\ln \Delta_{\eta_{V_1},x_\iota \psi;V_{1}}) x_\iota \psi
= Bx_\iota \psi \qquad B = \int ds \widetilde{g}_k(s) \Delta_{\eta_{V_1},x_\iota \psi;V_{1}}^{is} \Delta_{x_\iota\psi}^{-is} 
\end{equation}
where $\tilde{g}_k$ is the Fourier transform of $g_k$ and $B \in \mathcal{A}_{V_1}$. But we can estimate the norm of this operator $\|B\| \leq \int ds |\tilde{g}_k(s)|
= g_k(0)= 1$. So the domination condition between the induced states for $\mathcal{A}_{V_1}'$ gives $\ln \Delta_{\eta_{V_1},\chi_{\iota,k};V_1} \geq \ln \Delta_{\eta_{V_1},x_\iota \psi;V_1} $ which combines with \eqref{eq:combineswith}
to give \eqref{eq:remainsestablish}.
\end{proof}

\bibliography{entanglement}

\end{document}